\documentclass[prd,nofootinbib,onecolumn,preprint,11pt]{revtex4}

\usepackage{graphicx}
\usepackage{bm}
\usepackage{float}
\usepackage{subfigure}
\usepackage{amsmath}
\usepackage{amsfonts}
\usepackage{color}
\definecolor{darkblue}{rgb}{0.0, 0.0, 0.55}
\usepackage{slashed}
\usepackage{fancyhdr}
\usepackage{url}
\usepackage[colorlinks=true,linkcolor=black,citecolor=black,filecolor=black,urlcolor=black,unicode]{hyperref}

\newcommand{\GN}{G_{\rm N}}

\newcommand{\dd}{\text{d}}

\begin{document}

\title{The Transverse-Traceless Spin-2 Gravitational Wave Cannot Be A Standalone Observable Because It Is Acausal}

\author{Yi-Zen Chu$^{1,2}$ and Yen-Wei Liu$^1$}
\affiliation{
$\,^1$Department of Physics, National Central University, Chungli 32001, Taiwan \\
$\,^2$Center for High Energy and High Field Physics (CHiP), National Central University, Chungli 32001, Taiwan
}

\begin{abstract}
\noindent We show, through an explicit calculation of the relevant Green's functions, that the transverse-traceless (TT) portion of the gravitational perturbations of Minkowski spacetime and of spatially flat cosmologies with a constant equation-of-state $w$ receive contributions from their isolated matter source(s) outside the past null cone of the observer. This implies the TT gravitational wave (GW) cannot be a standalone observable -- despite widespread (apparent) claims in the gravitational wave literature to the contrary. About a Minkowski background, all 4 of the gauge-invariant variables -- the two scalars, one vector and tensor -- play crucial roles to ensure the spatial tidal forces encoded within the gauge-invariant linearized Riemann tensor are causal. These gravitational tidal forces do not depend solely on the TT graviton but rather on the causal portion of its acceleration. However, in the far zone radiative limit, the flat spacetime `TT' graviton Green's function does reduce to the causal `tt' ones, which are the ones commonly used to compute gravitational waveforms. Similar remarks apply to the spin-1 photon; for instance, the electric field does not depend solely on the photon, but is the causal part of its velocity. As is known within the quantum theory of photons and linearized gravitons, there are obstacles to the construction of simultaneously gauge-invariant and Lorentz-covariant descriptions of these massless spin-1 and spin-2 states. Our results transparently demonstrate that the quantum operators associated with the helicity-1 photon and helicity-2 linear graviton both violate micro-causality: namely, they do not commute outside the light cone in flat and cosmological spacetimes.
\end{abstract}

\maketitle

\tableofcontents

\section{Motivation}
\label{Section_Introduction}

Students of gravitational wave (GW) physics are taught that the key observable -- the fractional distortion of the arms of laser interferometers employed by detectors such as LIGO, Virgo, Kagra, etc. -- induced by the passage of a GW train generated by a distant astrophysical source, is directly proportional to the `transverse-traceless' portion of the metric perturbation. Specifically, in a weakly curved spacetime\footnote{The Greek indices $\mu, \nu,\ldots,$ run from $0$ to $d-1$, while the Latin ones $i, j, \ldots,$ run over spatial coordinates from $1$ to $d-1$, and the ``mostly plus'' sign convention for the metric is used, namely $\eta_{\mu\nu} = \text{diag}[-1,+1,\ldots,+1]$. Throughout this paper, the symmetrization and anti-symmetrization of indices are denoted by the symbols $(\ldots)$ and $[\ldots]$, respectively, e.g., $T_{(\mu\nu)} \equiv \frac12 \left( T_{\mu\nu} + T_{\nu\mu}\right)$ and $T_{[\mu\nu]} \equiv \frac12 \left( T_{\mu\nu}  -  T_{\nu\mu} \right)$.}
\begin{align}
\label{PerturbedMinkowski}
g_{\mu\nu}[t,\vec{x}] = \eta_{\mu\nu} + h_{\mu\nu}[t,\vec{x}],
\qquad\qquad
|h_{\mu\nu}| \ll 1 ;
\end{align}
if $X^i$ denotes the Cartesian coordinate vector joining one end of an interferometer arm to another, its change $\delta X^i$ due to a GW signal impinging upon the detector is often claimed to be\footnote{See, for example, eq.~(27.26) of Thorne and Blandford \cite{ThorneBlandfordBook}.}
\begin{align}
\label{FractionalDistortion_Textbooks}
\delta X^i = \frac{1}{2} h_{ij}^{\text{TT}} X^j ,
\end{align}
where $h_{ij}^{\text{TT}}$ is the `transverse-traceless' portion of the space-space components of $h_{\mu\nu}$ in eq.~\eqref{PerturbedMinkowski}. What does `transverse-traceless' really mean in this context? R\'{a}cz \cite{Racz:2009nq} and -- more recently -- Ashtekar and Bonga \cite{Ashtekar:2017wgq,Ashtekar:2017ydh} have pointed out, the GW literature erroneously uses two distinct notions of `transverse-traceless' interchangeably.\footnote{Frenkel and R\'{a}cz \cite{Frenkel:2014cra} have also pointed out a similar error within the electromagnetic context.} (We shall adopt Ashtekar and Bonga's notation of `TT' and `tt'.) On the one hand, there is one involving the divergence-free condition,
\begin{align}
\label{TT}
\partial_i h_{ij}^{\text{TT}} = \partial_i h_{ji}^{\text{TT}} = 0 = \delta^{ij} h_{ij}^{\text{TT}} ;
\end{align}
while on the other hand there is one involving a transverse-projection in position space,
\begin{align}
\label{tt}
h_{ij}^{\text{tt}} &\equiv P_{ij ab} h_{ab} .
\end{align}
The definition of transverse-projection in eq.~\eqref{tt} is based on the unit radial vector $\widehat{r} \equiv \vec{x}/|\vec{x}|$ pointing from the isolated astrophysical source centered at $\vec{0}$ to the observer at $\vec{x}$, namely
\begin{align}
\label{tt_Projector}
P_{ij ab} 	&\equiv P_{a(i} P_{j)b} - \frac{1}{d-2} P_{ij} P_{ab} , \\
P_{ij} 		&\equiv \delta_{ij} - \widehat{r}_i \widehat{r}_j
\label{t_Projector} .
\end{align}
Because the rank-2 object is a projector, in the sense that
\begin{align}
P_{ia} P_{aj} = P_{ij} ,
\end{align}
and is also transverse to the radial direction,
\begin{align}
\widehat{r}^i P_{ij} = 0 = P_{ij} \widehat{r}^j ,
\end{align}
we see that the `tt' GW in eq.~\eqref{tt} enjoys the same traceless condition as its `TT' counterpart in eq.~\eqref{TT} (i.e., $\delta^{ij} h_{ij}^{\text{tt}} = 0$) but is transverse to the unit radial vector
\begin{align}
\widehat{r}^i h_{ij}^{\text{tt}} = 0 = h_{ij}^{\text{tt}} \widehat{r}^j
\end{align}
instead of being divergence-free.

We believe the intent of much of the contemporary gravitational literature is to claim the TT GW, obeying eq.~\eqref{TT}, to be the
observable; while the tt one in eq.~\eqref{tt} to be only an approximate expression of the same gravitational signal when the observer is very far from the source.\footnote{The exception appears to be Thorne and Blandford \cite{ThorneBlandfordBook}, where they went straight to the tt form of the GW (see Box 27.2) without any discussion of gauge invariance whatsoever.} To our knowledge, the clearest enunciation of this stance may be found in the review by Flanagan and Hughes \cite{Flanagan:2005yc}. After describing how the TT piece of the gravitational perturbation of flat spacetime is the only gauge invariant portion that obeys a wave equation in \S 2.2 -- the remaining 2 scalars and one vector obey Poisson equations -- and after attempting to justify how the TT GW is the one appearing in eq.~\eqref{FractionalDistortion_Textbooks} (cf.~eq.~(3.12) of \cite{Flanagan:2005yc}) they went on in \S 4.2 to assert, albeit without justification, that the far zone version of this TT GW in fact reduces to the tt one. In eq.~(4.23), they then expressed the final GW quadrupole formula in the latter tt form.

Other pedagogical discussions of GWs usually begin with the homogeneous TT wave solutions in perturbed Minkowski spacetime completely devoid of matter:
\begin{align}
\label{TT_Spin2Waves}
h_{ij}^{\text{TT}}[t,\vec{x}]
&= \int_{\mathbb{R}^{d-1}} \frac{\dd^{d-1}\vec{k}}{(2\pi)^{d-1}} \left( \widehat{P}_{ij ab}[\vec{k}] \epsilon_{ab}[\vec{k}] e^{ik_\mu x^\mu} + \text{c.c.} \right) , \\
k_\mu &\equiv \left( - |\vec{k}|, \, k_i \right), \qquad\qquad
k^2 \equiv k_\sigma k^\sigma = 0 ;
\end{align}
\footnote{See, for instance, \S 35.2-35.4 of Misner, Thorne, Wheeler \cite{Misner:1974qy}; \S 10.2 of Weinberg \cite{Weinberg:1972kfs}; or \S 9.1 of Schutz \cite{Schutz:1985jx}. Briefly, one may start with the de Donder gauge condition $\partial^\mu h_{\mu\nu} = (1/2) \partial_\nu h$, where $h \equiv \eta^{\mu\nu} h_{\mu\nu}$, and solve the linearized vacuum Einstein's equations $\partial^2 (h_{\mu\nu} - (1/2) \eta_{\mu\nu} h) = 0$. By performing a gauge transformation in Fourier spacetime to set to zero the $h_{\mu 0} = h_{0 \mu}$ components, one would arrive at eq.~\eqref{TT_Spin2Waves}. Often, eq.~\eqref{TT} is called the `TT gauge' but texts often do not caution the reader that this gauge condition can no longer be imposed once matter is introduced into the setup, i.e., once the very source of GWs is present. (Maggiore \cite{Maggiore:1900zz} does note that `TT gauge' does not exist inside the source, but goes on to impose it anyway in the far zone.) We wish to reiterate the remarks already made in \S C of \cite{Chu:2016ngc}: if the `TT gauge' were to exist, that would mean the initially non-trivial gauge invariant scalars and vector variables may be coordinate-transformed to zero. Furthermore, the misleading use of the phrase `TT gauge' suggests one may even choose a different gauge to compute GW patterns -- after all, one ought to be able to use any desired coordinate system -- but this cannot be the case, as the GW pattern is an observable and must therefore yield a unique result.}where ``c.c." denotes the complex conjugate of the preceding term; $\epsilon_{ab}$ can be viewed as the purely spatial gravitational wave amplitude tensor; and the projector is now one in momentum/Fourier space,
\begin{align}
\label{GravitonProjector}
\widehat{P}_{ij ab}[\vec{k}]
&\equiv \widehat{P}_{a(i} \widehat{P}_{j)b} - \frac{1}{d-2} \widehat{P}_{ij} \widehat{P}_{ab} , \\
\widehat{P}_{ij}
&\equiv \delta_{ij} - \widehat{k}_i \widehat{k}_j,
\qquad\qquad \widehat{k}_i \equiv \frac{k_i}{|\vec{k}|} .
\label{TransverseProjector}
\end{align}
Because in Fourier space a spatial derivative is replaced with a momentum vector, $\partial_j \to i k_j$, and because of the transverse-traceless properties
\begin{align}
k^i \widehat{P}_{ij ab} = k^i \widehat{P}_{ji ab} = 0 = \delta^{ij} \widehat{P}_{ij ab} ;
\end{align}
the perturbations in eq.~\eqref{TT_Spin2Waves} do indeed satisfy the `TT' conditions in eq.~\eqref{TT}.

These GW discussions typically go on to justify eq.~\eqref{FractionalDistortion_Textbooks} in vacuum before, as opposed to after, solving the perturbations engendered by a non-trivial source. The excuse is that, one expects these perturbations from an isolated system to approach TT plane GWs in the asymptotic far zone limit. As we shall see below, the TT and tt GWs do indeed coincide in this $r \equiv |\vec{x}| \to \infty$ limit. Hence, one might reasonably question: why bother with the distinction at all? To this end, Ashtekar and Bonga point out that the tt GWs miss the `Coulombic aspects' that are contained in the TT ones. Moreover, in Quantum Field Theory, each mode of the superposition of TT GWs in eq.~\eqref{TT_Spin2Waves} and not those of $h_{ij}^{\text{tt}}$ -- because it is the former that is fully gauge-invariant -- would be regarded as an irreducible spin-2 graviton. Therefore, one may be led to a principled stance and insist that it is $h_{ij}^{\text{TT}}$ that is physical.

But is the TT GW really a standalone observable? One of us (YZC) has been confused by this issue since several years ago, when he began developing a program to explore novel ways to understand the causal structure of gravitational signals in curved spacetimes -- i.e., how they propagate both on and within the null cone. As highlighted in \cite{Flanagan:2005yc}, the gauge invariant TT GW is a nonlocal function of the metric perturbation $h_{\mu\nu}$ in eq.~\eqref{PerturbedMinkowski}, because the TT projection process takes place in Fourier space. Since, at least about a Minkowski background, the de Donder gauge gravitational perturbation depends on its matter source in a causal manner, this suggests the TT GW may thus depend on the same matter source in an acausal manner due to this nonlocal character. This in turn would render it unphysical, as no classical physical observable should arise from outside the past light cone of the observer.

In this paper, we wish to clarify how the gauge-invariant forms of the vector potential and metric perturbations of, respectively, electromagnetism and linearized gravitation contribute to the observables of these theories. This will include understanding how all their gauge-invariant field variables, not just the dynamical massless spin-1 and spin-2 ones, play crucial roles in ensuring that their physical observables depend on their progenitors -- namely, the electric current and matter stress tensor -- in a causal manner. Through a concrete evaluation of the massless spin-1 photon and spin-2 graviton Green's functions, in Minkowski and spatially flat cosmological spacetimes, we will show that they are indeed acausally dependent on these sources and therefore cannot be standalone observables. However, by ensuring the rest of the gauge-invariant variables are included in the computation of the electromagnetic Faraday tensor $F_{\mu\nu}$ as well as the $\delta_1 R_{0i0j}$ components of the linearized Riemann tensor in Minkowski and the $\delta_1 C^i{}_{0j0}$ components of the linearized Weyl tensor in spatially flat cosmologies, the electromagnetic and gravitational tidal forces become strictly causal ones. In particular, we gain the following insight into the gauge-invariant content of electromagnetism and linearized General Relativity. While the magnetic field $F_{ij}$ does depend only on the massless spin-1 photon, the electric field $F_{0i}$ depends on the causal portion of the velocity of the photon, with its acausal portion canceled by the gauge-invariant scalar of the vector potential, in all spacetime dimensions $d \geq 3$. For the gravitational case, tidal forces in a flat spacetime background are encoded within the causal part of the acceleration of the massless spin-2 graviton, with the acausal portion eliminated by the two gauge-invariant scalars and one vector potential for all $d \geq 4$. Additionally, about a cosmological background, if the Weyl tensor describes the dominant contributions to tidal forces, then the latter appear to depend on the causal portions of both the massless spin-2 and the two gauge-invariant Bardeen scalar potentials. We view this latter analysis as a first step towards an understanding of whether the two Bardeen gauge-invariant scalar potentials ought to be considered an integral portion of gravitational waves and their associated memories in cosmological settings -- even though the dynamics of General Relativity (in $3+1$ dimensions) is usually attributed exclusively to its two spin-2 degrees of freedom.

In \S \eqref{Chapter_Observables} we will define the electromagnetic and gravitational gauge invariant variables; and proceed to clarify what the relevant (classical) electromagnetic and gravitational observables are. In \S \eqref{WhyAcausal} we will use the non-local character of the transverse projection in momentum space to argue why these gauge-invariant variables are expected to be acausal. Following which, we begin in \S \eqref{MinkowskiSpacetime} to compute the explicit forms of the transverse-photon and TT graviton Minkowski Green's functions, confirming their acausal nature. We also compute the solutions to the gauge-invariant scalars and vectors; and combine the results to study how the electromagnetic Faraday tensor and gravitational linearized Riemann are causally dependent on their respective sources. The far zone and stationary limit are examined; and micro-causality violated is pointed out. In \S \eqref{SpatiallyFlatw}, we move on to study similar issues but in a cosmology dominated by a cosmological constant or driven by a relativistic fluid with equation of state $0 < w \leq 1$. Finally, we summarize our findings and outline future directions in \S \eqref{Section_Summary}.

\section{Gauge-Invariance and Observables}
\label{Chapter_Observables}

{\bf Setup} \qquad Throughout the rest of this paper, we will be studying the $d-$dimensional perturbed Friedmann-Lema\^{i}tre-Robertson-Walker (FLRW)-like metric
\begin{align}
\label{PerturbedFLRW}
g_{\mu\nu}[x] = a[\eta]^2 \left( \eta_{\mu\nu} + \chi_{\mu\nu}[x] \right) ,
\qquad\qquad
x^\mu \equiv (\eta,\vec{x}) ;
\end{align}
where $a=1$ for a flat background or
\begin{align}
\label{ScaleFactor}
a[\eta] = \left(\frac{\eta}{\eta_0}\right)^{\frac{2}{q_w}}, \qquad\qquad
q_w \equiv (d-3) + (d-1) w .
\end{align}
In eq.~\eqref{ScaleFactor}, if the perturbations $\chi_{\mu\nu}$ were not present, setting $w=-1$ with $\eta<0$ yields de Sitter spacetime and $0 \leq w \leq 1$ with $\eta>0$ a spatially flat cosmology driven by a single perfect fluid with equation-of-state $w$. The non-trivial perturbations $\chi_{\mu\nu}$ satisfy Einstein's equations coupled to the fluid plus a compact astrophysical system, linearized about the corresponding backgrounds. The detailed analysis can be found in \S III and \S IV in \cite{Chu:2016ngc}, and we will cite the relevant results below.

Let us consider an infinitesimal coordinate transformation
\begin{align}
\label{GaugeTransformation_Gravity}
x^\mu = x'^\mu + \xi^\mu[x'] ,
\end{align}
where $x' \equiv (\eta',\vec{x}')$ and $\xi^\mu$ is small in the same sense that $h_{\mu\nu}$ is small. Then up to first order in perturbations, the metric tensor in the primed coordinate system may be written as
\begin{align}
\label{PerturbedFLRW_Prime}
g_{\mu'\nu'}[\eta',\vec{x}'] = a[\eta']^2 \left( \eta_{\mu\nu} + \chi_{\mu'\nu'}[\eta',\vec{x}'] \right) ,
\end{align}
with all the coordinate transformation induced by eq.~\eqref{GaugeTransformation_Gravity} attributed to that of the metric perturbation in the following way:
\begin{align}
\label{GaugeTransformation_Perturbation}
\chi_{\mu'\nu'}[x']
&= \chi_{\mu\nu}[x'] + 2\eta_{\sigma(\mu}\partial_{\nu')} \xi^\sigma[x'] + 2\frac{\dot{a}[\eta']}{a[\eta']} \xi^0[x'] \eta_{\mu\nu} .
\end{align}
(The $\chi_{\mu\nu}[\eta',\vec{x}']$ and $\xi^{\mu}[x']$ on the right hand side of eq.~\eqref{GaugeTransformation_Perturbation} are the perturbation and gauge-transformation vector components in the `old' $x^\mu \equiv (\eta,\vec{x})$ coordinate basis, but with the replacement $x \to x'$.) Next, we perform a scalar-vector-tensor decomposition of both the metric perturbations\footnote{See, for example, \S IV in \cite{Chu:2016ngc} for a discussion of the scalar-vector-tensor decomposition and the gauge-invariant formalism of linearized gravitation.}
\begin{align}
\label{Bardeen_SVTDecomposition}
\chi_{00} &\equiv E, \qquad \qquad \chi_{0i} \equiv \partial_i F + F_i, \nonumber\\
\chi_{ij} &\equiv D_{ij} + \partial_{( i} D_{j) } + \frac{D}{d-1} \delta_{ij} + \left( \partial_i \partial_j - \frac{\delta_{ij}}{d-1} \vec{\nabla}^2 \right) K ,
\end{align}
as well as the astrophysical stress tensor
\begin{align}
\label{Astro_SVT_IofII}
\,^{(\text{a})}T_{00} & \equiv \rho, \qquad\qquad
\,^{(\text{a})}T_{0i} \equiv \Sigma_i + \partial_i \Sigma, \\
\label{Astro_SVT_IIofII}
\,^{(\text{a})}T_{ij} & \equiv \sigma_{ij} + \partial_{ (i } \sigma_{ j ) } + \frac{\sigma}{d-1} \delta_{ij}
+ \left( \partial_i \partial_j - \frac{\delta_{ij}}{d-1} \vec{\nabla}^2 \right) \Upsilon ;
\end{align}
where these variables subject to the following constraints
\begin{align}
\label{Bardeen_SVTConstraints}
\partial_i F_i &= \partial_i D_i = 0 = \delta^{ij} D_{ij} = \partial_i D_{ij} , \\
\label{AstroTmunu_SVTConstraints}
\partial_i \Sigma_i &= \partial_i \sigma_i = 0 = \delta^{ij} \sigma_{ij} = \partial_i \sigma_{ij} .
\end{align}
We may then gather the following are gauge-invariant at first order in perturbations:
\begin{align}
\label{Bardeen_Psi}
\Phi &\equiv -\frac E2 + \frac{1}{a} \partial_0 \left\{ a \left( F - \frac{\dot{K}}{2} \right) \right\} , \\
\label{Bardeen_Phi}
\Psi &\equiv -\frac{D - \vec{\nabla}^2 K}{2(d-1)} -  \frac{\dot{a}}{a} \left( F - \frac{\dot{K}}{2} \right) , \\
\label{Bardeen_VandDij}
V_i &\equiv F_i - \frac{\dot{D}_i}2
\qquad \qquad \text{ and }\qquad \qquad
D_{ij} \equiv \chi_{ij}^\text{TT} .
\end{align}
\footnote{Notice that the sign convention for the metric as well as certain gauge-invariant variables are defined differently in \cite{Chu:2016ngc}. To change the gauge-invariant notations into those employed in \cite{Chu:2016ngc}, we follow the conversions: $\Phi[\text{here}] \to \Psi[\text{\cite{Chu:2016ngc}}]/2$, $\Psi[\text{here}] \to \Phi[\text{\cite{Chu:2016ngc}}]/2$, $V_i[\text{here}] \to -V_i[\text{\cite{Chu:2016ngc}}]$, and $D_{ij}[\text{here}] \to - D_{ij}[\text{\cite{Chu:2016ngc}}]$.}Within the cosmological case, the solution of $D_{ij}$ can be found in eq.~(111), that of $V_i$ in eq.~(119) and those of $\Psi$ in equations (123), (125) and (130) of \cite{Chu:2016ngc}. $\Phi$ and $\Psi$ are related through
\begin{align}
\label{PhiPsiRelationship}
(d-3) \Psi - \Phi = 8\pi \GN \Upsilon .
\end{align}
Within the Minkowski case, on the other hand, eq.~\eqref{PhiPsiRelationship} still holds but the solution of $\Psi$ can be found in eq.~(A27); that of $V_i$ in eq.~(A28); and that of $D_{ij}$ in eq.~(A29) of \cite{Chu:2016ngc}. As already alluded to, of all the gauge invariant variables in a flat background, only the tensor $D_{ij}$ admits wave solutions.

Because of the TT constraints in eq.~\eqref{Bardeen_SVTConstraints}, note that the tensor mode $D_{ij}$ exists only for $d \geq 4$. The apparent physical importance of these field variables in equations \eqref{Bardeen_Psi}, \eqref{Bardeen_Phi}, and \eqref{Bardeen_VandDij} lies in the fact that, if some observable can be expressed in terms of them, then the same observable cannot be rendered trivial merely by a small change in coordinates since $\Phi$, $\Psi$, $V_i$ and $D_{ij}$ will remain invariant.

When dealing with electromagnetism, we will set to zero the perturbations $\chi_{\mu\nu}$ in eq.~\eqref{PerturbedFLRW} and proceed to solve Maxwell's equations
\begin{align}
\nabla_\nu F^{\mu\nu} = J^\mu,
\qquad\qquad
F_{\mu\nu} \equiv 2\partial_{[\mu} A_{\nu]} .
\label{Maxwell_equations}
\end{align}
Under the gauge transformation
\begin{align}
\label{Maxwell_GaugeTransformation}
A_\mu \to A_\mu - \partial_\mu \mathcal{C} ,
\end{align}
the vector potential $A_\mu$ leaves the Faraday tensor $F_{\mu\nu}$ invariant
for an arbitrary function $\mathcal{C}$. If we perform a scalar-vector decomposition of the vector potential
\begin{align}
A_i \equiv \alpha_i + \partial_i \alpha, \qquad\qquad \partial_i \alpha_i = 0 ,
\end{align}
and that of the electric current
\begin{align}
J_0 & \equiv -\rho , \\
J_i & \equiv \Gamma_i + \partial_i \Gamma, \qquad\qquad \partial_i \Gamma_i = 0 ;
\end{align}
we may proceed to identify the following gauge-invariant scalar $\Phi$ and transverse (helicity$-1$) photon $\alpha_i$:
\begin{align}
\label{Maxwell_GaugeInvariant_ScalarVector}
\Phi \equiv  \dot{\alpha} - A_0
\qquad \text{and} \qquad
\alpha_i \equiv A_i^\text{T} .
\end{align}
\footnote{We highlight that, in the electromagnetic case here, some conventions are defined differently relative to \cite{Chu:2016ngc}. For instance, when switching to the gauge-invariant variables used in \cite{Chu:2016ngc}, we may implement the following conversions: $\alpha_i[\text{here}] \to -\alpha_i[\text{\cite{Chu:2016ngc}}]$, $\Phi[\text{here}] \to \Phi[\text{\cite{Chu:2016ngc}}]$, $\rho[\text{here}] \to a^2 \rho[\text{\cite{Chu:2016ngc}}]$, and $\Gamma_i[\text{here}] \to - a^2 \Gamma_i[\text{\cite{Chu:2016ngc}}]$.}In terms of these variables, the Faraday tensor reads
\begin{align}
\label{Maxwell_FaradayTensor}
F_{0i} = \dot{\alpha}_i + \partial_i \Phi,
\qquad\qquad
F_{ij} = 2\partial_{[i} \alpha_{j]} .
\end{align}
We refer the reader to \S V of \cite{Chu:2016ngc} for further details; again, we will cite the relevant results below.

{\bf Electromagnetic Observables} \qquad In classical electromagnetism, it is the electric $F_{0i} = -F_{i0}$ and magnetic $F_{ij}$ fields in eq.~\eqref{Maxwell_FaradayTensor} that are regarded as observables, because they provide the forces on electrically charged systems. We believe the situation for gravity is more subtle, however.

{\bf Gravitational Observables: (Simplified) Weber bar} \qquad Let us begin with a small lump of freely falling material acting as a Weber-bar detector of GWs. In what follows, the assumption of freely falling detectors makes it technically advantageous to describe their trajectories using the synchronous gauge metric -- where the perturbations are purely spatial:
\begin{align}
\label{SynchronousGauge}
g_{\mu\nu} = a^2 \left( \eta_{\mu\nu} \dd x^\mu \dd x^\nu + \chi_{ij}^{\text{(Synch)}} \dd x^i \dd x^j \right) .
\end{align}
For, if the particles comprising the detectors experience negligible inter-particle forces, then the synchronous gauge in eq.~\eqref{SynchronousGauge} can be chosen such that each of them would in fact have constant spatial trajectories; specifically, for the $i$th particle its timelike geodesic reads
\begin{align}
Z_{(i)}^\mu = \left( \eta, \vec{Z}_{(i)} \right),
\qquad\qquad
\text{$\vec{Z}_{(i)}$ constant} .
\end{align}
The tidal forces due to a passing GW acting between an infinitesimally nearby pair of particles, whose worldlines are joined by $\ell^\mu$, is given by the geodesic deviation equation
\begin{align}
\label{GeodesicDeviation}
U^\alpha U^\beta \nabla_\alpha \nabla_\beta \ell^\mu
&= -R^\mu_{\phantom{\mu}\nu \alpha\beta} U^\nu \ell^\alpha U^\beta .
\end{align}
The $U^\mu = a^{-1} \delta^\mu_0$ in eq.~\eqref{GeodesicDeviation} is the timelike geodesic vector tangent to either one of the two worldlines. For the freely falling scenario at hand, it is in fact consistent to choose $\ell^\mu$ to be purely spatial, i.e., $\ell^0 = 0$, so that eq.~\eqref{GeodesicDeviation} becomes
\begin{align}
\label{GeodesicDeviation_FreeFall}
\nabla_0 \nabla_0 \ell^i
&= -R^i_{\phantom{\mu}0 j0} \ell^j .
\end{align}
{\it Minkowski analysis} \qquad In a flat background, $a=1$, we witness from eq.~\eqref{GeodesicDeviation_FreeFall} that the $\delta_1 R_{0i0j}$ components of the linearized Riemann tensor must therefore describe the first-order tidal forces between infinitesimally nearby particles within our idealized Weber bar. Any inter-particle (electromagnetic) forces that are responsible for holding the lump of material together must therefore counter the $\delta_1 R_{0i0j}$ on the right hand side of eq.~\eqref{GeodesicDeviation_FreeFall}. Moreover, as long as our Weber bar's proper size is very small compared to that of the GW wavelength, the physical pattern of rarefaction and compression of the material asserted by the GW's passage must also be encoded entirely within $\delta_1 R_{0i0j}$. Now, not only do these components carry physical meaning, the entire linearized Riemann tensor is in fact gauge invariant because the full Riemann tensor is zero when evaluated on the background $g_{\mu\nu} = \eta_{\mu\nu}$. These reasons explain why we will, in the coming sections, compute $\delta_1 R_{0i0j}$ within the gauge invariant formalism:
\begin{align}
\label{LinearizedRiemann_Minkowski_GaugeInvariant}
\delta_1 R_{0i 0j}
= \delta_{ij} \ddot{\Psi} + \partial_i \partial_j \Phi + \partial_{(i} \dot{V}_{j)} - \frac12 \ddot{D}_{ij}  ,
\qquad (a=1).
\end{align}

{\it Cosmological background} \qquad If our freely falling particles were in a cosmological background, the geodesic deviation equation \eqref{GeodesicDeviation_FreeFall} continues to be applicable. However, the linearized Riemann tensor is no longer gauge invariant because its background value, when $g_{\mu\nu}=a^2\eta_{\mu\nu}$, is no longer zero. This renders its physical interpretation more subtle. On the other hand, the traceless portion of the Riemann tensor, i.e., the Weyl tensor $C^\mu{}_{\nu \alpha\beta}$, is conformally invariant. This means $C^\mu{}_{\nu \alpha\beta}[g_{\mu\nu}=a^2\eta_{\mu\nu}]=0$ and the $\delta_1 C^i{}_{0j0}$ components of the linearized Weyl tensor is gauge invariant. It may be possible to argue that $\delta_1 C^i{}_{0j0}$ provides the dominant contribution to tidal forces in cosmology -- for, in flat spacetime, it is exactly equivalent to the Riemann tensor whenever the zero cosmological constant form of Einstein's equations holds and the Weber bar is in a vacuum region\footnote{See, for e.g., eq.~(14) of \cite{Chu:2016ngc} or eq.~\eqref{RiemannWeyl}.} -- but we shall leave the detailed analysis of this cosmological case to future work \cite{ChuLiuInPrep}.

{\bf Gravitational observable: (Simplified) Laser Interferometer} \qquad We move on to consider a toy model of a freely falling laser interferometer. If we assume the proper size of the interferometer is small compared to the GW wavelength, it is reasonable to then state the observed interference pattern will be proportional to the differences in its arm-lengths. As argued in \S C of \cite{Chu:2016ngc}, we may again employ the synchronous gauge in eq.~\eqref{SynchronousGauge} to compute the time dependent proper distance between two ends $(\eta,\vec{Y}_0)$ and $(\eta,\vec{Z}_0)$ of a single arm. (Remember the $\vec{Y}_0$ and $\vec{Z}_0$ here are constants.) Using Synge's world function, and assuming the interferometer is turned on at $\eta'$ but does not operate over cosmological timescales, the fractional distortion of this $\vec{Y} \leftrightarrow \vec{Z}$ arm is (cf.~eq.~(6) of \cite{Chu:2016ngc})
\begin{align}
\label{FractionalDistortion_AtConstantCosmicTime_FirstOrderHuman}
\left(\frac{\delta L}{L_0}\right)[\eta > \eta']
= \frac{\widehat{n}^i \widehat{n}^j}{2} \int_{0}^{1} \Delta\chi_{ij}^{\text{(Synch)}} \dd \lambda
+ \mathcal{O}\left[ \left(\chi_{mn}^{\text{(Synch)}}\right)^2,(\eta-\eta') \dot{a}[\eta']/a[\eta'] \right],
\end{align}
where $\widehat{n} \equiv (\vec{Z}_0 - \vec{Y}_0)/|\vec{Z}_0 - \vec{Y}_0|$ is the unit radial vector pointing from one mass to the other; and the $\lambda$-integral involves a Euclidean straight line between the two end points:
\begin{align}
\label{FractionalDistortion_AtConstantCosmicTime_Deltachiij}
\Delta \chi_{ij}^{\text{(Synch)}}
\equiv
\chi_{ij}^{\text{(Synch)}}\left[ \eta, \vec{Y}_0 + \lambda \left(\vec{Z}_0-\vec{Y}_0\right) \right]
- \chi_{ij}^{\text{(Synch)}}\left[ \eta', \vec{Y}_0 + \lambda \left(\vec{Z}_0-\vec{Y}_0\right) \right] .
\end{align}
{\it Minkowski analysis} \qquad Now, in the synchronous gauge of eq.~\eqref{SynchronousGauge}, the linearized Riemann tensor reads
\begin{align}
\label{LinearizedRiemann_Minkowski_SynchGauge}
\delta_1 R_{0i 0j}
= -\frac{1}{2} \ddot{\chi}_{ij}^{\text{(Synch)}} .
\end{align}
We may thus solve for the synchronous gauge perturbation needed in eq.~\eqref{FractionalDistortion_AtConstantCosmicTime_FirstOrderHuman} by first connecting it to the gauge-invariant variables. This, in turn, is achieved by exploiting the gauge-invariance of the linearized Riemann tensor in a Minkowski background. In other words, since equations \eqref{LinearizedRiemann_Minkowski_GaugeInvariant} and \eqref{LinearizedRiemann_Minkowski_SynchGauge} refer to the same object, we have
\begin{align}
\label{SynchronousGauge_to_GaugeInvariant}
\ddot{\chi}_{ij}^{\text{(Synch)}}
&= \ddot{D}_{ij} - 2\partial_{(i} \dot{V}_{j)} - 2\delta_{ij} \ddot{\Psi} - 2 \partial_i \partial_j \Phi,
\qquad (a=1) .
\end{align}
Equations \eqref{LinearizedRiemann_Minkowski_GaugeInvariant}, \eqref{LinearizedRiemann_Minkowski_SynchGauge} and \eqref{SynchronousGauge_to_GaugeInvariant} inform us that the fractional distortion formula in eq.~\eqref{FractionalDistortion_AtConstantCosmicTime_FirstOrderHuman} is therefore -- at least in principle -- related to the double time integral of the linearized Riemann tensor itself.\footnote{See, for example, equations (27.22) and (27.24) of \cite{ThorneBlandfordBook}.} In any event, as long as the GW detector is sufficiently far enough from the astrophysical source, we may take the far zone limit of the right hand side of eq.~\eqref{SynchronousGauge_to_GaugeInvariant}. Below, we will use methods different from those in \cite{Ashtekar:2017wgq,Ashtekar:2017ydh} to argue that, this far zone limit yields
\begin{align}
\label{SynchronousGauge_to_GaugeInvariant_StepI}
\left(\ddot{D}_{ij} - 2\partial_{(i} \dot{V}_{j)} - 2\delta_{ij} \ddot{\Psi} - 2 \partial_i \partial_j \Phi\right)_{\text{far zone}}
= \ddot{D}_{ij}[\text{far zone}]
= \ddot{\chi}_{ij}^{\text{tt}}[\text{far zone}] ,
\qquad
(a=1);
\end{align}
where $\chi_{ij}^{\text{tt}}$ is the tt projection of the de Donder gauge solution as $r \equiv |\vec{x}| \to \infty$. Comparing equations \eqref{SynchronousGauge_to_GaugeInvariant} and \eqref{SynchronousGauge_to_GaugeInvariant_StepI} allows us to deduce:
\begin{align}
\label{SynchronousGauge_to_GaugeInvariant_FullAnswer}
\chi_{ij}^{(\text{Synch})}[\eta,\vec{x}]
&= \chi_{ij}^{\text{tt}}[\eta,\vec{x}]
+ (\eta-\eta') \mathcal V_{ij}[\eta',\vec{x}]
+ \mathcal W_{ij}[\eta',\vec{x}],
\end{align}
where $\mathcal V_{ij}$ and $\mathcal W_{ij}$ are the two undetermined initial conditions at $\eta'$, and, on both sides, the far zone limit has been taken. One issue that is often not addressed is, why the initial conditions -- the last two terms on the right hand side of eq.~\eqref{SynchronousGauge_to_GaugeInvariant_FullAnswer} -- are usually neglected. We will take the perspective that realistic GW detectors are sensitive to waves within a limited bandwidth, and since the second and third terms of eq.~\eqref{SynchronousGauge_to_GaugeInvariant_FullAnswer} are zero-frequency `waves' one may ignore their contributions. More explicitly, the ($\omega-$)frequency transform of eq.~\eqref{SynchronousGauge_to_GaugeInvariant_FullAnswer} becomes
\begin{align}
\label{SynchronousGauge_to_GaugeInvariant_FullAnswer_FreqSpace}
\widetilde{\chi}_{ij}^{(\text{Synch})}[\omega,\vec{x}]
&= \widetilde{\chi}_{ij}^{\text{tt}}[\omega,\vec{x}]
- (2\pi i) \delta'[\omega]e^{i\omega\eta'} \mathcal V_{ij}[\eta',\vec{x}]
+ (2\pi) \delta[\omega] \mathcal W_{ij}[\eta',\vec{x}] ,
\end{align}
where $\delta[\omega]$ and $\delta'[\omega]$ are the Dirac delta function and its derivative.

We close this section by providing an expedited method to obtain the synchronous gauge metric perturbation from the gauge-invariance of the linearized Riemann tensor, at least in $(3+1)$ dimensional Minkowski spacetime. The key is that the de Donder gauge $\partial^\mu \chi_{\mu\nu} = (1/2) \partial_\nu \chi$, with $\chi \equiv \eta^{\rho\sigma} \chi_{\rho\sigma}$, allows us to solve $\chi_{\mu\nu}$ rather easily with the massless scalar Green's function. In terms of the `trace-reversed' variable
\begin{align}
\overline{\chi}_{\mu\nu}
&\equiv \chi_{\mu\nu} - \frac{1}{2} \eta_{\mu\nu} \chi,
\qquad
\overline{\chi} = \eta^{\rho\sigma} \overline{\chi}_{\rho\sigma} ;
\end{align}
the linearized Einstein's equations lead us to the following far zone solution:
\begin{align}
\label{deDonder_4D_FarZone}
\overline{\chi}_{\mu\nu}[\eta,\vec{x}]
&= \frac{4\GN}{r} \int_{\mathbb{R}^3} \dd^3 \vec{x}'
T_{\mu\nu}[\eta-r+\vec{x}'\cdot\widehat{r},\vec{x}'] ,
\qquad
\widehat{r} \equiv \frac{\vec{x}}{|\vec{x}|} .
\end{align}
(Note: we have chosen $\vec{0}$ to lie within the astrophysical system.) We also have, in de Donder gauge, the following components of the linearized Riemann tensor
\begin{align}
\label{LinearizedR0i0j_deDonder}
\delta_1 R_{0i0j}
&= \frac{1}{2} \left( 2\partial_l \partial_{(i} \overline{\chi}_{j)l} - \partial_0^2 \left( \overline{\chi}_{ij} - \frac{1}{2} \delta_{ij} \overline{\chi}_{ll} \right) - \frac{1}{2} \delta_{ij} \partial_l \partial_m \overline{\chi}_{lm} - \frac{1}{2} \partial_i \partial_j \overline{\chi}_{ll} - \frac{1}{2} \partial_i \partial_j \overline{\chi}_{00} \right) .
\end{align}
Upon inserting eq.~\eqref{deDonder_4D_FarZone} into eq.~\eqref{LinearizedR0i0j_deDonder}, one may recognize that every spatial derivative acting on $\overline{\chi}_{\mu\nu}$ may be replaced with a time derivative, via
\begin{align}
\partial_i \to -\widehat{r}^i \partial_0 .
\end{align}
In the far zone limit we are working in, the error incurred by this replacement scales as (time-scale of source)/(observer-source distance) or (characteristic size of source)/(observer-source distance), both of which are small by assumption. At this point, one would find that eq.~\eqref{LinearizedR0i0j_deDonder} has been massaged into
\begin{align}
\label{LinearizedR0i0j_deDonder_FarZone}
\delta_1 R_{0i0j}
&= -\frac{1}{2} \ddot{\chi}_{ij}^{\text{tt}} ,
\end{align}
where the tt perturbation is the following projection (cf.~eq.~\eqref{tt_Projector}) of the de Donder gauge solution
\begin{align}
\chi_{ij}^{\text{tt}} = P_{ij ab} \overline\chi_{ab}[\text{de Donder}] .
\end{align}
Comparing equations \eqref{LinearizedRiemann_Minkowski_SynchGauge} and \eqref{LinearizedR0i0j_deDonder_FarZone} now hands us, for finite frequency $\omega$,
\begin{align}
\widetilde{\chi}_{ij}^{(\text{Synch})}[\omega,\vec{x}]
= \widetilde{\chi}_{ij}^{\text{tt}}[\omega,\vec{x}] \qquad
(\text{Far zone}) .
\end{align}
Finally, we once again leave to future work \cite{ChuLiuInPrep} the connection between the cosmological synchronous gauge metric perturbation to its gauge-invariant counterparts.

\section{Why are the massless spin-1 photon and spin-2 graviton acausal?}
\label{WhyAcausal}

Before we proceed to tackle the computations of the massless spin-1 and spin-2 Green's functions, let us first explain why acausality is to be expected. Since this section is meant to be heuristic, we shall be content to work strictly in a Minkowski background.

\begin{figure}[!ht]
\begin{center}
\includegraphics[width=5.5in]{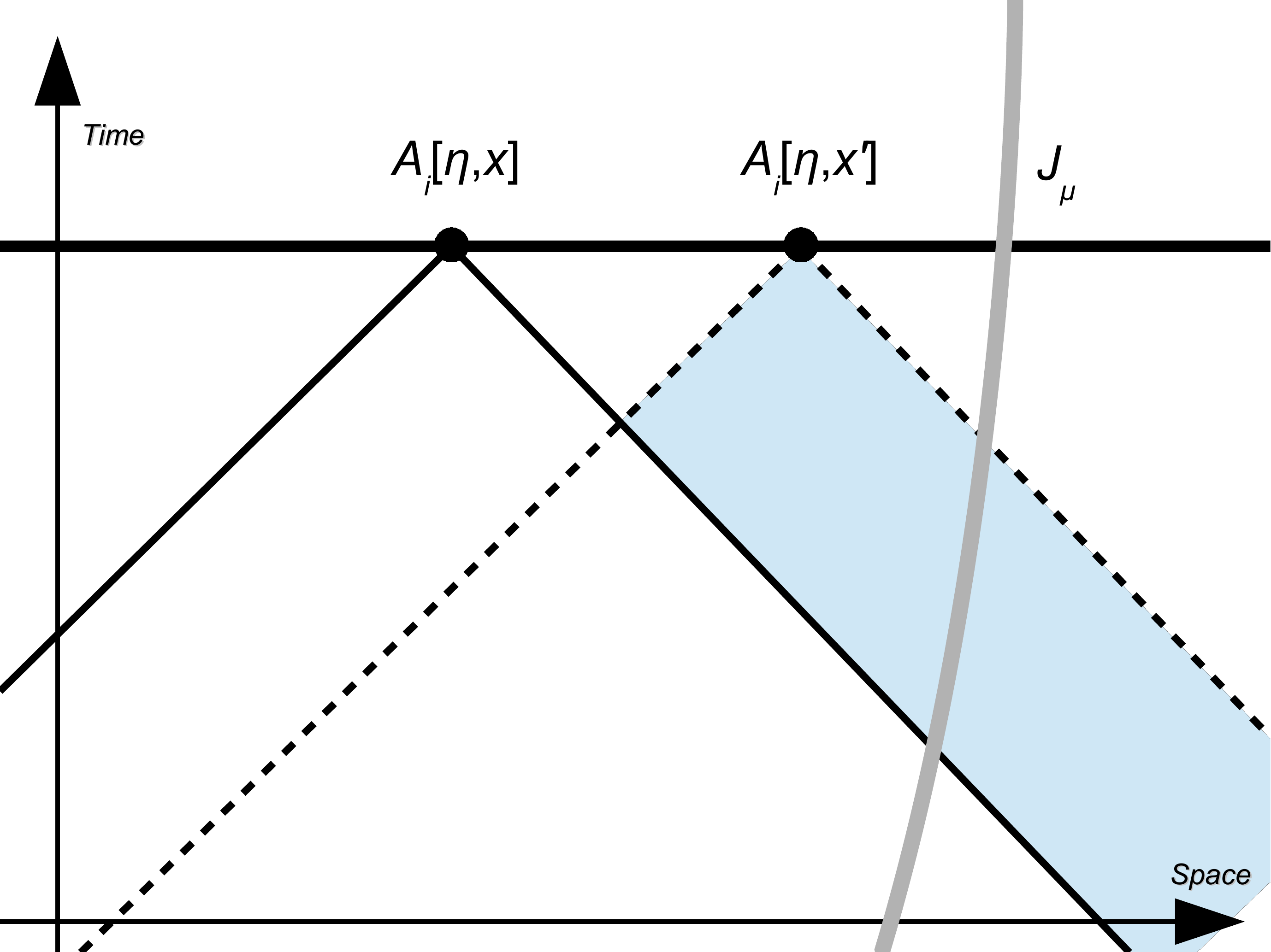}
\caption{The $A_i[\eta,\vec{x}]$ is the Lorenz gauge photon vector potential at the observer's location $(\eta,\vec{x})$. The $A_i[\eta,\vec{x}']$ is the Lorenz gauge potential at some other spatial location but at the same time $\eta$. The solid cone is the past null cone of the observer at $(\eta,\vec{x})$; while the dotted one is that of $(\eta,\vec{x}')$. The gray thick line denotes the worldtube of the electric current. (See Brill and Goodman \cite{BrillGoodman} for a closely related discussion, but from the Coulomb gauge perspective.)}
\label{WhySpin1IsAcausal}
\end{center}
\end{figure}

{\bf Spin-1 Photons} \qquad We begin by recalling the fact that the Lorenz gauge vector potential, which obeys $\partial^\mu A_\mu = 0$ and $\partial^2 A_\mu = - J_\mu$, is causally dependent on the electric current $J_\mu$. If $G^+_d$ denotes the retarded Green's function of the massless scalar, we have
\begin{align}
A_\mu[x]
= - \int_{\mathbb{R}^{d-1,1}} \dd^d x' G^+_d[x-x'] J_\mu[x'].
\end{align}
In even dimensions, where $G^+_d$ propagates signals strictly on the null cone, $A_\mu$ is the field due to the electric current lying on the past light cone of the observer. In odd dimensions, where $G^+_d$ propagates signals strictly inside the null cone (at least for timelike sources \cite{Chu:2016ngc}), $A_\mu$ is the field due to the electric current lying within the past light cone of the observer. On the other hand, the transverse spin-1 photon can be constructed from the Lorenz gauge photon via the following Fourier-space projection involving the spatial Fourier transform of the spatial components of the vector potential $\widetilde{A}_j[\eta,\vec{k}]$:
\begin{align}
\label{Spin1Photon_Minkowski}
\alpha_i[\eta,\vec{x}]
&= \int_{\mathbb{R}^{d-1}} \frac{\dd^{d-1} \vec{k}}{(2\pi)^{d-1}} \widehat{P}_{ij}[\vec{k}] \widetilde{A}_j[\eta,\vec{k}] e^{i\vec{k}\cdot\vec{x}}  ,
\end{align}
where the transverse projector $\widehat{P}_{ij}[\vec k]$ is defined in eq.~\eqref{TransverseProjector}. Now, in Fourier space, $-1/\vec{k}^2$ is simply the Euclidean Green's function
\begin{align}
\label{EuclideanGreenFunction_Fourier}
G^{(\text{E})}_{d}[\vec{x}-\vec{x}']
= \frac{1}{\partial_i\partial_i}[\vec{x}-\vec{x}']
= \int_{\mathbb{R}^{d-1}} \frac{\dd^{d-1}\vec{k}}{(2\pi)^{d-1}} \frac{e^{i\vec{k}\cdot(\vec{x}-\vec{x}')}}{-\vec{k}^2} ,
\end{align}
with the concrete expressions
{\allowdisplaybreaks
\begin{align}
G^{(\text{E,reg})}_{3+2\epsilon}[\vec{x}-\vec{x}'] & =  -\frac1{4\pi} \left( \frac1\epsilon  -  \gamma - \ln[\pi] -  2\ln [\mu R] \right),
\label{EuclideanGreenFunction_3D}\\
G^{(\text{E})}_{d\geq4}[\vec{x}-\vec{x}']
& = - \frac{\Gamma[\frac{d-3}{2}]}{4\pi^{\frac{d-1}{2}} |\vec{x}-\vec{x}'|^{d-3}} ;
\label{EuclideanGreenFunction}
\end{align}}%
where $G^{(\text{E})}_{3}$ has been dimensional-regularized in eq.~\eqref{EuclideanGreenFunction_3D} with an arbitrary mass scale $\mu$ introduced and $\gamma$ being the Euler-Mascheroni constant, and $-\partial_i \partial_j$ is replaced with $k_i k_j$. Utilizing eqs. \eqref{EuclideanGreenFunction_Fourier}, \eqref{EuclideanGreenFunction_3D} and \eqref{EuclideanGreenFunction} in eq.~\eqref{Spin1Photon_Minkowski} informs us that the transverse photon itself must therefore be related to its Lorenz gauge counterpart through the subtraction of the latter's longitudinal piece:
\begin{align}
\label{TransversePhoton_Decomposed}
\alpha_i[\eta,\vec{x}]
&= A_i[\eta,\vec{x}] + A^\parallel_i[\eta,\vec{x}] , \\
\label{TransversePhoton_AcausalPart_3D}
A^\parallel_i[\eta,\vec{x}]
&\equiv -\partial_i \partial_j \int_{\mathbb{R}^{2}} \dd^{2}\vec{x}'  \, \frac{\ln \big[|\vec{x}-\vec{x}'|\big]}{2\pi} A_j[\eta,\vec{x}'] \qquad\qquad (d = 3) \\
&\equiv \partial_i \partial_j \int_{\mathbb{R}^{d-1}} \dd^{d-1}\vec{x}' \frac{\Gamma\left[\frac{d-3}{2}\right]}{4\pi^{\frac{d-1}{2}} |\vec{x}-\vec{x}'|^{d-3}} A_j[\eta,\vec{x}'] \qquad\qquad (d \geq 4).
\label{TransversePhoton_AcausalPart}
\end{align}
We will now explain how the second term $A^\parallel_i[\eta,\vec{x}]$ in eq.~\eqref{TransversePhoton_Decomposed} is most likely acausal, because it is essentially the causal Lorenz gauge vector potential but smeared over all space, weighted by the Euclidean Green's function in eqs.~\eqref{EuclideanGreenFunction_3D} and \eqref{EuclideanGreenFunction}. Referring to Fig.~\eqref{WhySpin1IsAcausal}, we see that eqs.~\eqref{TransversePhoton_AcausalPart_3D} and \eqref{TransversePhoton_AcausalPart} are the weighted superposition of $A_i[\eta,\vec{x}']$ over all $\vec{x}'$, which -- for a fixed $\vec{x}'$ -- receives signals from the electric current from the past light cone of $(\eta,\vec{x}')$ (for even dimensions) or within it (for odd dimensions). But from the perspective of the observer at $(\eta,\vec{x})$, this means $A^\parallel_i$ is getting a signal from the portion of the source residing within the shaded (blue) region, which lies outside its past null cone.

{\bf Spin-2 Gravitons} \qquad The de Donder gauge $\partial^\mu \chi_{\mu\nu} = (1/2) \partial_\nu \chi$, with $\chi \equiv \eta^{\rho\sigma} \chi_{\rho\sigma}$, like the Lorenz gauge for photons, yields gravitational perturbations that are causally sourced by their matter $T_{\mu\nu}$. Equation (A39) of \cite{Chu:2016ngc} tells us
\begin{align}
\overline{\chi}_{\mu\nu}[x]
&= -16\pi \GN \int_{\mathbb{R}^{d-1,1}} \dd^d x'
G^+_d[x-x'] T_{\mu\nu}[x'].
\end{align}
The transverse-traceless spin-2 graviton is gotten from its de Donder gauge counterpart via the Fourier space projection
\begin{align}
\chi_{ij}^{\text{TT}}[\eta,\vec{x}]
&= \int_{\mathbb{R}^{d-1}} \frac{\dd^{d-1}\vec{k}}{(2\pi)^{d-1}} \widehat P_{ij mn}[\vec{k}] \, \widetilde{\overline\chi}_{mn}[\eta,\vec{k}] e^{i\vec{k}\cdot\vec{x}},
\end{align}
where the projection tensor $\widehat P_{ijmn}[\vec{k}]$ is given in eq.~\eqref{GravitonProjector}, which can also be expressed explicitly in terms of $\vec k$,
\begin{align}
\widehat P_{ij mn}[\vec{k}]
&= \widehat P_{m(i} \widehat P_{j)n} - \frac{1}{d-2} \widehat P_{ij} \widehat P_{mn} , \notag\\
&=
\delta_{m(i} \delta_{j)n}
- \frac{\delta_{ij} \delta_{mn}}{d-2} - \frac{\left( \delta_{m(i} k_{j)} k_n + \delta_{n(i} k_{j)} k_m \right)}{\vec k^2} \notag\\
&\qquad\qquad
+ \frac{\delta_{ij} k_m k_n + \delta_{mn} k_i k_j}{(d-2) \vec{k}^2}
+ \left(\frac{d-3}{d-2}\right) \frac{k_i k_j k_m k_n}{\vec{k}^4} .
\label{GravitonProjector_k}
\end{align}
The same sort of arguments made for the spin-1 photon would apply here to tell us the spin-2 graviton receives signals from $T_{\mu\nu}$ from outside the observer's past light cone. For instance, the third and fourth group of terms in the second equality in eq.~\eqref{GravitonProjector_k} involves two spatial derivatives acting on the weighted superposition of the de Donder GW over all space but at the same observer time $\eta$, namely
\begin{align}
\sim \partial_a \partial_b \int_{\mathbb{R}^{d-1}} \dd^{d-1}\vec{x}'
\frac{\overline\chi_{ij}[\eta,\vec{x}']}{|\vec{x}-\vec{x}'|^{d-3}} ;
\end{align}
whereas the last group of terms in eq.~\eqref{GravitonProjector_k} involves four spatial derivatives acting on a different weighted superposition of the same:
\begin{align}
\sim \partial_a \partial_b \partial_c \partial_e \int_{\mathbb{R}^{d-1}}
\dd^{d-1}\vec{x}' \frac{\overline\chi_{ij}[\eta,\vec{x}']}{|\vec{x}-\vec{x}'|^{d-5}} .
\end{align}

\section{Minkowski Spacetime}  \label{MinkowskiSpacetime}

As demonstrated in the previous section, both massless spin-1 and spin-2 fields are expected to contain the acausal information from their isolated sources. To quantify this acausality, we will in this section perform a detailed analysis of the effective Minkowski spacetime Green's functions of the electromagnetic and gravitational gauge-invariant variables.

\subsection{Electromagnetism}

We will begin with electromagnetism in all spacetime dimensions equal to or higher than three, $d\geq3$; for $d=3$ corresponds to the lowest physical dimension for spin-1 photons to exist. In addition, the discussion for the electromagnetic field here will provide a useful guide for us to tackle the more complicated and subtle case of linearized gravitation.

{\bf Field Equations} \qquad  In terms of the gauge-invariant variables \eqref{Maxwell_GaugeInvariant_ScalarVector} and provided the electric current $J^\mu$ is conserved, the non-redundant portions of Maxwell's equations \eqref{Maxwell_equations} -- see \cite{Chu:2016ngc} for a discussion -- are the dynamical wave equation for the transverse spin-1 photon $\alpha_i$,
\begin{align}
\partial^2\alpha_i&=-\Gamma_i,
\label{Spin1_WaveEq_Minkowski}
\end{align}
with $\partial^2\equiv\eta^{\mu\nu}\partial_\mu\partial_\nu$, and a Poisson's equation obeyed by the gauge-invariant scalar $\Phi$,
\begin{align}
\vec\nabla^2\Phi=-\rho.
\label{EMPoisson_Minkowski}
\end{align}
Notice both equations involve non-locality. The source $\Gamma_i$ on the right hand side of eq.~\eqref{Spin1_WaveEq_Minkowski} is the transverse component $J_i$, which (recalling arguments from the previous section) is thus a non-local functional of $J_i$. Whereas the Poisson equation obeyed by $\Phi[\eta,\vec{x}]$ means it is sensitive to the charge density $\rho[\eta,\vec{x}]$ on the right hand side of eq.~\eqref{EMPoisson_Minkowski} at the same instant $\eta$. As we will show later in this section, only when $\alpha_i$ and $\Phi$ are both involved, do the physical observables -- i.e., the field strength $F_{\mu\nu}$ -- become causally dependent on the electromagnetic current $J_\mu$.

{\bf Spin-1 Photons } \qquad  To solve for $\alpha_i$ in eq.~\eqref{Spin1_WaveEq_Minkowski} through its effective Green's function convolved against its localized sources, it is convenient to first go to the Fourier space, where the transverse property is implemented through a projection of the Fourier transform of the current $\widetilde{J}_i$. For all $\vec k\neq\vec0$, the $\alpha_i$ can be written as the superposition
\begin{align}
\alpha_i[\eta,\vec x]=-\int_{\mathbb{R}} \dd\eta'\int_{\mathbb{R}^{d-1}} \frac{\dd^{d-1}\vec{k}}{(2\pi)^{d-1}}\,\widetilde G^{+}_d [\eta,\eta';\vec k]\left(\delta_{ij}-\frac{k_ik_j}{\vec k^2}\right)
\widetilde{J}_j[\eta',\vec{k}]\, e^{i\vec{k}\cdot\vec{x}},
\label{Spin1_Fourier}
\end{align}
where $\widetilde G^{+}_d$ denotes the Fourier transform of the retarded Green's function of the massless scalar. In flat spacetime, it enjoys time-translation symmetry and reads
\begin{align}
\widetilde G^{+}_d [\eta,\eta';\vec k]=-\Theta[T]\frac{\sin |\vec k|T}{|\vec k|},
\label{FourierG+_Minkowski}
\end{align}
with $T\equiv\eta-\eta'$. As we have seen before, $k_ik_j$ in momentum space can be pulled out with the replacement $k_ik_j\to-\partial_i\partial_j$ acting on the Fourier integral. Therefore, the expression \eqref{Spin1_Fourier} can be re-cast into the convolution of the spin-1 effective Green's function $G^+_{ij}$ against the local electromagnetic current,
\begin{align}
\alpha_i[x]=-\int_{\mathbb{R}^{d-1,1}} \dd^dx'\,G^+_{ij}[T,\vec R]\,J_j[x'],
\label{Spin1_Convolution_Minkowski}
\end{align}
where $\vec R\equiv\vec x-\vec x'$ and the $G^+_{ij}$ takes the form
\begin{align}
G^+_{ij}[T,\vec R]
&\equiv -\Theta[T]C_{ij}[T,\vec R], \notag \\
C_{ij}[T,\vec R]
&=\delta_{ij}C_{1,d}[T,R]+\partial_i\partial_jC_{2,d}[T,R].
\label{G+ij_Minkowski}
\end{align}
The $C_{1,d}$ and $C_{2,d}$ are respectively defined to be two scalar Fourier integrals
\begin{align}
C_{1,d}[T,R]&\equiv\int_{\mathbb{R}^{d-1}}\frac{\dd^{d-1}\vec{k}}{(2\pi)^{d-1}}\frac{\sin |\vec k|T}{|\vec k|}e^{i\vec{k}\cdot\vec{R}},\label{C1_Minkowski} \\
C_{2,d}[T,R]&\equiv\int_{\mathbb{R}^{d-1}}\frac{\dd^{d-1}\vec{k}}{(2\pi)^{d-1}}\frac{\sin |\vec k|T}{|\vec k|^3}e^{i\vec{k}\cdot\vec{R}},
\label{C2_Minkowski}
\end{align}
with the observer-source spatial distance denoted as $R\equiv|\vec x-\vec x'|$. Hence, the effective Green's function $G^+_{ij}$ can be gotten explicitly through eq.~\eqref{G+ij_Minkowski} once $C_{1,d}$ and $C_{2,d}$ are known. One of the advantages of using $C_{1,d}$ and $C_{2,d}$ to compute $G^+_{ij}$ is that our calculations can be simplified by exploiting the fact that they both obey the homogeneous wave equation $\partial^2C_{1,d}=0$ and $\partial^2C_{2,d}=0$ with the initial conditions
\begin{align}
\label{C12_Minkowski_T=0}
C_{1,d}[T=0,R]=0 \qquad \text{ and } \qquad C_{2,d}[T=0,R]=0 ;
\end{align}
as well as the initial velocities
\begin{align}
\label{C12dot_Minkowski_T=0}
\dot{C}_{1,d}[T=0,R] = \delta^{(d-1)}[\vec{x}-\vec{x}']
\qquad \text{ and } \qquad
\dot{C}_{2,d}[T=0,R] = -G^{(\text{E})}_d[R] ,
\end{align}
where the overdot denotes the time derivative with respect to $\eta$. In addition, $C_{1,d}$ and $C_{2,d}$ are connected via the spatial Laplacian operator or double-time derivatives:
\begin{align}
\label{C1ddot_Minkowski}
\ddot{C}_{2,d}[T,R] 			&= -C_{1,d}[T,R], \\
\label{C1Laplacian_Minkowski}
\vec{\nabla}^2 C_{2,d}[T,R] 	&= -C_{1,d}[T,R].
\end{align}
(Equations \eqref{C12_Minkowski_T=0} through \eqref{C1Laplacian_Minkowski} conditions follow readily from the Fourier representations in equations \eqref{EuclideanGreenFunction_Fourier}, \eqref{C1_Minkowski} and \eqref{C2_Minkowski}.) Moreover, note that
\begin{align}
G^\pm_d [x-x'] = \mp \Theta[\pm T] C_{1,d}[T,R]
\end{align}
are the retarded $(+)$/advanced $(-)$ massless scalar Green's functions; obeying
\begin{align}
\label{MasslessScalarG_Minkowski}
\partial^2 G^\pm_d = \delta^{(d)}[x-x'] .
\end{align}
In other words, $-C_{1,d}$ itself is the retarded minus advanced Green's function:
\begin{align}
\label{C1_Minkowski_RetardedMinusAdvanced}
-C_{1,d} = G^+_d[x-x'] - G^-_d[x-x'] .
\end{align}
Likewise, for the spin-1 case, the $-C_{ij}$ in eq.~\eqref{G+ij_Minkowski} is the difference between the retarded Green's function in eq.~\eqref{G+ij_Minkowski} and that of its advanced counterpart:
\begin{align}
\label{Cij_Minkowski_RetardedMinusAdvanced}
-C_{ij} = G_{ij}^+[x-x'] - G_{ij}^-[x-x'] .
\end{align}
In Quantum Field Theory, the $C_{1,d}$ in eq.~\eqref{C1_Minkowski_RetardedMinusAdvanced} is proportional to the commutator of massless scalar fields. In turn, $C_{ij}$ is proportional to the commutator of (spin-1) photon fields. Therefore, the elucidation of the (classical) causal structure of $C_{ij}$ will also lead to insights regarding the quantization of the associated spin-1 photons.

Before moving on to the analytic solutions, let us show that the source of $G^+_{ij}$ in eq.~\eqref{G+ij_Minkowski} is an extended one, as opposed to the usual spacetime point source of, say, the massless scalar Green's function. Applying the wave operator to the expression \eqref{G+ij_Minkowski} for $G^+_{ij}$ hands us
\begin{align}
\label{Gij_WaveEquation_Minkowski}
\partial^2G^+_{ij}[T,\vec R]
&=\delta_{ij}\delta^{(d)}[x-x']-\delta[T]\partial_i\partial_jG^{\mathrm{(E)}}_d[R] \\
&=\delta[T] \left( \delta_{ij} \delta^{(d-1)}[\vec{x}-\vec{x}'] - \partial_i\partial_j G^{\mathrm{(E)}}_d[R] \right) ;
\end{align}
where the $G^{(\text{E})}_d$ is the Euclidean Green's function of eq.~\eqref{EuclideanGreenFunction_Fourier} and the relations in eq.~\eqref{C12dot_Minkowski_T=0} were employed. We may view eq.~\eqref{Gij_WaveEquation_Minkowski} as a $(d-1) \times (d-1)$ matrix of massless scalar wave equations. That $-\partial_i \partial_j G^{(\text{E})}_d[|\vec{x}-\vec{x}'|]$ is non-zero everywhere in space at $T=0$ tells us, for a fixed pair of indices $ij$, the (retarded) signal it generates likely fills all of spacetime to the future of $\eta'$. This is to be contrasted against the massless scalar Green's function equation itself in eq.~\eqref{MasslessScalarG_Minkowski}; where, because the source at $x'$ is point-like, the signal it generates propagates only on and/or within its future light cone. If the observer at $x$ lies outside the light cone of $x'$, the signal $G^+_d[x-x']$ will be zero and causality respected. Returning to eq.~\eqref{Gij_WaveEquation_Minkowski}, if one continues to insist on viewing $G^+_{ij}[x-x']$ as the signal at $x$ generated at $x'$, since it is non-zero throughout all $\vec{x}$ whenever $\eta>\eta'$, once $x$ lies outside the light cone of $x'$ the observer at $x$ would be led to conclude the signal is acausal.

{\it Recursion relations} \qquad In both Minkowski and spatially flat cosmologies, we are aided by the spatial-translation and spatial-parity invariance of the underlying spacetimes. In particular, these symmetries allow us to solve for $C_{1,d}$ and $C_{2,d}$ for all dimensions once we know their $3-$ and $4-$dimensional solutions. This is because the higher-dimensional ones can be generated through the ``dimension-raising operator''
\begin{align}
\mathcal D_R\equiv - \frac1{2\pi R} \frac{\partial}{\partial R}.
\end{align}
(See appendix (E) of \cite{Chu:2016ngc} for a detailed discussion.) In brief, any bi-scalar function $f_d$ that depends on space solely through $R \equiv |\vec{x}-\vec{x}'|$ and takes the same Fourier integral form
\begin{align}
f_d\left[R\right]
&= \int \frac{\dd^{d-1}\vec{k}}{(2\pi)^{d-1}} \widetilde{f}_d\big[|\vec{k}|\big]
e^{i\vec{k}\cdot(\vec{x}-\vec{x}')}
\end{align}
for all relevant spacetime dimensions $d$, obeys the recursion relation
\begin{align}
\label{RecursionRelation}
f_{d+2}[R] = - \frac{1}{2 \pi R} \frac{\partial f_d[R]}{\partial R} .
\end{align}
This remark applies to both $C_{1,d}$ and $C_{2,d}$; specifically, we only need $C_{1,4}$ and $C_{2,4}$ to determine their counterparts in all even $d \geq 4$:
\begin{align}
\label{C1_EvenMinkowski_Recursion}
C_{1,\text{even $d$} \geq 4} &= \mathcal{D}_R^{\frac{d-4}{2}} C_{1,4} , \\
\label{C2_EvenMinkowski_Recursion}
C_{2,\text{even $d$} \geq 4} &= \mathcal{D}_R^{\frac{d-4}{2}} C_{2,4} .
\end{align}
Likewise, we only need $C_{1,3}$ and $C_{2,3}$ to obtain their counterparts in all odd $d \geq 3$:
\begin{align}
\label{C1_OddMinkowski_Recursion}
C_{1,\text{odd $d$} \geq 3} &= \mathcal{D}_R^{\frac{d-3}{2}} C_{1,3} , \\
\label{C2_OddMinkowski_Recursion}
C_{2,\text{odd $d$} \geq 3} &= \mathcal{D}_R^{\frac{d-3}{2}} C_{2,3} .
\end{align}
Also notice that, by counting the powers $|\vec k|$ in the integrals \eqref{C1_Minkowski} and \eqref{C2_Minkowski} as $|\vec k|\to0$, $C_{1,d}$ is finite for all $d$, while $C_{2,d}$ is expected to diverge when $d\leq3$. However, on physical grounds, the full effective Green's function $G^+_{ij}$ should converge for all spacetime dimensions $d \geq 3$. This suggests that, for $d=3$, the two spatial derivatives acting on $C_{2,3}$ in eq.~\eqref{G+ij_Minkowski} will eliminate the divergence completely.

{\it Time integral method} \qquad According to eq.~\eqref{C1_Minkowski_RetardedMinusAdvanced}, $-\Theta[T] C_{1,d}$ is the retarded Green's function $G^+_d$ of the massless scalar. Because eq.~\eqref{C1_Minkowski_RetardedMinusAdvanced} will continue to hold even in cosmology and because the analytic position spacetime solutions to $C_{1,d}$ and $G^+_d$ are known in all Minkowski and constant equation-of-state universes \cite{Chu:2016ngc}, we shall introduce a `time-integral' method here that will allow us to solve the (retarded part of) $C_{2,d}$ in terms of time integrals of $C_{1,d}$. We first recall that eq.~\eqref{C1ddot_Minkowski} provides us a ordinary differential equation (ODE) relating $C_{2,d}$ to $C_{1,d}$. Integrating it twice with respect to time, and taking into account the initial conditions in equations \eqref{C12_Minkowski_T=0} and \eqref{C12dot_Minkowski_T=0},
\begin{align}
C_{2,d}[T,R]
&= -\int_{0}^{T} \dd \tau_2 \int_{0}^{\tau_2} \dd \tau_1 C_{1,d}[\tau_1,R]
+ T \dot{C}_{2,d}[T=0,R] + C_{2,d}[T=0,R] \\
\label{TimeIntegralMethod_C2d}
&= -\int_{0}^{T} \dd \tau_2 \int_{0}^{\tau_2} \dd \tau_1 C_{1,d}[\tau_1,R]
- T G^{(\text{E})}_d[R] .
\end{align}
Now, any casual quantity $\mathfrak{Q}^+[\eta,\vec{x};\eta',\vec{x}']$ -- which we define as one that is non-zero only when $T \geq R \geq 0$ -- may be multiplied by $\Theta[T-R^-]$. While any anti-causal expression $\mathfrak{Q}^-[\eta,\vec{x};\eta',\vec{x}']$ -- which we define as one that is non-zero only when $T \leq -R \leq 0$ -- may be multiplied by $\Theta[-T-R^-]$.\footnote{The $R^-$ guarantees that signals on the null cone proportional to $\delta[T-R]$ and its derivatives are included.} If we then consider
\begin{align}
\int_{\eta'}^{\eta' + \tau} \dd \eta \, \mathfrak{Q}^+[\eta,\vec{x};\eta',\vec{x}']
= \int_{\eta'}^{\eta' + \tau} \dd \eta \,
\Theta[\eta-\eta'-R^-] \mathfrak{Q}^+
= \int_{0}^{\tau} \dd \tau' \Theta[\tau'-R^-] \mathfrak{Q}^+ .
\end{align}
We see that, whenever $\tau > R^-$, we may set $\Theta[\tau'-R^-]$ to one for $\tau'\in (R^-, \tau)$; whereas whenever $\tau < R^-$, the latter is to be set to zero.
\begin{align}
\int_{0}^{\tau} \dd \tau' \Theta[\tau'-R^-] \mathfrak{Q}^+
= \Theta[\tau-R^-] \left. \int_{R^-}^{\tau} \dd \tau' \mathfrak{Q}^+ \right\vert_{\tau' = \eta-\eta'}
\end{align}
Iterating this reasoning, we may deduce that one or multiple nested integrals of a causal quantity would return another causal one:
\begin{align}
\int_{0}^{T} \dd \tau_n \dots \int_{0}^{\tau_3} \dd \tau_2 \int_{0}^{\tau_2} \dd \tau_1 \Theta[\tau_1-R^-] \mathfrak{Q}^+
&= \Theta[T-R^-] \left. \int_{R^-}^{T} \dd \tau_n \dots \int_{R^-}^{\tau_3} \dd \tau_2 \int_{R^-}^{\tau_2} \dd \tau_1 \mathfrak{Q}^+ \right\vert_{\tau_1 = \eta-\eta'} .
\end{align}
Similarly, if we instead consider
\begin{align}
\int_{\eta' + \tau}^{\eta'} \dd \eta \, \mathfrak{Q}^-[\eta,\vec{x};\eta',\vec{x}']
= \int_{\eta' + \tau}^{\eta'} \dd \eta \,
\Theta[-(\eta-\eta')-R^-] \mathfrak{Q}^-
= \int_{\tau}^0 \dd \tau' \Theta[-\tau'-R^-] \mathfrak{Q}^- .
\end{align}
We see that, whenever $\tau < -R^-$, we may set $\Theta[-\tau'-R^-]$ to one for $\tau'\in (\tau, -R^-)$; whereas whenever $\tau \geq -R^-$, the latter is to be set to zero.
\begin{align}
\int_{0}^{\tau} \dd \tau' \Theta[-\tau'-R^-] \mathfrak{Q}^-
= \Theta[-\tau-R^-] \left. \int_{-R^-}^{\tau} \dd \tau' \mathfrak{Q}^- \right\vert_{\tau' = \eta-\eta'}
\end{align}
Iterating this reasoning, we may deduce that one or multiple nested integrals of an anti-causal quantity would return another anti-causal one:
\begin{align}
\int_{0}^{T} \dd \tau_n \dots \int_{0}^{\tau_3} \dd \tau_2 \int_{0}^{\tau_2} \dd \tau_1 \Theta[-\tau_1-R^-] \mathfrak{Q}^-
&= \Theta[-T-R^-] \left. \int_{-R^-}^{T} \dd \tau_n \dots \int_{-R^-}^{\tau_3} \dd \tau_2 \int_{-R^-}^{\tau_2} \dd \tau_1 \mathfrak{Q}^- \right\vert_{\tau_1 = \eta-\eta'} .
\end{align}
This discussion implies the integral of the difference between the causal Green's function and its anti-causal counterpart -- recall eq.~\eqref{C1_Minkowski_RetardedMinusAdvanced} -- namely $-C_{1,d}$, returns a causal minus anti-causal object:
\begin{align}
\int_{0}^{T} \dd \tau' & \big( G^+_d[\tau',R] - G^-_d[\tau',R] \big)   \notag\\
& \qquad = \Theta[T-R^-] \int_{R^-}^{T} \dd \tau' \,  G^+_d[\tau',R] - \Theta[-T-R^-]  \int_{-R^-}^{T} \dd \tau' \, G^-_d[\tau',R]  .
\end{align}
Furthermore, referring to equations \eqref{C1_Minkowski_RetardedMinusAdvanced} and \eqref{TimeIntegralMethod_C2d}, we see that the retarded portion of $C_{2,d}$ -- which is what we need -- is gotten by integrating the retarded portion of $C_{1,d}$:
\begin{align}
\label{TimeIntegralMethod_C2dRet}
C_{2,d}^{+}[T,R]
&= \Theta[T-R^-] \int_{R^-}^{T} \dd \tau_2 \int_{R^-}^{\tau_2} \dd \tau_1 \, G_d^+[\tau_1,R]
- \Theta[T] \cdot T G^{(\text{E})}_d[R] ,
\end{align}
where $C_{2,d}^{+}\equiv \Theta[T] C_{2,d}$. Observe that the first term on the right hand side is strictly causal, whereas the second term arising from the initial condition is retarded but acausal because it contributes a non-zero signal outside the past null cone. As additional Minkowski and cosmological examples below will further corroborate, the `time-integral' method not only allows us to compute (up to quadrature) the retarded part of $C_{2,d}$ from the known solutions of the massless scalar causal Green's functions, it provides a clean separation between the strictly causal versus the retarded-but-acausal terms arising from the initial conditions -- even if the time-integrals themselves cannot be performed analytically.

{\it Exact solutions in even dimensions $d\geq4$} \qquad The retarded and advanced Green's functions of a massless scalar in 4D Minkowski are
\begin{align}
G^\pm_4 [x-x'] =  - \frac{\delta[T \mp R]}{4\pi R} .
\end{align}
The $\delta$-function teaches us that $G^+_4$ propagates signals strictly on the forward null cone; and the $G^-_4$ strictly on the backward null cone. From eq.~\eqref{C1_Minkowski_RetardedMinusAdvanced}, $C_{1,4}$ thus reads:
\begin{align}
C_{1,4}[T,R]
=\frac1{4\pi R}\left(\delta[T-R] - \delta[T+R]\vphantom{\dot T}\right) .
\label{C14_Minkowski}
\end{align}
Of course, $C_{1,4}$ can be worked out straightforwardly from eq.~\eqref{C1_Minkowski} by setting $d=4$.

To compute $C_{2,4}$, on the other hand, we insert eq.~\eqref{C14_Minkowski} into eq.~\eqref{TimeIntegralMethod_C2d} and obtain
\begin{align}
C_{2,4}[T,R]&=\frac1{4\pi}\left(\Theta[T-R]-\Theta[-T-R]\vphantom{\dot T}\right)+\frac T{4\pi R}\left(\Theta[T]\Theta[-T+R]+\Theta[-T]\Theta[T+R]\vphantom{\dot T}\right),
\label{C24_Minkowski}
\end{align}
We may check this result by tackling eq.~\eqref{C2_Minkowski} directly. After integrating over the angular coordinates in $\vec{k}-$space,
\begin{align}
C_{2,4}[T,R]
&= \int_{-\infty}^{+\infty} \frac{\dd k}{(2\pi)^{2}}
\frac{\sin kT }{k} \frac{\sin kR }{kR} .
\end{align}
The sines can be converted into exponentials; and because there are no singularities the contour on the real line may be displaced slightly toward the positive or negative imaginary $k-$axis near $k=0$. The resulting expression would consist of $4$ terms, each of which would now be amendable to the residue theorem by closing the contour appropriately in the lower or upper half complex $k$ plane.

From the retarded portion of eq.~\eqref{C24_Minkowski}, we find that the contribution of $C_{2,4}$ comes from both inside and outside the past light cone of the observer; however, the signal that resides within the light cone -- its ``tail" -- is a spacetime constant and will therefore be removed by the spatial derivatives in eq.~\eqref{G+ij_Minkowski}. In contrast, the acausal one with $T<R$ still remains and does contribute to the 4D effective Green's function $G_{ij}^+$, along with some additional light-cone contributions from differentiating the step functions in $C_{2,4}$.
\begin{align}
G_{ij}^{(+,\text{4D})}[T, \vec R]
&= - \left( \delta_{ij} - \widehat R_i \widehat R_j \right) \frac{\delta[T-R]}{4\pi R}
- \Theta[T] \Theta[-T+R] \partial_i \partial_j \left( \frac{T}{4\pi R}  \right) .
\end{align}
To sum: the 4D effective Green's function $G^{(+,\text{4D})}_{ij}[x-x']$ propagates signals on and outside the forward light cone of the source at $x'$ -- namely, it is acausal.

With the 4D solutions in eqs.~\eqref{C14_Minkowski} and \eqref{C24_Minkowski} at hand, we may employ eqs. \eqref{C1_EvenMinkowski_Recursion} and \eqref{C2_EvenMinkowski_Recursion} to state:
\begin{align}
C_{1, \text{even }d\geq4}^+[T,R]&=\mathcal D^{\frac{d-4}2}_R\left(\frac{\delta[T-R]}{4\pi R}\right) \label{C1_Even_Minkowski} ,
\end{align}
where $C_{1,d}^+ \equiv \Theta[T] C_{1,d}$, and
\begin{align}
C_{2,\text{even }d\geq4}^+[T,R]&=\mathcal D^{\frac{d-4}2}_R \left( \Theta[T-R]\,\frac1{4\pi} + \Theta[T]\Theta[-T+R]\,\frac T{4\pi R} \right) ;
\label{C2_Even_Minkowski}
\end{align}
where only the retarded contributions are shown, and we highlight that the tail portion of eq.~\eqref{C24_Minkowski} in higher dimensions $d\geq6$ partially cancels the acausal part of it upon differentiation. Like the 4D case, the $T \mathcal{D}_R^{\frac{d-4}{2}} (4\pi R)^{-1}$ part of $C_{2,\text{even }d\geq4}^+$ indicates the latter continues to receive acausal contribution from outside the light cone.

{\it Exact solutions in odd dimensions $d\geq3$} \qquad  Odd-dimensional solutions differ from even ones due to the presence of inside-the-null-cone propagation -- ``tail" signals. In $(2+1)-$dimensions, the retarded Green's function of the massless scalar is
\begin{align}
C^+_{1,3}[T,R]=-G^+_{3}[T,R]=\frac{\Theta[T-R]}{2\pi\sqrt{T^2-R^2}},
\label{C13_Minkowski}
\end{align}
which, unlike the 4D case, is pure tail. We now turn to solving the integral $C_{2,3}$, which, as we have reasoned earlier, is expected to blow up when considered alone, but its divergent piece does not really enter the physical spin-1 Green's function $G^+_{ij}$, as it will be eliminated by the two spatial derivatives $\partial_i \partial_j$ in eq.~\eqref{G+ij_Minkowski}. Despite being divergent, $C_{2,3}$ can nonetheless be regularized to a finite expression in the time-integral method, where the divergence only takes place on the initial condition. Within dimensional regularization, the resulting regularized form of it, $C^{(+,\text{reg})}_{2,3+2\epsilon}$, is given by
\begin{align}
C_{2,3+2\epsilon}^{(+,\text{reg})}[T,R]= \Theta[T-R]\, &  \frac{-T\ln\left[\mu \left(T+\sqrt{T^2-R^2}\right)\right]+\sqrt{T^2-R^2}}{2\pi} \notag\\
& \qquad \qquad   +\Theta[T]\Theta[-T+R]\,\frac{-T\ln [\mu R]}{2\pi}  + \Theta[T] \, \frac T{4\pi} \left( \frac1\epsilon - \gamma - \ln[\pi] \right).
\label{C23_Minkowski}
\end{align}
By referring to eq.~\eqref{G+ij_Minkowski}, we see that both tail and the acausal parts of $C_{2,3+2\epsilon}^{\mathrm{(reg)}}$ contribute to the three-dimensional $G^+_{ij}$, with no pure light-cone signals involved.

To further justify the validity of eq.~\eqref{C23_Minkowski}, we independently computed finite $C_{2,5}$ using its Fourier representation in eq.~\eqref{C2_Minkowski}:
\begin{align}
C^+_{2,5}[T,R]&=\Theta[T-R]\,\frac{T-\sqrt{T^2-R^2}}{4\pi^2 R^2}  + \Theta[T]\Theta[-T+R]\,\frac T{4\pi^2R^2}.
\label{C25_Minkowski}
\end{align}
This then allows us to verify $C^+_{2,5}=\mathcal D_RC_{2,3+2\epsilon}^{(+,\text{reg})}$ in eq.~\eqref{C2_OddMinkowski_Recursion}. Higher odd-dimensional results follow from eqs.~\eqref{C1_OddMinkowski_Recursion}, \eqref{C2_OddMinkowski_Recursion}, \eqref{C13_Minkowski}, and \eqref{C23_Minkowski}, where we can simply drop the last term of eq.~\eqref{C23_Minkowski} and set the mass scale $\mu$ to one, since they will be removed in $G^+_{ij}$ by the spatial derivatives in eq.~\eqref{G+ij_Minkowski},
{\allowdisplaybreaks
\begin{align}
C^+_{1,\text{odd } d\geq3}[T,R]&=\mathcal D^{\frac{d-3}2}_R\left(\frac{\Theta[T-R]}{2\pi\sqrt{T^2-R^2}}\right),
\label{C1_Odd_Minkowski}\\
C^+_{2,\text{odd } d\geq3}[T,R]&=\mathcal D^{\frac{d-3}2}_R \Bigg(\Theta[T-R]\,  \frac{-T\ln\left[ \left(T+\sqrt{T^2-R^2}\right)\right]+\sqrt{T^2-R^2}}{2\pi}   +\Theta[T]\Theta[-T+R]\,\frac{-T\ln  R}{2\pi}  \Bigg) .
\label{C2_Odd_Minkowski}
\end{align}}%
With eqs.~\eqref{C1_Odd_Minkowski} and \eqref{C2_Odd_Minkowski} plugged into eq.~\eqref{G+ij_Minkowski}, we now have the explicit spin-1 Green's function $G^+_{ij}$ for all odd dimensions $d\geq3$. These analytic solutions reveal that, in odd dimensions, the spin-1 photon receives not only the causal tail signals from both $C_{1,d}$ and $C_{2,d}$, with no strictly $\delta$-function light-cone counterpart, but also the acausal contribution from $C_{2,d}[0\leq T<R]$. As a result, we have explicitly shown that, in the presence of the local electromagnetic source, the spin-1 photon being acausal turns out to be a generic feature in any spacetime dimensions $d\geq3$.

{\bf Scalar } \qquad  The scalar solution for the gauge-invariant $\Phi$, which obeys the Poisson's equation \eqref{EMPoisson_Minkowski}, is given by a Coulomb-type form,
\begin{align}
\Phi[\eta,\vec x]=\int_{\mathbb{R}^{d-1}} \dd^{d-1}\vec{x}'\,G^{(\mathrm{E})}_{d}[R]\, J_0[\eta,\vec x'],
\label{EMPhi_Minkowski}
\end{align}
where we recall that $G^{(\mathrm{E})}_{d}$ takes the form of eq.~\eqref{EuclideanGreenFunction_3D} for $d=3$ and eq.~\eqref{EuclideanGreenFunction} for $d\geq4$. Clearly, this solution manifestly violates causality, in the sense that the scalar $\Phi$ is instantaneously sourced by the local charge density $\rho$. Therefore, neither the spin-1 photon $\alpha_i$ nor the scalar $\Phi$ can be a standalone observable in classical electromagnetism, which then leads us to pose the question: how do these gauge-invariant variables enter the key observable -- the Faraday tensor -- such that the result is causally dependent on their corresponding sources?

{\bf  Faraday Tensor} \qquad  The causal nature of $F_{\mu\nu}$ can be seen from its own wave equation, derived by taking the divergence of the identity $\partial_{[\mu}F_{\nu\sigma]}=0$ and imposing the Maxwell eqautions \eqref{Maxwell_equations},
\begin{align}
\Box F_{\mu\nu} + R_{\rho\sigma\mu\nu}F^{\rho\sigma}+2R^{\sigma}{}_{[\mu}F_{\nu]\sigma}=-2\nabla_{[\mu}J_{\nu]}, \qquad\qquad \Box \equiv \nabla_\sigma\nabla^\sigma .
\label{WaveEq_FaradayTensor}
\end{align}
In Minkowski spacetime, the geometric tensors vanish and the electromagnetic fields encoded within $F_{\mu\nu}$ are thus given by the massless scalar Green's function convoluted against the first derivatives of the electromagnetic sources,
\begin{align}
F_{\mu\nu}[x]=-2\int_{\mathbb{R}^{d-1,1}} \dd^dx'\,\partial_{[\mu}G_d^+J_{\nu]}[x'],
\label{FieldStrength_Sol_Minkowski}
\end{align}
Here, we have dropped the surface terms at infinity when integrating by parts, which can be justified by the causal structures of eqs.~\eqref{C1_Even_Minkowski} and \eqref{C1_Odd_Minkowski} as well as the fact that those at past infinity ($T\to\infty$) in odd dimensions are negligible (see eq.~\eqref{C1_Odd_Minkowski}).

Let us now recover eq.~\eqref{FieldStrength_Sol_Minkowski} within the gauge-invariant formalism. We first make use of the conservation law for the electromagnetic current, $\partial_{j}J_{j}=\dot{J}_0$, to re-write the spin-1 expression \eqref{Spin1_Convolution_Minkowski},
\begin{align}
\alpha_i[x]=-\int_{\mathbb{R}^{d-1,1}} \dd^dx'\,\left(G^+_{d}J_i[x']-\Theta[T]\,\partial_i\dot{C}_{2,d}J_0[x']\right),
\label{Spin1_Convolution2_Minkowski}
\end{align}
where the second term is now the convolution with the charge density $J_0$. The surface terms from integration by parts -- namely, $\int \dd\eta'\,\dd^{d-2}\vec x'\,\Theta[T]\partial_iC_{2,d}J_{j}$ evaluated at spatial infinity and $\int \dd^{d-1}\vec{x}'\,\Theta[T]\partial_iC_{2,d}J_0$ at past infinity -- have been neglected, as the former falls off as $R\to \infty$ in both even and odd dimensions, and the electric current is assumed to be isolated; whereas the latter, when $T\to\infty$, has zero contribution in even dimensions and becomes negligible in odd dimensions (see eqs.~\eqref{C2_Even_Minkowski} and \eqref{C2_Odd_Minkowski}). The magnetic field, according to eq.~\eqref{Maxwell_FaradayTensor}, is therefore consistent with eq.~\eqref{FieldStrength_Sol_Minkowski}:
\begin{align}
F_{ij}[x] = 2\partial_{[i} \alpha_{j]}=-2\int_{\mathbb{R}^{d-1,1}}  \dd^dx'\,\partial_{[i}G^+_d\,J_{j]}[x'].
\label{Fij_Minkowski}
\end{align}
This calculation shows that, despite $\alpha_i$ being acausal, taking the curl of the spin-1 field ends up removing its acausal information encoded in the second term of eq.~\eqref{Spin1_Convolution2_Minkowski}.

According to eq.~\eqref{Maxwell_FaradayTensor}, the electric field $F_{0i}$ is the sum of $\dot{\alpha}_i$ and $\partial_i \Phi$. Employing $\dot C_{2,d}\big|_{T=0}=-G^{\mathrm{(E)}}_d$ and $\ddot{C}_{2,d}=-C_{1,d}$ in eqs.~\eqref{C12dot_Minkowski_T=0} and \eqref{C1ddot_Minkowski}, the time derivative of eq.~\eqref{Spin1_Convolution2_Minkowski} is
\begin{align}
\dot\alpha_i[x]=-\int_{\mathbb{R}^{d-1,1}} \dd^dx'\left(\dot G^+_{d}J_i[x']-\partial_iG^+_{d}J_0[x']\right)-\int_{\mathbb{R}^{d-1}} \dd^{d-1}\vec{x}'\,\partial_i G^{(\mathrm{E})}_{d}J_0[\eta,\vec x'] .
\label{DotSpin1_Minkowski}
\end{align}
We see that the acausal term containing $G^{(\mathrm{E})}_{d}$ in eq.~\eqref{DotSpin1_Minkowski} is canceled by adding to $\dot{\alpha}_i$ the spatial gradient of eq.~\eqref{EMPhi_Minkowski}. The result is
\begin{align}
F_{0i}[x]=\dot\alpha_i[x]+\partial_i\Phi[x]=-\int_{\mathbb{R}^{d-1,1}} \dd^dx'\,\left(\dot G^+_{d}J_i[x']-\partial_iG^+_{d}J_0[x']\right) .
\label{F0i_Minkowski}
\end{align}
which, again, agrees exactly with eq.~\eqref{FieldStrength_Sol_Minkowski}. To sum: the spin-1 photon $\alpha_i$ contains all relevant electromagnetic information, but is acausal. On the other hand, the primary role of $\Phi$ is to cancel the acausal part of $\dot\alpha_i$, rendering the electric field $F_{0i}$ strictly causal. In other words, the electric field turns out to be determined by the causal portion of the velocity of the transverse spin-1 field,
\begin{align}
F_{0i}=\big(\dot{\alpha}_i\big)_{\text{causal}}=\big(\dot A_i^\text{T}\big)_{\text{causal}} .
\label{F0i_CausalSpin1_Minkowski}
\end{align}
Next, we move on to investigate how the spin-1 field and the Faraday tensor behave under certain physically interesting limits.

{\bf Stationary Limit and $\Phi$} \qquad  That the electric field in eq.~\eqref{F0i_CausalSpin1_Minkowski} is the causal piece of $\dot{\alpha}_i$ reveals a subtlety in the stationary limit, where the electric current is time independent. For, the first term containing $\dot{G}_d^+$ in eq.~\eqref{DotSpin1_Minkowski} integrates to zero, which then informs us that
\begin{align}
\dot\alpha_i[\vec{x}]
=\int_{\mathbb{R}^{d-1}} \dd^{d-1}\vec{x}'\partial_i G^{(\mathrm{E})}_{d}[\vec{x}-\vec{x}'] J_0[\vec{x}']-\int_{\mathbb{R}^{d-1}} \dd^{d-1}\vec{x}'\,\partial_i G^{(\mathrm{E})}_{d}[\vec{x}-\vec{x}']J_0[\vec x'] = 0 ;
\label{alphadot_GJ-GJ}
\end{align}
because in the second term of eq.~\eqref{DotSpin1_Minkowski},
\begin{align}
\int_{\mathbb{R}} \dd \eta' G^+_{d}[\eta-\eta',\vec{x}-\vec{x}']
= G^{(\mathrm{E})}_{d}[\vec{x}-\vec{x}'] .
\end{align}
In words: within the stationary limit, the causal structure of $\dot{\alpha}_i$ itself becomes degenerate -- the otherwise causal and acausal terms in eq.~\eqref{DotSpin1_Minkowski} cancel one another.

At first sight, eq.~\eqref{EMPoisson_Minkowski} appears to tell us $\Phi$ is the Coulomb potential of a static charge distribution. This seems to be further reinforced by the fact that $\partial_i \Phi$ from eq.~\eqref{EMPhi_Minkowski} is the sole contribution to the electric field $F_{0i}$ in eq.~\eqref{F0i_Minkowski}, since $\dot{\alpha}_i = 0$. But the interpretation that $\partial_i \Phi$ is (the dominant piece of) the electric force becomes erroneous once there is the mildest non-trivial time dependence in the electric current -- as already pointed out -- because $\Phi$ is purely acausal and hence cannot be a standalone physical observable. Instead, the gradient of eq.~\eqref{EMPhi_Minkowski} cancels the (normally acausal) second term in the first equality of eq.~\eqref{alphadot_GJ-GJ} and thus eq.~\eqref{F0i_CausalSpin1_Minkowski} continues to hold:
\begin{align}
F_{0i}[\vec x] = \big(\dot{\alpha}_i \big)_\text{causal}
= \int_{\mathbb{R}^{d-1}} \dd^{d-1}\vec{x}'\partial_i G^{(\mathrm{E})}_{d}[\vec{x}-\vec{x}'] J_0[\vec{x}'] .
\end{align}
{\bf Far-Zone Limit} \qquad  Provided that the observer is very far from the isolated sources, the leading-order term of the field, which scales as $1/r^{\frac d2-1}$, corresponds to the radiative piece that is capable of carrying energy-momentum to infinity. To extract the leading contribution of the spin-1 field in such a regime, firstly, we re-express the spin-1 commutator $C_{ij}$, by explicitly carrying out the spatial derivatives assuming $R\neq0$, in the following form\footnote{At $R=0$, when two spatial derivatives $\partial_i\partial_j$ act on $1/R^{d-3}$, terms involving $\delta^{(d-1)}[\vec x-\vec x']$ could arise. However, if the observer is away from the source, then those local terms will not contribute to the effective Green's function, and therefore, we can simply ignore them in the calculation.}
\begin{align}
C_{ij}[T,\vec R]= P_{ij}[\vec R]\,C_{1,d}[T,R]+\Pi_{ij}[\vec R]\,2\pi C_{2,d+2}[T,R],
\label{Cij_FarZone_Minkowski}
\end{align}
where the projection tensor $P_{ij}$ and $\Pi_{ij}$, respectively, are defined as
\begin{align}
 P_{ij}[\vec R]&\equiv\delta_{ij}-\widehat{R}_i\widehat{R}_j,
 \label{TransverseProjector_Position}  \\
\Pi_{ij}[\vec R]&\equiv -\delta_{ij}+(d-1)\widehat{R}_i\widehat{R}_j ;
\end{align}
the $\widehat R \equiv (\vec x-\vec x')/|\vec x-\vec x'|$ is pointing from the source location $\vec x'$ to the observer $\vec x$; and $C_{2,d+2}$ in the second term of eqs.~\eqref{Cij_FarZone_Minkowski} is the $(d+2)$-dimensional form of eqs.~\eqref{C2_EvenMinkowski_Recursion} and \eqref{C2_OddMinkowski_Recursion} but its $R$ is the one in $d-1$ spatial dimensions. Note also that, to reach eq.~\eqref{Cij_FarZone_Minkowski}, we have employed the homogeneous wave equation, $\partial^2C_{2,d}=0$, with $R\neq0$, as well as the conversion $\ddot C_{2,d}=-C_{1,d}$, to relate its second spatial derivatives to $C_{1,d}$. Also, it can be checked directly that the expression \eqref{Cij_FarZone_Minkowski} for $R\neq0$ is indeed divergenceless. Altogether, as long as the observer is away from the source, we have an alternative expression for the spin-1 effective Green's function by inserting eq.~\eqref{Cij_FarZone_Minkowski} into $G^+_{ij}=-\Theta[T]C_{ij}$. The purpose of putting $C_{ij}$ in this form is that, the dominant far-zone contribution of the field can be extracted simply by comparing $C_{1,d}$ and $C_{2,d+2}$. Furthermore, each term in eq.~\eqref{Cij_FarZone_Minkowski} is manifestly finite for all spacetime dimensions in which photons exist, since there is no divergence incurred in $C_{1,d}$ and $C_{2,d+2}$ for $d\geq3$.

If $\tau_c$ and $r_c$ are respectively the characteristic time scale and proper size of the source, and $r$ is the observer-source distance, the far zone is defined as the limits $\tau_c/r \ll 1$ and $r_c/r \ll 1$. To perform this limit on the Green's function, we will work in frequency $\omega-$space. Specifically, the far zone then translates into $|\omega| r\gg1$. We shall be content in extracting the leading expressions in the limit $|\omega| R\gg1$.

In terms of the superposition of individual frequencies, the spin-1 field $\alpha_i$ can be written as
\begin{align}
\label{FrequencyTransformOfGij}
\alpha_i[\eta,\vec x]=-\int_{\mathbb{R}^{d-1}} \dd^{d-1}\vec{x}'\int_{\mathbb{R}}  \frac{\dd\omega}{2\pi}\,\widetilde{G}^+_{ij}[\omega,\vec R]e^{-i\omega \eta}\widetilde{J}_j[\omega,\vec{x}'],
\end{align}
with $\widetilde{G}^+_{ij}[\omega,R]$ being the frequency transform of the spin-1 effective Green's function,
\begin{align}
\widetilde{G}^+_{ij}[\omega,\vec R]&=\int_{\mathbb{R}}  \dd T\,G^+_{ij}[T,\vec R]\,e^{i\omega T}  \notag\\
&= P_{ij}[\vec R]\,\widetilde{G}^+_{d}[\omega,R]  -  \Pi_{ij}[\vec R]\,2\pi\widetilde{C}^+_{2,d+2}[\omega,R],
\label{Gij_FrequencyTransform}
\end{align}
where we have assumed $R\neq0$ and used the expression \eqref{Cij_FarZone_Minkowski}. The $\widetilde{G}^+_{d}[\omega,R]$ and $\widetilde{C}^+_{2,d+2}[\omega,R]$, respectively, denote the frequency transforms of the massless scalar Green's function $G^+_d$ and $C^+_{2,d+2}$.\footnote{The $\widetilde{G}_d^+[\omega,R]$ in eq.~\eqref{Gij_FrequencyTransform} is the frequency transform of $G_{d}^+[\eta-\eta';R]$; this is to be distinguished from $\widetilde{G}_d^+[\eta,\eta';\vec{k}]$ in eq.~\eqref{Spin1_Fourier}, which is the spatial-Fourier transform of the same $G_{d}^+[\eta-\eta';R]$.}

{\it Spin-1 photons in even dimensions $d\geq4$} \qquad  A direct calculation starting from eqs.~\eqref{C1_Even_Minkowski} and \eqref{C2_Even_Minkowski} tells us, in all even $d = 4 + 2n \geq 4$,
\begin{align}
\widetilde{G}^+_{4+2n}[\omega,R]& = -\mathcal{D}_R^n \left( \frac{e^{i\omega R}}{4\pi R} \right) = -\frac{i\omega^{2n+1}}{2(2\pi)^{n+1}(\omega R)^n}h^{(1)}_n[\omega R],
\label{C1_Even_FrequencySpace} \\
\widetilde{C}^+_{2,6+2n}[\omega,R]&=\mathcal D^n_R\left(\frac{e^{i\omega R}}{8\pi^2 R^2(i\omega)}-\frac{e^{i\omega R}}{8\pi^2 R^3(i\omega)^2}\right)  -  \frac{(2n+1)!!\,\omega^{2n+1}}{2(2\pi)^{n+2}(\omega R)^{2n+3}},
\label{C2_Even_FrequencySpace}
\end{align}
where $h^{(1)}_n$ is the spherical Hankel function of the first kind. Notice that the last term in eq.~\eqref{C2_Even_FrequencySpace} does not contain the factor $e^{i\omega R}$; it describes the non-propagating portion of the signal in frequency space, which in turn arises from the acausal effect found in position space. In the limit $|\omega|R\gg1$, eqs.~\eqref{C1_Even_FrequencySpace} and \eqref{C2_Even_FrequencySpace} behave asymptotically as
\begin{align}
\widetilde{G}^+_{4+2n}[\omega,R]&= \frac{(-1)^{n+1}i^n\omega^{2n+1}}{2(2\pi\omega R)^{n+1}}\,e^{i\omega R}\left(1+\mathcal{O}\bigg[\frac1{\omega R}\bigg]\right),
\label{C1_Even_FarZone_FrequencySpace}\\
\widetilde{C}^+_{2,6+2n}[\omega,R]&=\frac{(-1)^{n+1}i^n\omega^{2n+1}}{2(2\pi\omega R)^{n+1}}\,e^{i\omega R}\cdot\frac{i}{2\pi \omega R}\left(1+\mathcal{O}\left[\frac1{\omega R}\right]\right),
\label{C2_Even_FarZone_FrequencySpace}
\end{align}
which reveals that, for any fixed dimension $d=4+2n$ and at leading order, the acausal $\widetilde{C}^+_{2,6+2n}[\omega,R]$ term is suppressed as $1/(\omega R)$ relative to $\widetilde{G}^+_{4+2n}$.\footnote{Strictly speaking, eq.~\eqref{C1_Even_FarZone_FrequencySpace} applies only for $n > 0$. There are no $1/(\omega R)$ corrections in (3+1)-dimensions, because eq.~\eqref{C1_Even_FrequencySpace} informs us that $\widetilde{G}^+_{4}[\omega,R] = - e^{i\omega R}/(4\pi R)$.} Therefore, at leading $1/(\omega R)$ order, the effective Green's function $\widetilde{G}^+_{ij}[\omega,\vec R]$, in frequency space, is exclusively dependent on $\widetilde{G}^+_{d}[\omega,R]$; moreover, with the assumption $r_c/r\ll1$, its far-zone leading contribution can be extracted from the first term of eq.~\eqref{Gij_FrequencyTransform}, which is given by
\begin{align}
\widetilde{G}^+_{ij}[\omega,\vec R]&=P_{ij}\widetilde{G}^{(+,\,\text{fz})}_{4+2n}[\omega;\vec x,\vec x']\left(1+\mathcal{O}\left[\frac1{\omega r},\frac{r_c}{r}\right]\right),
\end{align}
where $P_{ij}$ is the far-zone spatial projector defined in eq.~\eqref{t_Projector} and
\begin{align}
\widetilde{G}^{(+,\,\text{fz})}_{4+2n}[\omega;\vec x,\vec x']\equiv\frac{(-1)^{n+1}(i\omega)^n}{2(2\pi)^{n+1}}\frac{e^{i\omega (r-\vec{x}'\cdot\widehat r)}}{r^{n+1}} .
\label{G+_FrequencySpace_Even_FarZone}
\end{align}
By performing the inverse frequency transform, in the far-zone radiative limit, the transverse spin-1 photon $\alpha_i=A^{\text T}_i$ reduces to a transverse projection in space:
\begin{align}
\lim_{r\to\infty} \alpha_i \to A_i^\text t, \qquad  A_i^\text t[x]\equiv P_{ij}\left(-\int_{\mathbb{R}^{d-1,1}} \dd^{d}x'\,G^{(+,\,\text{fz})}_{d}[T;\vec x,\vec x']J_j[x']\right),
\label{At_FarZone}
\end{align}
where $G^{(+,\,\text{fz})}_{d}[T;\vec x,\vec x']$ is the far-zone contribution of the massless scalar Green's function,
\begin{align}
G^{(+,\,\text{fz})}_{d}[T;\vec x,\vec x']&=\int_{\mathbb{R}}  \frac{\dd\omega}{2\pi}\,\widetilde{G}^{(+,\,\text{fz})}_{d}[\omega;\vec x,\vec x']e^{-i\omega T} \notag\\
&=-\frac{1}{2(2\pi r)^{\frac{d-2}2}}\left(\frac{\partial}{\partial \eta}\right)^\frac{d-4}2\!\delta\big[T-r+\vec{x}'\cdot\widehat r\big]. \quad(\text{even }d)
\label{G+_Even_FarZone}
\end{align}

{\it Spin-1 photons in odd dimensions $d\geq3$} \qquad  For $d=3+2n$, we can frequency transform the retarded position-space solutions \eqref{C1_Odd_Minkowski} and \eqref{C2_Odd_Minkowski} to obtain
\begin{align}
\widetilde{G}^+_{3+2n}[\omega>0,R]&=-\frac{i\omega^{2n}}{4(2\pi\omega R)^n}H^{(1)}_n[\omega R],
\label{C1_Odd_FrequencySpace} \\
\widetilde{C}^+_{2,5+2n}[\omega>0,R]&= \frac{i\omega^{2n}}{4(2\pi\omega R)^{n+1}}H^{(1)}_{n+1}[\omega R] - \frac{2^nn!\,\omega^{2n}}{(2\pi)^{n+2}(\omega R)^{2n+2}},
\label{C2_Odd_FrequencySpace}
\end{align}
where $H^{(1)}_n$ is the Hankel function of the first kind and its differential recursion relation has been employed, and these expressions are only valid for positive frequencies $\omega>0$; however, since $G^+_{d}$ and $C^+_{2,d+2}$ are real, the negative-frequency modes can be expressed in terms of the complex conjugates of eqs.~\eqref{C1_Odd_FrequencySpace} and \eqref{C2_Odd_FrequencySpace}, $\widetilde{G}^+_{d}[-\omega,R]=\widetilde{G}^{+*}_{d}[\omega,R]$ and $\widetilde{C}^+_{2,d+2}[-\omega,R]=\widetilde{C}^{+*}_{2,d+2}[\omega,R]$, where the asterisk ``$*$'' denotes complex conjugation. As in the even-dimensional case, the non-propagating piece of signals also shows up in the second term of eq.~\eqref{C2_Odd_FrequencySpace}. At leading $1/(\omega R)$ order, the Hankel function goes asymptotically to $H^{(1)}_n[\omega R]=\sqrt{2/\pi\omega R}\,e^{i(\omega R-n\pi/2-\pi/4)}+\mathcal{O}\left[1/(\omega R)^{3/2}\right]$, so
we can read off the leading-order pieces of eqs.~\eqref{C1_Odd_FrequencySpace} and \eqref{C2_Odd_FrequencySpace} accordingly,
\begin{align}
\widetilde{G}^+_{3+2n}[\omega>0,R]&=-\frac{i\omega^{2n}}{2(2\pi\omega R)^{n+\frac12}}e^{i\left(\omega R-\frac{n\pi}2-\frac{\pi}4\right)}\left(1+\mathcal{O}\bigg[\frac1{\omega R}\bigg]\right),
\label{G+_Odd}\\
\widetilde{C}^+_{2,5+2n}[\omega>0,R]&=-\frac{i\omega^{2n}}{2(2\pi\omega R)^{n+\frac12}}e^{i\left(\omega R-\frac{n\pi}2-\frac{\pi}4\right)}\cdot\frac i{2\pi \omega R}\Bigg(1+\mathcal{O}\left[\frac1{\omega R}\right]+\mathcal O\left[\frac1{(\omega R)^{n+\frac12}}\right]\Bigg),
\label{C2+_Odd}
\end{align}
from which we infer that, unlike the even-dimensional results, amplitudes of these tail signals contain fractional powers of frequencies. And, these asymptotic behaviors tell us that, in the far-zone regime $|\omega| R\gg1$, the massless scalar Green's function $\widetilde{G}^+_{d}[\omega,R]$ still dominates over $\widetilde{C}^+_{2,d+2}[\omega,R]$ here. Hence, we can extract the far-zone leading order in $1/r$ piece of $\widetilde{G}^+_{ij}[\omega,\vec R]$ in the same way,
\begin{align}
\label{Gij+_FrequencySpace_Odd_FarZone}
\widetilde{G}^+_{ij}[\omega>0,\vec R]=P_{ij}\widetilde{G}^{(+,\,\text{fz})}_{3+2n}[\omega;\vec x,\vec x']\left(1+\mathcal{O}\left[\frac1{\omega r},\frac{r_c}{r}\right]\right),
\end{align}
with $\widetilde{G}^{(+,\,\text{fz})}_{3+2n}$ defined by
\begin{align}
\widetilde{G}^{(+,\,\text{fz})}_{3+2n}[\omega;\vec x,\vec x']\equiv-\frac{i\omega^{n-\frac12}}{2(2\pi)^{n+\frac12}}\frac{e^{i\left(\omega (r-\vec{x}'\cdot\widehat r)-\frac{n\pi}2-\frac{\pi}4\right)}}{r^{n+\frac12}}.
\label{G+_FrequencySpace_Odd_FarZone}
\end{align}
Consequently, in the radiative limit $r\to\infty$, we reach the same conclusion stated in eq.~\eqref{At_FarZone} for odd dimensions $d=3+2n$ as well, where $G^{(+,\,\text{fz})}_{d}$ is given instead by
\begin{align}
G^{(+,\,\text{fz})}_{d}[T;\vec x,\vec x']=-\frac{1}{(2\pi r)^{\frac{d-2}2}}\,\mathrm{Re}\left[\int^{\infty}_{0} \frac{\dd\omega}{2\pi}\,i\omega^{\frac{d-4}2}e^{i\left(\omega (r-\vec{x}'\cdot\widehat r-T)-\frac{(d-2)\pi}4\right)}\right]. \quad (\text{odd }d)
\label{G+_Odd_FarZone}
\end{align}
Thus, based on the frequency-space analysis, the fact that $\alpha_i\equiv A_i^\text T\to A_i^\text t$ as $r\to\infty$ holds generically in any spacetime dimensions $d\geq3$, clearly demonstrating that the acausal portion of the spin-1 field actually contributes negligibly to the far-zone signals.

{\it Summary: Far Zone Transverse Green's Functions} \qquad To sum, the massless spin-1 transverse photon $\alpha_i$ in the radiative regime will coincide with another notion of the ``transverse'' vector potential $A^\text t_i$. While $\alpha_i$ in eq.~\eqref{Spin1Photon_Minkowski} involves a transverse projection in Fourier space, the $A_i^\text{t}$ in eq.~\eqref{At_FarZone} is a local-in-space transverse projection of the far-zone Lorenz-gauge causal solution for the vector potential, i.e., $\widehat{r}^i A_i^\text{t} = 0$, which consists solely of the light-cone signals in even dimensions. In 4D Minkowski spacetime, that the two different notions of the transverse vector potential overlap in the far zone has already been pointed out in Refs.~\cite{Ashtekar:2017wgq,Ashtekar:2017ydh}. The method used here allows us to generalize the conclusion to all dimensions.

{\it Faraday tensor} \qquad Since we have already shown the spin-1 photons $\alpha_i$, for all $d\geq3$, reduce asymptotically to causal $A_i^\text t$ in the far zone, we then expect in this regime the magnetic and electric fields, eqs.~\eqref{Fij_Minkowski} and \eqref{F0i_CausalSpin1_Minkowski}, to become
\begin{align}
F_{ij}&\approx -2\,\widehat r_{[i}\dot A^{\text t}_{j]}, \label{Fij_FarZone}\\
F_{0i}&\approx \dot A^{\text t}_i .
\label{F0i_FarZone}
\end{align}
Here, the far-zone limit has been taken on both sides of the equations, and we have also used the results in eqs. \eqref{C1_Even_FarZone_FrequencySpace}, \eqref{C2_Even_FarZone_FrequencySpace}, \eqref{G+_Odd} and \eqref{C2+_Odd} to deduce -- as far as the leading-order contribution is concerned -- the replacement rule $\partial_i=-\partial_{i'}=-\widehat R_i\partial_0$ holds at the leading $1/(\omega R)$ level, and after which the dominant far-zone contribution in terms of $r$ can be extracted readily. In addition, the far-zone expressions \eqref{Fij_FarZone} and \eqref{F0i_FarZone} can also be checked for consistency through eq.~\eqref{FieldStrength_Sol_Minkowski}, by using the replacement rule as well as the conservation law for the charge current.

{\bf Commutator of Spin-1 Photons} \qquad As already alluded to, the results for the retarded Green's function of the massless spin-1 $\alpha_i$ are intimately related to the commutator of these photon operators in Quantum Field Theory. Let us first consider a free scalar field $\phi$ as a simple example. Its commutator is related to $C_{1,d}$ in the following manner:
\begin{align}
\big[\phi[x],\phi[x']\big]=-iC_{1,d}[T,R].
\end{align}
According to eq.~\eqref{C1_Minkowski_RetardedMinusAdvanced}, since the retarded/advanced Green's functions on the right hand side are strictly zero outside the null cone, this $C_{1,d}$ consists of only causal information -- i.e., it too is zero whenever the two spacetime points are spacelike: $(x-x')^2>0$. In contrast, because the spin-1 Green's functions are non-zero outside the light cone, according to eq.~\eqref{Cij_Minkowski_RetardedMinusAdvanced}, the non-interacting spin-1 commutator is therefore acausal:
\begin{align}
\big[\alpha_i[x],\alpha_j[x']\big]=-iC_{ij}[T,\vec R].
\label{Spin1_Commutator}
\end{align}
In Quantum Field Theory, operators that commute outside the light cone are said to obey micro-causality. Free spin-1 photons are therefore seen to violate micro-causality. It is likely that this acausal character of their commutator is a manifestation of the known tension between Lorentz covariance and gauge invariance when constructing massless helicity-1 theories in flat spacetime.

\subsection{Linearized Gravitation}

We now turn to the linearized theory of General Relativity in a Minkowski background, as described in \S\eqref{Chapter_Observables}. The relevant Green's functions will be computed analytically for all spacetime dimensions $d\geq4$; i.e., excluding those without spin-2 degrees of freedom.

{\bf Field Equations} \qquad  The gauge-invariant form of the linearized Einstein's equations can be expressed in terms of the variables defined in eqs.~\eqref{Bardeen_Psi}, \eqref{Bardeen_Phi}, and \eqref{Bardeen_VandDij}, where, as a constrained system, the full set of gauge-invariant field equations can be reduced to four fundamental ones, i.e., eq.~\eqref{PhiPsiRelationship} and the following three \cite{Chu:2016ngc}:
{\allowdisplaybreaks
\begin{align}
(d-2)\vec\nabla^2\Psi&=8\pi G_\text N\rho, \label{Psi_Poisson_Minkowski}\\
\vec\nabla^2 V_i&=-16\pi G_\text N \Sigma_i, \label{Vi_Poisson_Minkowski}\\
\partial^2D_{ij}&=-16\pi G_\text N\sigma_{ij}; \label{Spin-2_WaveEquation_Minkowski}
\end{align}}%
where the source terms $\rho$, $\Sigma_i$, and $\sigma_{ij}$ refer to different parts of the scalar-vector-tensor decomposition of the astrophysical stress-energy tensor $^{(\text{a})}T_{\mu\nu}$ (cf.~eqs.~\eqref{Astro_SVT_IofII} and \eqref{Astro_SVT_IIofII}). These four independent equations, along with the law of energy-momentum conservation, $\partial^\mu\, ^{(\text{a})}T_{\mu\nu}=0$, already imply the other three remaining ones in the linearized Einstein's equations. We see that only the spin-2 graviton field, $D_{ij}\equiv \chi^{\text{TT}}_{ij}$, admits dynamical wave solutions, sourced by the TT portion of $^{(\text{a})}T_{\mu\nu}$, while the Bardeen scalar potential $\Psi$, as well as the vector mode $V_i$, obey Poisson-type ones. This set of equations appear to be similar to their electromagnetic counterparts, and thus, by the same arguments used earlier, we already expect these gauge-invariant variables to be acausal in nature once the GW sources are taken into account. In this sense, none of these gauge-invariant variables -- including the spin-2 $D_{ij}$ -- may be regarded as a standalone observable. Indeed, as we will see in the subsequent discussion, the linearized Riemann tensor $\delta_1 R_{0i 0j}$, discussed in \S\eqref{Chapter_Observables}, in close analogy to the field strength $F_{\mu\nu}$ for electromagnetism, does require all their contributions to become a causal object.

{\bf Spin-2 Gravitons} \qquad  The analytic solutions for the effective Green's functions are crucially important for capturing the propagation of wave signals in this linearized system. Here, we start with the massless spin-2 field $D_{ij}$, obeying the wave equation \eqref{Spin-2_WaveEquation_Minkowski}. Since the TT projection of the source takes place locally in Fourier space, as long as $\vec k\neq\vec0$, we can firstly express $D_{ij}$ as
\begin{align}
D_{ij}[\eta,\vec x]=-16\pi G_\text N\int_{\mathbb{R}} \dd\eta'\int_{\mathbb{R}^{d-1}} \frac{\dd^{d-1}\vec{k}}{(2\pi)^{d-1}}\,\widetilde G^+_{d}[\eta,\eta';\vec k]\,\widehat P_{ijmn}[\vec k]\,
^{(\text{a})}\widetilde{T}_{mn}[\eta',\vec{k}]\, e^{i\vec{k}\cdot\vec{x}},
\label{Spin2_Fourier}
\end{align}
where $\widetilde G^+_d [\eta,\eta';\vec k ]$ is given in eq.~\eqref{FourierG+_Minkowski}, and the spin-2 TT projector $\widehat P_{ijmn}[\vec k]$ is defined in eq.~\eqref{GravitonProjector}. Then, the expression \eqref{Spin2_Fourier}, with each $k_j$ in eq.~\eqref{GravitonProjector_k} replaced by a spatial derivative via $\partial_j \to i k_j$, can be re-written as the spin-2 effective Green's function $G^+_{ijmn}$ convolved against the local stress-energy tensor of the source,
\begin{align}
D_{ij}[x]=-16\pi G_\text N\int_{\mathbb{R}^{d-1,1}} \dd^dx'G^+_{ijmn}[T,\vec R]\,^{(\text{a})}T_{mn}[x'],
\label{Spin2_Convolution_Minkowski}
\end{align}
where the spin-2 effective Green's function $G^+_{ijmn}$ is given by the following tensor structure,
\begin{align}
G^+_{ijmn}[T,\vec R]=\,&-\Theta[T]C_{ijmn}[T,\vec R],  \notag\\
C_{ijmn}[T,\vec R]=\,&\bigg(\delta_{m(i}\delta_{j)n}-\frac{\delta_{ij}\delta_{mn}}{d-2}\bigg)C_{1,d}[T,R]+\bigg(\delta_{m(i}\partial_{j)}\partial_n+
\delta_{n(i}\partial_{j)}\partial_m     \notag\\
& \qquad \qquad  -\frac{\delta_{ij}\partial_m\partial_n-\delta_{mn}\partial_i\partial_j}{d-2}
\bigg)C_{2,d}[T,R] +\left(\frac{d-3}{d-2}\right)\partial_i\partial_j\partial_m\partial_nC_{3,d}[T,R],
\label{Cijmn_Minkowski}
\end{align}
with $C_{1,d}$ and $C_{2,d}$ defined previously in eqs.~\eqref{C1_Minkowski} and \eqref{C2_Minkowski}, and $C_{3,d}$ defined by
\begin{align}
C_{3,d}[T,R]\equiv\int_{\mathbb{R}^{d-1}}\frac{\dd^{d-1}\vec{k}}{(2\pi)^{d-1}}\frac{\sin |\vec k|T}{|\vec k|^5}e^{i\vec{k}\cdot\vec{R}}.
\label{C3_Minkowski}
\end{align}
Compared with the spin-1 photon case, even though the whole tensor structure of $C_{ijmn}$ here is very different than that of $C_{ij}$ (cf.~eq.~\eqref{G+ij_Minkowski}), the first two terms have structural similarity to $C_{ij}$, and the scalar Fourier integrals $C_{1,d}$ and $C_{2,d}$ have already been dealt with analytically; the only new term that remains to be computed is $C_{3,d}$. Moreover, it can be checked readily that, by employing the relations $\vec\nabla^2 C_{2,d}=-C_{1,d}$ and $\vec\nabla^2 C_{3,d}=-C_{2,d}$ (cf.~eqs.~\eqref{C1_Minkowski}, \eqref{C2_Minkowski}, and \eqref{C3_Minkowski}), the expression \eqref{Cijmn_Minkowski} indeed satisfies the TT conditions $\delta^{ij}G^+_{ijmn}=0$ and $\partial_iG^+_{ijmn}=\partial_jG^+_{ijmn}=0$. We shall soon deploy the time-integral method, which amounts to solving
\begin{align}
\ddot{C}_{3,d} = -C_{2,d} .
\label{C3ddot_Minkowski}
\end{align}
By integrating eq.~\eqref{C3ddot_Minkowski} twice, followed by recalling eq.~\eqref{TimeIntegralMethod_C2d}, we have
{\allowdisplaybreaks
\begin{align}
C_{3,d}[T,R] &= -\int_{0}^{T} \dd \tau_2 \int_{0}^{\tau_2} \dd \tau_1 \, C_{2,d}[\tau_1,R]
+ T \dot{C}_{3,d}[T=0,R]
+ C_{3,d}[T=0,R], \\
&=\int^T_0 \dd\tau_4\int^{\tau_4}_0 \dd\tau_3\,\int^{\tau_3}_0 \dd\tau_2\int^{\tau_2}_0 \dd\tau_1\,C_{1,d}[\tau_1,R]+\frac{T^3}6G^{\mathrm{(E)}}_d[R]  +  TD_d[R],
\label{TimeIntegralMethod_C3d}
\end{align}}%
where the initial conditions $C_{3,d}[T=0,R]=0$ and $\dot C_{3,d}[T=0,R]=D_d[R]$ have been employed (cf.~eq.~\eqref{C3_Minkowski}), with $D_{d}$ defined by
\begin{align}
D_{d}[R] & \equiv \int_{\mathbb{R}^{d-1}}\frac{\dd^{d-1}\vec{k}}{(2\pi)^{d-1}}\frac{e^{i\vec{k}\cdot\vec{R}}}{\vec k^4},
\label{Dd}
\end{align}
whose concrete position-space expressions read
{\allowdisplaybreaks
\begin{align}
D^{(\text{reg})}_{3+2\epsilon}[R] & =  -\frac{R^2}{16\pi}  \left(  \frac1\epsilon - \big(\gamma + \ln[\pi] -1\big)  -  2 \ln[\mu R]     \right)  ,
\label{D3}       \\
D^{(\text{reg})}_{4}[R] & =  -\frac{R}{8\pi}      ,
\label{D4}  \\
D^{(\text{reg})}_{5+2\epsilon}[R] & =   \frac1{16\pi^2}  \left( \frac1\epsilon -\gamma - \ln[\pi] - 2\ln[\mu R] \right)    ,
\label{D5}   \\
D_{d\geq6}[R] & = \frac{\Gamma\left[\frac{d-5}2\right]}{16\pi^{\frac{d-1}{2}}R^{d-5}} .
\label{Dd_Expression}
\end{align}}%
Note that $D_3$, $D_4$, and $D_5$ are in the dimensional-regularized forms. Finally, parallel to eq.~\eqref{Gij_WaveEquation_Minkowski} in the photon case, acausality encoded in $G^+_{ijmn}$ can be seen at the level of its wave equation,
\begin{align}
\partial^2G^+_{ijmn}[T,\vec R]=\,&  \delta[T] \Bigg\{ \bigg(\delta_{m(i}\delta_{j)n}-\frac{\delta_{ij}\delta_{mn}}{d-2}\bigg)\delta^{(d-1)}[\vec x-\vec x']-
\bigg(\delta_{m(i}\partial_{j)}\partial_n+\delta_{n(i}\partial_{j)}\partial_m\notag\\
& \qquad \qquad \qquad \quad -\frac{\delta_{ij}
\partial_m\partial_n-\delta_{mn}\partial_i\partial_j}{d-2}\bigg)G^{\text{(E)}}_{d}[R] +\left(\frac{d-3}{d-2}\right)\partial_i\partial_j\partial_m\partial_nD_{d}[R] \Bigg\}.
\label{Gijmn_WaveEquation_Minkowski}
\end{align}
The last two terms in eq.~\eqref{Gijmn_WaveEquation_Minkowski} correspond to the acausal contributions to the signal attributed to $x' \equiv (\eta',\vec{x}')$, but arising from a non-zero source smeared over the rest of the equal-time spatial hypersurface at $\eta = \eta'$.

To compute $C_{3,d}$ in eq.~\eqref{C3_Minkowski}, we first notice that the integral itself diverges when $d\leq5$, inferred from the power of $|\vec k|$ in the limit $|\vec k|\to0$. However, in the physical spacetime dimensions where spin-2 gravitons exist, namely $d\geq4$, those divergences in $C_{3,4}$ and $C_{3,5}$ are expected to be removed by the multiple spatial derivatives in the spin-2 effective Green's function \eqref{Cijmn_Minkowski}. Furthermore, since the dimension-raising operator still applies to $C_{3,d}$, in either even or odd spacetime dimensions, the lowest-dimensional form is already adequate for us to generate all the higher-dimensional ones.

{\it Exact solutions in even dimensions $d\geq4$} \qquad  Since the analytic forms of $C_{1,d}$ and $C_{2,d}$, for all even $d\geq4$, have been obtained in eqs.~\eqref{C1_Even_Minkowski} and \eqref{C2_Even_Minkowski}, we now focus on the $C_{3,d}$ term needed in eq.~\eqref{Cijmn_Minkowski}. Similar methods used to compute $C_{2,d}$ will be performed to tackle this new integral. Here, we start with $C_{3,4}$, the lowest even dimension for the spin-2 graviton to exist. Even though $C_{3,4}$ itself is a divergent integral, we can still extract its regularized finite contribution through the time-integral method, as we did for $C_{2,3}$ in the spin-1 calculation. Utilizing dimensional-regularization, the result turns out to be finite:
\begin{align}
C_{3,4}^{(+,\text{reg})}[T,R]&=\Theta[T-R]\,\frac{-R^2-3T^2}{24\pi}+ \Theta[T] \Theta[-T+R]\,\frac{-3R^2T-T^3}{24\pi R}.
\label{C34_Minkowski}
\end{align}
Similar to $C_{2,4}$, the tail function in eq.~\eqref{C34_Minkowski}, when plugged into eq.~\eqref{Cijmn_Minkowski}, will be eliminated by the spatial derivatives in $G^+_{ijmn}$. Whereas the acausal portion of the signals, from the second term of eq.~\eqref{C34_Minkowski}, will still remain. In addition, this regularized form can be justified by checking whether this expression, with one dimension-rasing operator acting on it, coincides with $C^+_{3,6}$ obtained by a direct contour-integral calculation:
\begin{align}
C^+_{3,6}[T,R]&=\Theta[T-R]\,\frac1{24\pi^2}+\Theta[T]\Theta[-T+R]\,\frac{3R^2T-T^3}{48\pi^2R^3}.
\label{C36_Minkowski}
\end{align}
(One may check, indeed, that $ C_{3,6} = \mathcal{D}_R C^{(+,\text{reg})}_{3,4}$.) With this solution at hand, applying dimension-raising operators to it then produces all the higher even-dimensional results,
\begin{align}
C^+_{3,\text{even }d\geq4}[T,R]&= \mathcal D^{\frac{d-4}2}_R  \left( \Theta[T-R]\,\frac{-R^2-3T^2}{24\pi}+ \Theta[T] \Theta[-T+R]\,\frac{-3R^2T-T^3}{24\pi R}  \right)  .
\label{C3_Even_Minkowski}
\end{align}
Now, plugging eq.~\eqref{C3_Even_Minkowski} for $C_{3,d}$, along with known $C_{1,d}$ and $C_{2,d}$, into the spin-2 effective Green's function \eqref{Cijmn_Minkowski}, we find that the spin-2 causal structure is analogous to that of the spin-1 field. More precisely, for all even dimensions $d\geq4$, no tail signals exist in $G^+_{ijmn}$, but the spin-2 graviton receives not only the light-cone signals but also the acausal ones from both $C_{2,d}$ and $C_{3,d}$.

{\it Exact solutions in odd dimensions $d\geq5$} \qquad For odd spacetime dimensions, we begin with $d=5$, since TT gravitons are non-existent in lower odd dimensions. To calculate $C_{3,5}$, we can make use of the time-integral method to first extract the regularized $C_{3,3}$, namely inserting eq.~\eqref{C13_Minkowski} into eq.~\eqref{TimeIntegralMethod_C3d} along with the dimensional-regularized initial conditions:
\begin{align}
C^{(+,\text{reg})}_{3,3+2\epsilon}[T,R] =\,&\Theta[T-R]\,\frac1{8\pi} \bigg(-\big(T^2-R^2\big)^{\frac32}-\frac1{9}\big(13R^2+2T^2\big)
\sqrt{T^2-R^2} + T\bigg(R^2+\frac23T^2\bigg) \notag\\
& \times \ln\left[\mu\left(T+\sqrt{T^2-R^2}\right)\right]\bigg)+ \Theta[T]\Theta[-T+R] \, \frac{T \ln[\mu R]}{8\pi}  \left(  R^2   + \frac{2}3 T^2 \right) \notag\\
&  - \Theta[T]\frac T{16\pi} \bigg( \bigg( R^2  +  \frac23 T^2 \bigg) \left(\frac1\epsilon - \gamma - \ln[\pi] \right) + R^2 \bigg).
\end{align}
By applying the raising operator $\mathcal D_R$ once,
\begin{align}
C^{(+,\text{reg})}_{3,5+2\epsilon}[T,R]=\,&\Theta[T-R] \, \frac1{24\pi^2} \left( \left(2+\frac{T^2}{R^2}\right)\sqrt{T^2-R^2}  -   3T\ln\left[\mu\left(
T+\sqrt{T^2-R^2} \right) \right]-\frac{T^3}{R^2}\right) \notag\\
&+\Theta[T]\Theta[-T+R]  \,  \frac1{24\pi^2} \left(-\frac{T^3}{R^2}  -  3T\ln[\mu R]\right) + \Theta[T] \, \frac{T}{16\pi^2}\left( \frac1\epsilon -\gamma  - \ln[\pi] \right),
\label{C35_Minkowski}
\end{align}
which is still a regularized expression of the divergent integral $C_{3,5}$, and again, whose validity can be justified in the same way. Through a direct computation of the finite integral $C_{3,7}$, as we did for $C_{2,5}$, we can show that the resulting expression of $C_{3,7}$, namely
\begin{align}
C^+_{3,7}[T,R]=\Theta[T-R]\,\frac{2\left(T^2-R^2\right)^{\frac32}+3R^2T-2T^3}{48\pi^3 R^4}+\Theta[T]\Theta[-T+R]\,\frac{3R^2T-2T^3}{48\pi^3 R^4},
\label{C37_Minkowski}
\end{align}
is simply $\mathcal{D}_R C^{(+,\text{reg})}_{3,5+2\epsilon}$. Following through the same procedures, we can extend the analytic solution \eqref{C35_Minkowski} to all odd dimensions $d\geq5$ via dimension-rasing operators, where, similar to the odd-dimensional photon case, we may discard the last term of eq.~\eqref{C35_Minkowski} and set $\mu=1$ as they will be eliminated in the physical $G^+_{ijmn}$ (cf.~eq.~\eqref{Cijmn_Minkowski}),
\begin{align}
C^+_{3,\text{odd }d\geq5}[T,R]=\,&\mathcal D_R^{\frac{d-5}2}\Bigg(  \Theta[T-R] \, \frac1{24\pi^2} \left( \left(2+\frac{T^2}{R^2}\right)\sqrt{T^2-R^2}  -   3T\ln\left[ \left(
T+\sqrt{T^2-R^2} \right) \right]-\frac{T^3}{R^2}\right) \notag\\
&+\Theta[T]\Theta[-T+R]  \,  \frac1{24\pi^2} \left(-\frac{T^3}{R^2}  -  3T\ln R\right)  \Bigg).
\label{C3_Odd_Minkowski}
\end{align}
As with spin-1 photons in odd dimensions, the spin-2 effective Green's function \eqref{Cijmn_Minkowski} in this case explicitly reveals that, besides pure tails from $C_{1,d}$, the spin-2 graviton receives extra tail and acausal contributions from both $C_{2,d}$ and $C_{3,d}$ for all odd $d\geq5$; and moreover, no signals traveling strictly on its past light cone -- namely, no $\delta$-function light-cone contributions.

Therefore, the acausal nature of the TT spin-2 field in all relevant spacetime dimensions is explicitly confirmed by the analytic solutions of $G^+_{ijmn}$ obtained in this section. Now, we proceed to solve for other gauge-invariant variables involved in this system.

{\bf Bardeen Scalars} \qquad  In a flat background, one of the Bardeen scalar potentials $\Psi$ obeys the Poisson's equation \eqref{Psi_Poisson_Minkowski}, which then leads to a Coulomb-type solution,
\begin{align}
\Psi[\eta,\vec x]=\frac{8\pi G_\text N}{d-2}  \int_{\mathbb{R}^{d-1}} \dd^{d-1}\vec{x}'\,   G^{(\mathrm{E})}_{d}[R] {}^{(\text{a})}T_{00}[\eta,\vec x'].
\label{BardeenPsi_Minkowski}
\end{align}
This transparently shows the acausal character of $\Psi$, since it is instantaneously sourced by the local matter energy density. The other Bardeen potential $\Phi$ is related to $\Psi$ via eq.~\eqref{PhiPsiRelationship}; recall that $\Upsilon$ is the nonlocal scalar portion of decomposed $^{(\text{a})}T_{ij}$ (cf.~eq.~\eqref{Astro_SVT_IIofII}), which, in Fourier space with $\vec k\neq\vec0$, is given by the following local projection \cite{Chu:2016ngc},
\begin{align}
\widetilde{\Upsilon}[\eta,\vec k]&=-\left(\frac{d-1}{d-2}\right)\frac1{\vec k^4}\bigg(k_ik_j-\frac{\vec k^2}{d-1}\delta_{ij}\bigg){}^{(\text{a})}\widetilde{T}_{ij}[\eta,\vec k].
\label{Upsilon_Fourier_Minkowski}
\end{align}
By virtue of this local decomposition, we can inverse Fourier transform eq.~\eqref{Upsilon_Fourier_Minkowski} to yield
\begin{align}
\Phi[\eta,\vec x]=\,&(d-3)\Psi[\eta,\vec x] \notag\\
&\quad+ \frac{8\pi G_\text N}{d-2} \int_{\mathbb{R}^{d-1}}  \dd^{d-1}\vec{x}'  \bigg(
G^{\mathrm{(E)}}_{d}[R]{}^{(\text{a})}T_{ll}[\eta,\vec{x}']
-(d-1)\partial_i\partial_jD_d[R]{}^{(\text{a})}T_{ij}[\eta,\vec{x}']\bigg),
\label{PhiPsiRelationship_Convolution}
\end{align}
where $^{(\text{a})}T_{ll}$ denotes the spatial trace of the $(ij)$ components of the matter stress-energy tensor, $^{(\text{a})}T_{ll}\equiv\delta^{ij}\,^{(\text{a})}T_{ij}$, and $D_d[R]$ is defined in eq.~\eqref{Dd}. Inserting eq.~\eqref{BardeenPsi_Minkowski} into eq.~\eqref{PhiPsiRelationship_Convolution}, we can express $\Phi$ itself in terms of the following convolution,
\begin{align}
\Phi[\eta,\vec x]=\,&\frac{8\pi G_\text N}{d-2}\int_{\mathbb{R}^{d-1}} \dd^{d-1}\vec{x}'\,\bigg((d-3)G^{(\mathrm{E})}_d[R]{}^{(\text{a})}T_{00}[\eta,\vec{x}']+G^{\mathrm{(E)}}_{d}[R]{}^{(\text{a})}T_{ll}[\eta,\vec{x}']
\notag\\
&-(d-1)\partial_i\partial_jD_d[R] {}^{(\text{a})}T_{ij}[\eta,\vec{x}'] \bigg),
\label{BardeenPhi_Minkowski}
\end{align}
where we see that $\Phi$ is effectively dependent on different components of the matter stress-energy tensor, weighted either by $G^{\mathrm{(E)}}_{d}$ or $D_d$ on the instantaneous $\eta=\eta'$ surface.

{\bf Vector Potential} \qquad  The gauge-invariant vector mode $V_i$, in linearized gravity, also obeys the Poisson's equation \eqref{Vi_Poisson_Minkowski}, but is instead sourced by the nonlocal transverse part of $^{(\text{a})}T_{0i}$ (cf.~eq.\eqref{Astro_SVT_IofII}). As before, since the decomposition is local in momentum space,
\begin{align}
\widetilde\Sigma_i[\eta,\vec k]&=\left(\delta_{ij}-\frac{k_ik_j}{\vec k^2}\right)\,^{(\text{a})}\widetilde T_{0j}[\eta,\vec k],
\label{Sigmai_Fourier_Minkowski}
\end{align}
the solution of eq.~\eqref{Vi_Poisson_Minkowski} can first be cast into a Fourier form using eq.~\eqref{Sigmai_Fourier_Minkowski}, and then translated back to the position space,
\begin{align}
V_i[\eta,\vec x]=16\pi G_\text N\int_{\mathbb{R}^{d-1}}   \dd^{d-1}\vec{x}'\left(\partial_i\partial_jD_d[R]
\,^{(\text{a})}T_{0j}[\eta,\vec{x}']-G^{\mathrm{(E)}}_d[R]{}^{(\text{a})}T_{0i}[\eta,\vec{x}']\right),
\label{Vi_Minkowski}
\end{align}
which is yet again an instantaneous acausal signal. Therefore, in Minkowski background, other than the spin-2 graviton field $D_{ij}$, the rest of the gauge-invariant variables depend exclusively on the weighted superposition of the matter sources evaluated at the instantaneous observer time $\eta$.

{\bf Linearized Riemann Tensor} \qquad In our discussion of gravitational observables in \S\eqref{Chapter_Observables}, we have argued that,  in a free-falling synchronous-gauge setup, the $\delta_1R_{0i0j}$ components of the linearized Riemann tensor encode the gravitational tidal forces exerted upon the neighboring test particles in the geodesic deviation equation (cf.~eq.~\eqref{GeodesicDeviation_FreeFall}). And, being also gauge-invariant in Minkowski spacetime, it would reasonably be regarded as a classical physical observable and expected to be strictly causal as well.

As with the Faraday tensor in electromagnetism, it can be directly shown via its second-order wave equation that the linearized Riemann tensor is causal with respect to the flat background. Firstly, by taking the divergence of the Bianchi identity obeyed by the exact Riemann tensor, followed by imposing Einstein's equations, one may obtain
\begin{align}
\label{Riemann_WaveEq}
\Box  R^{\rho\sigma}{}_{\mu\nu}  +  \big[\nabla^\lambda,\nabla^{[\rho}\big]R^{\sigma]}{}_{\lambda\mu\nu}  =  32\pi G_\text N\nabla^{[\rho}\nabla_{[\mu} \left(T_{\nu]}{}^{\sigma]}-\delta_{\nu]}^{\sigma]}\frac{T}{(d-2)}\right) .
\end{align}
The linearized version in Minkowski background is thus
\begin{align}
\partial^2\delta_1R^{\rho\sigma}{}_{\mu\nu}&=32\pi G_\text N\,\partial^{[\rho}\partial_{[\mu}\bigg({}^{(\text{a})}T_{\nu]}{}^{\sigma]}-\delta_{\nu]}^{\sigma]}\frac{^{(\text{a})}T}{(d-2)}\bigg),
\label{LinearizedRiemann_WaveEq_Minkowski}
\end{align}
where $^{(\text{a})}T$ is now the trace of the matter stress tensor in flat spacetime, namely $^{(\text{a})}T\equiv\eta^{\mu\nu}\,^{(\text{a})}T_{\mu\nu}$.\footnote{Strictly speaking, the matter stress tensor in eq.~\eqref{Riemann_WaveEq}, when perturbed around a background Minkowski spacetime $g_{\mu\nu} = \eta_{\mu\nu} + \chi_{\mu\nu}$, would typically admit an infinite series in $\chi_{\mu\nu}$. Whereas the matter stress tensor in eq.~\eqref{LinearizedRiemann_WaveEq_Minkowski} does not contain $\chi_{\mu\nu}$. The matter stress tensor appearing everywhere else in this paper denotes the latter.} The wave equation \eqref{LinearizedRiemann_WaveEq_Minkowski} then leads immediately to the fact that the linearized Riemann tensor is causally sourced by the second derivatives of the astrophysical stress-energy tensor, from which its $\delta_1R_{0i0j}$ components can be expressed explicitly as
\begin{align}
\delta_1R_{0i0j}[x]=\,&8\pi G_\text N\int_{\mathbb{R}^{d-1,1}}  \dd^dx'\Bigg\{\ddot{G}^+_{d}\bigg({}^{(\text{a})}T_{ij}[x']+\frac{\delta_{ij}}{(d-2)}\left({}^{(\text{a})}T_{00}[x']-{}^{(\text{a})}T_{ll}[x']
\right)\bigg)-2\partial_{(i}\dot G_{d}^+\,^{(\text{a})}T_{j)0}[x']\notag\\
&+\frac1{(d-2)}\partial_i\partial_jG^+_{d}\left((d-3){}^{(\text{a})}T_{00}[x']+{}^{(\text{a})}T_{ll}[x']\right)\Bigg\}.
\label{LinearizedRiemann_Sol_Minkowski}
\end{align}
To arrive at eq.~\eqref{LinearizedRiemann_Sol_Minkowski}, we have integrated by parts and dropped the boundary contributions evaluated at the spatial and past infinity. Not only does eq.~\eqref{LinearizedRiemann_Sol_Minkowski} show $\delta_1R_{0i0j}$ is completely causal for all $d\geq4$, it also provides a check for our calculations in the gauge-invariant approach.

As we have shown earlier, all the gauge-invariant variables -- the two scalars $\Psi$ and $\Phi$, one vector $V_i$, and one tensor $D_{ij}$ -- are acausal. From the similar issue encountered in the electromagnetic case, we would expect that, in describing the physical GW observables, a mutual cancellation of the acausal contributions must occur among these variables. Let us now check this statement more carefully using their analytic solutions. Given in eq.~\eqref{LinearizedRiemann_Minkowski_GaugeInvariant} is the linearized Riemann tensor $\delta_1R_{0i0j}$ expressed in terms of four gauge-invariant variables in Minkowski spacetime, all of which are non-trivial whenever matter sources are present. Notice that the spin-2 graviton field $D_{ij}$ enters $\delta_1R_{0i0j}$ through its acceleration; before taking time derivatives, the spin-2 field given in eq.~\eqref{Spin2_Convolution_Minkowski} can firstly be re-cast into another convolution,
{\allowdisplaybreaks
\begin{align}
D_{ij}[\eta,\vec x]=\,&-16\pi G_\text N\int_{\mathbb{R}^{d-1,1}} \dd^dx'\,\Bigg\{G^+_{d}\left({}^{(\text{a})}T_{ij}[x']+\frac{\delta_{ij}}{d-2}\left({}^{(\text{a})}T_{00}[x']-{}^{(\text{a})}T_{ll}[x']
\right)\right)
-2\,\Theta[T]\partial_{(i}\dot C_{2,d}{}^{(\text{a})}T_{j)0}[x']\notag\\
&+\frac{1}{d-2}\Theta[T]\partial_i\partial_jC_{2,d}\left((d-3){}^{(\text{a})}T_{00}[x']+{}^{(\text{a})}T_{ll}[x']\right)\Bigg\} \notag\\
&+ \frac{16\pi G_\text N}{d-2}\int_{\mathbb{R}^{d-1}}  \dd^{d-1}\vec x'  \bigg(\delta_{ij}G^{\mathrm{(E)}}_d{} ^{(\text{a})}T_{00}[\eta,\vec x']+(d-3)\partial_i\partial_jD_d{}^{(\text{a})}T_{00}[\eta,\vec x']\bigg),
\label{Spin2_Convolution2_Minkowski}
\end{align}}%
where we have utilized the conservation of the matter stress-energy tensor, ${}^{(\text{a})}\dot{T}_{0j}=\partial_{i}{}^{(\text{a})}T_{ij}$ and ${}^{(\text{a})}\ddot{T}_{00}=\partial_{i}\partial_{j}{}^{(\text{a})}T_{ij}$, the conversion properties $\ddot C_{2,d}=-C_{1,d}$ and $\ddot C_{3,d}=-C_{2,d}$, and the initial conditions $\dot C_{2,d}\big|_{T=0}=-G^{\mathrm{(E)}}_d$ and $\dot C_{3,d}\big|_{T=0}=D_d$. Also, as in the spin-1 case, the boundary terms from integrations by parts all vanish, which can be justified using the analytic expressions of $C_{2,d}$ and $C_{3,d}$ obtained above. Note also that, in eq.~\eqref{Spin2_Convolution2_Minkowski}, the second integral is performed on the equal-time surface, which is clearly acausal, but the whole expression for $D_{ij}$ is not yet a clean separation based on causality, since $C_{2,d}$ in the first integral still contains both causal and acausal pieces. Furthermore, we highlight that the energy-momentum conservation law is always assumed throughout this paper. However, this is no longer true in a self-gravitating system, such as the in-spiraling pairs of black holes/neutron stars whose GWs LIGO have detected to date. Conceptually speaking, to make the theory self-consistent, non-linear corrections of gravity must be incorporated into the right-hand side of the linearized wave equation, so that the conservation of total stress tensor remains valid at linear level. We hope to address this subtlety more systematically in future work.

Now, we proceed to take double-time derivative of the spin-2 field in eq.~\eqref{Spin2_Convolution2_Minkowski},
{\allowdisplaybreaks
\begin{align}
\ddot D_{ij}[\eta,\vec x]&=-16\pi G_\text N\int_{\mathbb{R}^{d-1,1}} \dd^dx'\,\Bigg\{\ddot G^+_{d}\left(^{(\text{a})}T_{ij}[x']+\frac{\delta_{ij}}{d-2}\left({}^{(\text{a})}T_{00}[x']-{}^{(\text{a})}T_{ll}[x']\right)\right)
-2\partial_{(i}\dot{G}^+_{d}{}^{(\text{a})}T_{j)0}[x']\notag\\
&+\frac{1}{d-2}\partial_i\partial_jG^+_{d}\left((d-3)\,{}^{(\text{a})}T_{00}[x']+{}^{(\text{a})}T_{ll}[x']\right)\Bigg\} \notag\\
&+16\pi G_\text N\int_{\mathbb{R}^{d-1}} \dd^{d-1}\vec{x}'\,\Bigg\{\! -2\partial_m\partial_{(i}G^{\mathrm{(E)}}_d
{}^{(\text{a})}T_{j)m}[\eta,\vec{x}']+\frac1{d-2}\bigg(\delta_{ij}\partial_m\partial_nG^{\mathrm{(E)}}_d{}^{(\text{a})}T_{mn}[\eta,\vec{x}'] \notag\\
&+\partial_i\partial_jG^{\mathrm{(E)}}_d\left((d-3){}^{(\text{a})}T_{00}[\eta,\vec{x}']+{}^{(\text{a})}T_{ll}[\eta,\vec{x}']\right)\bigg)
+\left(\frac{d-3}{d-2}\right)\partial_i\partial_j\partial_m\partial_nD_d{}^{(\text{a})}T_{mn}[\eta,\vec{x}']\Bigg\},
\label{DDotSpin2_Minkowski}
\end{align}}%
where the same properties used in the previous calculation have been employed to carry out the differentiation\footnote{Whenever a second time derivative acts on the expression involving a step function $\Theta[T]$, we make use of the following simplification for any function $F[\eta,\eta']$,
\begin{align}
\partial_\eta^2\left(\Theta[T]F[\eta,\eta']\vphantom{\dot A}\right)=\Theta[T]\ddot F[\eta,\eta']+\dot\delta[T]F[\eta',\eta']+\delta[T]\dot F[\eta,\eta']\big|_{\eta=\eta'}, \notag
\end{align}
where we have made the replacement $T\dot\delta[T]\to-\delta[T]$, which results from differentiating the identity $T\delta[T]=0$ with respect to $\eta$. Notice that the last two terms only contribute at $\eta=\eta'$, and this property will also be utilized later on in the cosmological case.
}, and we observe that the first integral in eq.~\eqref{DDotSpin2_Minkowski} is completely causal and is exactly $-2$ times the expression in eq.~\eqref{LinearizedRiemann_Sol_Minkowski}, whereas the second one is acausally performed over the equal-time hypersurface, which is therefore expected to connect to the other gauge-invariant variables. According to eq.~\eqref{LinearizedRiemann_Minkowski_GaugeInvariant}, the scalar and vector contributions to $\delta_1R_{0i0j}$ come from $\ddot\Psi$, $\partial_i\partial_j\Phi$, and $\partial_{(i} \dot{V}_{j)}$, the explicit forms of which can be readily deduced from eqs.~\eqref{BardeenPsi_Minkowski}, \eqref{BardeenPhi_Minkowski}, and \eqref{Vi_Minkowski},
{\allowdisplaybreaks
\begin{align}
\ddot \Psi[\eta,\vec x]=\,& \frac{8\pi G_\text N}{d-2}  \int_{\mathbb{R}^{d-1}} \dd^{d-1}\vec{x}'\,   \partial_m \partial_n G^{(\mathrm{E})}_{d} {}^{(\text{a})}T_{mn}[\eta,\vec x'],   \\
\partial_i\partial_j\Phi[\eta,\vec x]=\,&\frac{8\pi G_\text N}{d-2}\int_{\mathbb{R}^{d-1}} \dd^{d-1}\vec{x}'\,\bigg((d-3)\partial_i\partial_jG^{(\mathrm{E})}_d{}^{(\text{a})}T_{00}[\eta,\vec{x}']+
\partial_i\partial_jG^{\mathrm{(E)}}_{d}{}^{(\text{a})}T_{ll}[\eta,\vec{x}']
\notag\\
&-(d-1)\partial_i\partial_j\partial_m\partial_nD_d {}^{(\text{a})}T_{mn}[\eta,\vec{x}'] \bigg), \\
\partial_{(i}\dot{V}_{j)}[\eta,\vec x]  =\,&  -16\pi G_\text N \int_{\mathbb{R}^{d-1}} \dd^{d-1}\vec{x}'\left( \partial_m\partial_{(i}G^{\mathrm{(E)}}_d  {}^{(\text{a})}T_{j)m}[\eta,\vec{x}']
-\partial_i\partial_j\partial_m\partial_n
D_d {}^{(\text{a})}T_{mn}[\eta,\vec{x}']\right),
\end{align}}%
where the conservation law of $^{(\text{a})}T_{\mu\nu}$ allows us to switch between different components of the stress-energy tensor. As it turns out, the scalar and vector contributions, added together in accord with eq.~\eqref{LinearizedRiemann_Minkowski_GaugeInvariant}, do conspire to cancel the acausal portion of the acceleration of the spin-2 TT graviton completely, i.e., the second integral of eq.~\eqref{DDotSpin2_Minkowski}. As a result, the remaining part of eq.~\eqref{LinearizedRiemann_Minkowski_GaugeInvariant} is then strictly causal and exactly consistent with eq.~\eqref{LinearizedRiemann_Sol_Minkowski},
\begin{align}
\delta_1 R_{0i 0j}=\delta_{ij}\ddot{\Psi}+ \partial_i\partial_j\Phi+\partial_{(i}\dot{V}_{j)}-\frac12\ddot{D}_{ij}=-\frac12\left(\ddot D_{ij}\right)_{\text{causal}}=-\frac12\left(\vphantom{\dot D}\ddot \chi^{\text{TT}}_{ij}\right)_{\text{causal}},
\label{LinearizedRiemann_CausalSpin2_Minkowski}
\end{align}
which is valid in general weak-field situations. The physical insight gained from the gauge-invariant formalism is that the information regarding gravitational tidal forces is exclusively encoded within the causal part of the acceleration of the spin-2 field; whereas the acausal portion of $D_{ij}$ is completely canceled by the gauge-invariant Bardeen scalars and vector mode. This situation is very similar to that of the spin-1 field describing the electric field in electromagnetism.

{\bf Stationary Limit and $\Phi$} \qquad Like the photon case, the limit where the stress tensor ${}^{(\text a)}T_{\mu\nu}$ becomes time-independent leads to a degenerate causal structure for the acceleration of the spin-2 graviton; namely, its otherwise causal and acausal pieces cancel. In such a situation, one may further verify from eqs. \eqref{BardeenPsi_Minkowski}, \eqref{Vi_Minkowski}, and \eqref{DDotSpin2_Minkowski} that $\ddot{\Psi} = \dot{V}_i = \ddot{D}_{ij} = 0$, leaving the tidal forces to depend only on $\Phi$:
\begin{align}
\delta_1 R_{0i0j} = \partial_i \partial_j \Phi .
\end{align}
Since eq.~\eqref{LinearizedRiemann_CausalSpin2_Minkowski} holds in general, we may maintain that, despite appearances, it is really the acausal pieces of $\ddot{D}_{ij}$ -- which are equal in magnitude but opposite in sign to the causal ones -- that are canceling the $\partial_i \partial_j \Phi$. This interpretation ensures that causality is respected once there is the slightest time-dependence in the ${}^{(\text a)}T_{\mu\nu}$.
\begin{align}
\delta_1 R_{0i 0j}[\vec x]=-\frac12\left(\ddot D_{ij}\right)_{\text{causal}}=\frac{8\pi G_\text N}{d-2}\int_{\mathbb{R}^{d-1}} \dd^{d-1}\vec x'\,\partial_i\partial_jG^{(\mathrm{E})}_{d}\left((d-3){}^{(\text{a})}T_{00}[\vec x']+{}^{(\text{a})}T_{ll}[\vec x']\right).
\end{align}
{\bf Far-Zone Limit} \qquad  To extract the far-zone GW signals generated by the isolated astrophysical systems, we perform the same frequency space analysis for the spin-2 effective Green's function here as for its spin-1 counterpart. Before taking the far-zone limit, we first re-cast the spin-2 effective Green's function \eqref{Cijmn_Minkowski} into the one analogous to eq.~\eqref{Cij_FarZone_Minkowski} for spin-1 photons, by carrying out all the spatial derivatives involved in $C_{ijmn}$ while avoiding the point $R=0$,
\begin{align}
C_{ijmn}[T,\vec R]&= P_{ijmn}[\vec R]\,C_{1,d}[T,R]+\Pi_{ijmn}[\vec R]\,2\pi C_{2,d+2}[T,R]+\Xi_{ijmn}[\vec R]\,4\pi^2C_{3,d+4}[T,R],
\label{Cijmn_FarZone_Minkowski}
\end{align}
where $ P_{ijmn}[\vec R]$ denotes the TT spatial projector based on the unit vector $\widehat R$,
\begin{align}
P_{ijmn}[\vec R] &\equiv  P_{m(i}[\vec R] \, P_{j)n}[\vec R]-\frac{1}{d-2}  P_{ij}[\vec R] \, P_{mn}[\vec R],
\end{align}
with $P_{ij}[\vec R]$ given in eq.~\eqref{TransverseProjector_Position}, and the other symmetric tensor structures $\Pi_{ijmn}[\vec R]$ and $\Xi_{ijmn}[\vec R]$, respectively, are defined as
{\allowdisplaybreaks
\begin{align}
\Pi_{ijmn}[\vec R]=\,&-2 P_{ijmn}[\vec R]+\frac{d(d-3)}{d-2}\left(\delta_{m(i}\widehat R_{j)}\widehat R_{n}+\delta_{n(i}\widehat R_{j)}\widehat R_{m}
-2\widehat R_i\widehat R_j\widehat R_m\widehat R_n \right)\\
\Xi_{ijmn}[\vec R]=\,&\left(\frac{d-3}{d-2}\right)\bigg(\delta_{ij}\delta_{mn}+2\delta_{m(i}\delta_{j)n}-(d+1)\left(\delta_{ij}\widehat R_m\widehat R_n+\delta_{mn}\widehat R_i\widehat R_j+2\delta_{m(i}\widehat R_{j)}\widehat R_n\right. \notag\\
&\left.+2\delta_{n(i}\widehat R_{j)}\widehat R_m\right)+(d+1)(d+3)\widehat R_i\widehat R_j\widehat R_m\widehat R_n\bigg) .
\end{align}}%
We have taken advantage of the homogeneous wave equation obeyed by both $C_{2,d}$ and $C_{3,d}$, along with the properties $\ddot C_{2,d}=-C_{1,d}$ and $\ddot C_{3,d}=-C_{2,d}$, to relate different scalar functions. Here, we highlight again that, for a fixed $d$, the notations $C_{2,d+2}$ and $C_{3,d+4}$ used in eq.~\eqref{Cijmn_FarZone_Minkowski} represent their corresponding $(d+2)$ and $(d+4)$-dimensional functional forms, but the $R$ is the one in $d-1$ spatial dimensions. Essentially, as long as $R\neq0$, eq.~\eqref{Cijmn_FarZone_Minkowski} is equivalent to its original expression \eqref{Cijmn_Minkowski}, and, as explained in the similar spin-1 situation, this form is useful for the far-zone analysis and is manifestly finite in all relevant spacetime dimensions, because $C_{1,d}$, $C_{2,d+2}$, and $C_{3,d+4}$ all converge for $d\geq4$. As a consistency check of eq.~\eqref{Cijmn_FarZone_Minkowski}, its TT properties can be shown explicitly by a direct calculation.

The leading contribution of GWs, responsible for the far-zone tidal forces, can be extracted from the spin-2 effective Green's function using eq.~\eqref{Cijmn_FarZone_Minkowski} in frequency space. The relative amplitudes of the three scalar functions can in turn be directly compared in the limit $|\omega|R\gg1$. To begin, we express the spin-2 field $D_{ij}$ in terms of the superposition of monochromatic modes,
\begin{align}
D_{ij}[\eta,\vec x]=-16\pi G_\text N \int_{\mathbb{R}^{d-1}}  \dd^{d-1}\vec{x}' \int_{\mathbb{R}} \frac{\dd\omega}{2\pi}\,\widetilde{G}^+_{ijmn}[\omega,\vec R]\,e^{-i\omega\eta}\,{}^{(\text{a})}\widetilde{T}_{mn}[\omega,\vec{x}'],
\end{align}
where $\widetilde{G}^+_{ijmn}[\omega,\vec R]$ is the frequency transform of the spin-2 effective Green's function assuming $R\neq0$,
\begin{align}
\widetilde{G}^+_{ijmn}[\omega,\vec R]&=\int_{\mathbb{R}} \dd T\,G^+_{ijmn}[T,\vec R]\,e^{i\omega T} \notag\\
&= P_{ijmn}[\vec R]\,\widetilde{G}^+_{d}[\omega,R]  -  \Pi_{ijmn}[\vec R]\,2\pi\widetilde{C}^+_{2,d+2}[\omega,R]  -  \Xi_{ijmn}[\vec R]\,4\pi^2\widetilde{C}^+_{3,d+4}[\omega,R],
\label{Gijmn_FrequencyTransform}
\end{align}
with $\widetilde{G}^+_{d}[\omega,R]$, $\widetilde{C}^+_{2,d+2}[\omega,R]$, and $\widetilde{C}^+_{3,d+4}[\omega,R]$, respectively, defined to be the frequency transforms of their real-space counterparts, $G^+_d$, $C^+_{2,d+2}$, and $C^+_{3,d+4}$. As we did for the far-zone spin-1 waves, we now take the limit of eq.~\eqref{Gijmn_FrequencyTransform} as $|\omega|R\to \infty$, from which to extract the dominant spin-2 GWs in the radiative regime. Since we have calculated $\widetilde{G}^+_{d}[\omega,R]$ and $\widetilde{C}^+_{2,d+2}[\omega,R]$ earlier in the photon case, $\widetilde{C}^+_{3,d+4}[\omega,R]$ is the only term left to evaluate here.

{\it Spin-2 gravitons in even dimensions $d\geq4$} \qquad  For even-dimensional spacetimes, $\widetilde{G}^+_{d}[\omega,R]$ and $\widetilde{C}^+_{2,d+2}[\omega,R]$, for $d=4+2n$, have been obtained in eqs.~\eqref{C1_Even_FrequencySpace} and \eqref{C2_Even_FrequencySpace}. In the same way, with the analytic expression of $C_{3,d}$ given in eq.~\eqref{C3_Even_Minkowski}, its frequency transform $\widetilde{C}^+_{3,d+4}[\omega,R]$ can be computed straightforwardly,
\begin{align}
\widetilde{C}^+_{3,8+2n}[\omega,R]=\,&  \mathcal D^n_R\left(\frac{e^{i\omega R}}{16\pi^3 R^3(i\omega)^2}-\frac{3e^{i\omega R}}{16\pi^3 R^4(i\omega)^3}+\frac{3e^{i\omega R}}{16\pi^3 R^5(i\omega)^4}\right)  -  \frac{(2n+1)!!\,\omega^{2n+1}}{4(2\pi)^{n+3}(\omega R)^{2n+3}} \notag\\
&-  \frac{(2n+3)!!\,\omega^{2n+1}}{2(2\pi)^{n+3}(\omega R)^{2n+5}},
\label{C3_Even_FrequencySpace}
\end{align}
which, analogous to $\widetilde{C}^+_{2,d+2}[\omega,R]$, comprises the non-propagating modes associated with the acausal effect, as well as the propagating ones with the $e^{i\omega R}$ factor. Moreover, $\widetilde{C}^+_{3,8+2n}[\omega,R]$ turns out to be suppressed relative to $\widetilde{G}^+_{d}[\omega,R]$ and $\widetilde{C}^+_{2,d+2}[\omega,R]$ when $|\omega|R\to\infty$ (cf.~eqs.~\eqref{C1_Even_FrequencySpace} and \eqref{C2_Even_FrequencySpace}). More explicitly, at leading $1/(\omega R)$ order, it behaves like
\begin{align}
\widetilde{C}^+_{3,8+2n}[\omega,R]=\frac{(-1)^{n+1}i^n\omega^{2n+1}}{2(2\pi\omega R)^{n+1}}\,e^{i\omega R}  \cdot  \frac{1}{(2\pi \omega R)^2}\left(1+\mathcal{O}\left[\frac1{\omega R}\right]\right).
\label{C3_Even_FarZone_FrequencySpace}
\end{align}
Hence, as inferred from the asymptotic behaviors of three scalar functions \eqref{C1_Even_FarZone_FrequencySpace}, \eqref{C2_Even_FarZone_FrequencySpace}, and \eqref{C3_Even_FarZone_FrequencySpace}, the pure causal one, $\widetilde{G}^+_{d}[\omega,R]$, is still the dominant contribution to the spin-2 effective Green's function in the limit $|\omega|R\gg1$. In close analogy with the spin-1 case, the spin-2 GWs in the radiative zone is dominated by the first term of eq.~\eqref{Gijmn_FrequencyTransform}. That is, under the far-zone assumptions $|\omega|r\gg1$ and $r_c/r\ll1$, the leading $1/r$ piece of $\widetilde{G}^+_{ijmn}$ is given by
\begin{align}
\widetilde{G}^+_{ijmn}[\omega,\vec R]&=P_{ijmn}\widetilde{G}^{(+,\,\text{fz})}_{4+2n}[\omega;\vec x,\vec x']\left(1+\mathcal{O}\left[\frac1{\omega r},\frac{r_c}{r}\right]\right),
\label{Gijmn_Even_FrequencyTransform_FarZone}
\end{align}
where $P_{ijmn}$, the far-zone ``tt'' projector, is defined in eq.~\eqref{tt_Projector} and $\widetilde{G}^{(+,\,\text{fz})}_{4+2n}$ given in eq.~\eqref{G+_FrequencySpace_Even_FarZone}. Accordingly, as already alluded to in \S\eqref{Chapter_Observables}, the spin-2 TT graviton $D_{ij}$, in the far-zone radiative regime ($r\to\infty$), reduces to the causal ``tt'' GWs,
\begin{align}
\lim_{r\to\infty} D_{ij}\to \chi_{ij}^\text{tt}, \qquad  \chi_{ij}^\text{tt}[x]\equiv P_{ijmn}\left(-16\pi G_\text N\int_{\mathbb{R}^{d-1,1}} \dd^{d}x'\,G^{(+,\,\text{fz})}_{d}[T;\vec x,\vec x']{}^{(\text{a})}T_{mn}[x']\right),
\label{htt_FarZone}
\end{align}
where $G^{(+,\,\text{fz})}_{d}[T;\vec x,\vec x']$, as before, denotes the far-zone version of the massless scalar Green's function, which, for even $d\geq4$, is given in eq.~\eqref{G+_Even_FarZone} consisting of pure light-cone signals. This $\chi_{ij}^\text{tt}$ is thus the tt projection of the far-zone de Donder-gauge solution of the metric perturbations, $\chi_{ij}^\text{tt}=P_{ijmn}\overline\chi_{mn}[\text{de Donder}]$. In other words, like the consequence of the far-zone spin-1 field \eqref{At_FarZone}, the two distinct notions of ``transverse-traceless'' metric perturbations, $D_{ij}=\chi_{ij}^\text{TT}$ and $\chi_{ij}^\text{tt}$, are shown to coincide as $r\to\infty$, where the acausal effect in $D_{ij}$ becomes sufficiently insignificant.

{\it Spin-2 gravitons in odd dimensions $d\geq5$} \qquad Following the similar procedures, we are able to extract the far-zone portion of the spin-2 GWs for odd dimensions as well. Odd-dimensional $\widetilde{G}^+_{d}[\omega,R]$ and $\widetilde{C}^+_{2,d+2}[\omega,R]$ for $d=5+2n$ and positive frequencies can be obtained simply by replacing $n\to n+1$ in eqs.~\eqref{C1_Odd_FrequencySpace} and \eqref{C2_Odd_FrequencySpace}. And, given eq.~\eqref{C3_Odd_Minkowski}, $\widetilde{C}^+_{3,d+4}[\omega,R]$ can be tackled similarly to $\widetilde{C}^+_{2,d+2}[\omega,R]$,
\begin{align}
\widetilde{C}^+_{3,9+2n}[\omega>0,R]=\frac{i\omega^{2n+2}}{4(2\pi\omega R)^{n+3}}H^{(1)}_{n+3}[\omega R] - \frac{2^{n}(n+1)!\,\omega^{2n+2}}{(2\pi)^{n+4}(\omega R)^{2n+4}}  -  \frac{2^{n+2}(n+2)!\,\omega^{2n+2}}{(2\pi)^{n+4}(\omega R)^{2n+6}},
\end{align}
which, as in the even-dimensional case, resembles the structure of $\widetilde{C}^+_{2,d+2}[\omega,R]$ in eq.~\eqref{C2_Odd_FrequencySpace}, and tends to be more suppressed than both $\widetilde{G}^+_{d}[\omega,R]$ and $\widetilde{C}^+_{2,d+2}[\omega,R]$ at leading $1/(\omega R)$ order. That is, as $|\omega| R\gg1$, the asymptotic behavior of $\widetilde{C}^+_{3,d+4}[\omega,R]$ reads
\begin{align}
\widetilde{C}^+_{3,9+2n}[\omega>0,R]=\frac{-i\omega^{2n+2}}{2(2\pi\omega R)^{n+\frac32}}e^{i\left(\omega R-\frac{(n+1)\pi}2-\frac{\pi}4\right)} \cdot \frac{1}{(2\pi\omega R)^2}\Bigg(1+\mathcal{O}\left[\frac1{\omega R}\right]+\mathcal O\left[\frac1{(\omega R)^{n+\frac12}}\right]\Bigg),
\end{align}
where the expression has been factorized into the leading $1/(\omega R)$ piece of $\widetilde G^+_{5+2n}[\omega,R]$ times the suppression factor. Likewise, among the three scalar functions in eq.~\eqref{Gijmn_FrequencyTransform}, $\widetilde G^+_{d}[\omega,R]$ continues to be the dominant contribution in the limit $|\omega| R\to\infty$. As a result, the far-zone behavior of $\widetilde{G}^+_{ijmn}[\omega,R]$ here admits the same ``tt'' structure as eq.~\eqref{Gijmn_Even_FrequencyTransform_FarZone} for even $d$,
\begin{align}
\widetilde{G}^+_{ijmn}[\omega>0,\vec R]=P_{ijmn}\widetilde{G}^{(+,\,\text{fz})}_{5+2n}[\omega;\vec x,\vec x']\left(1+\mathcal{O}\left[\frac1{\omega r},\frac{r_c}{r}\right]\right),
\end{align}
where $\widetilde{G}^{(+,\,\text{fz})}_{5+2n}$ is given in eq.~\eqref{G+_FrequencySpace_Odd_FarZone} with $n$ replaced by $n+1$; and $P_{ijmn}$ in eq.~\eqref{tt_Projector}. A similar line of arguments then reveals that the spin-2 TT graviton $D_{ij}$, in odd dimensions $d\geq5$, also reduces to $\chi_{ij}^\text{tt}$ as $r\to \infty$, where the acausal nature of $D_{ij}$ becomes trivial; namely, the feature \eqref{htt_FarZone} still holds here, with odd-dimensional $G^{(+,\,\text{fz})}_{d}$ given in eq.~\eqref{G+_Odd_FarZone}.

{\it Linearized Riemann tensor} \qquad   Through the analysis of the spin-2 effective Green's function, we have just shown that, in the radiative limit, the spin-2 TT GWs in fact coincide with the tt ones, $D_{ij}\to \chi_{ij}^\text{tt}$, for all spacetime dimensions $d\geq4$. For this reason, the far-zone version of the tidal forces \eqref{LinearizedR0i0j_deDonder_FarZone} for all $d\geq4$, as well as the statement \eqref{SynchronousGauge_to_GaugeInvariant_StepI}, follows immediately from eq.~\eqref{LinearizedRiemann_CausalSpin2_Minkowski} and the fact that $\chi_{ij}^\text{tt}$ is completely causal. This result can alternatively be derived from the expression \eqref{LinearizedRiemann_Sol_Minkowski} for $\delta_1R_{0i0j}$, by repeatedly employing the replacement rule $\partial_i=-\partial_{i'}=-\widehat R_i\partial_0$ and the conservation of the matter stress tensor in the intermediate steps before reaching the final far-zone expression. Furthermore, the far-zone connection between $\chi_{ij}^\text{tt}$ and the synchronous-gauge metric $\chi_{ij}^{(\text{Synch})}$ can be made via eq.~\eqref{SynchronousGauge_to_GaugeInvariant_FullAnswer}, where as explained in \S\eqref{Chapter_Observables} the initial conditions could be dropped for GW detectors sensitive to only finite frequencies. In particular, the fractional distortion spin-2 pattern of the laser interferometer, described in eq.~\eqref{FractionalDistortion_AtConstantCosmicTime_FirstOrderHuman} are exclusively attributed to the causal $\chi_{ij}^\text{tt}$. Such a characterization of the GW observables in terms of $\chi_{ij}^{\text{tt}}$, however, is legitimate only when the GW detector is sufficiently far away from the matter sources.

{\bf Commutator of Spin-2 Gravitons} \qquad  Earlier in this section, we have shown micro-causality is violated for the massless spin-1 photons (cf.~eq.~\eqref{Spin1_Commutator}). A similar line of reasoning then reveals that the massless spin-2 graviton field violates micro-causality too. For, the tensor structure $C_{ijmn}$ in eq.~\eqref{Cijmn_Minkowski} is related to the commutator of the corresponding quantum operators via the relationship
\begin{align}
\big[D_{ij}[x],D_{mn}[x']\big]=-iC_{ijmn}[T,\vec R].
\label{Spin2_Commutator}
\end{align}
The acausal nature of $C_{ijmn}$ immediately tells us that the massless spin-2 gravitons do not commute at spacelike separations. Once again, it is likely that this violation of micro-causality is linked to the tension between gauge invariance and Lorentz covariance when constructing massless helicity-2 quantum Fock states.

\section{Spatially Flat Cosmologies with Constant $w$}
\label{SpatiallyFlatw}
We now move on to the spatially-flat cosmological background, driven by a perfect fluid with a constant equation-of-state $w$. Again, we will consider both the electromagnetism and linearized gravitation cases, where the dynamics of the linearized gravitational system, unlike that of electromagnetism, has non-trivial dependence on $w$. In cosmology, there is no longer a time-translation symmetry, and the full analytic expressions for spin-1 and spin-2 effective Green's functions may generally be difficult to attain. On the other hand, the background space-translation symmetry is still preserved, so the similar Fourier-space analysis exploited in Minkowski spacetime continues to apply in the cosmological setup. At a more technical level, the translation symmetry in space would still allow us to utilize the time-integral method to express the spin-1 and spin-2 effective Green's functions in terms of the analytic solutions found in \cite{Chu:2016ngc}, so that the corresponding causal structures can be analyzed.

\subsection{Electromagnetism}

We start with the electromagnetic field in the cosmological background spacetime, described by the metric \eqref{PerturbedFLRW} with $\chi_{\mu\nu}$ set to zero, and our focus here will be the causal structure of the theory in the gauge-invariant content for all $d\geq3$.

{\bf Field Equations} \qquad  In spatially-flat cosmologies, the Maxwell's equations \eqref{Maxwell_equations}, in terms of the gauge-invariant variables \eqref{Maxwell_GaugeInvariant_ScalarVector}, are translated into a set of two independent field equations,
\begin{align}
-\frac{\partial_0\left(a^{d-4}\,\dot\alpha_i\right)}{a^{d-4}}+\vec\nabla^2 \alpha_i&=-a^2\Gamma_i,
\label{Spin1_WaveEq_Cosmology} \\
\vec\nabla^2\Phi&=-a^2\rho.
\label{EMPoisson_Cosmology}
\end{align}
Notice that the spatial components of eq.~\eqref{Maxwell_equations} encode not only eq.~\eqref{Spin1_WaveEq_Cosmology} but also the equation $\partial_{0}\left(a^{d-4}\,\Phi\right)/a^{d-4}=a^2\Gamma$, which is in fact redundant, as is already implied by the Poisson's equation \eqref{EMPoisson_Cosmology} inserted into the conservation law of the charge current $\nabla_{\mu}J^{\mu}=0$,
\begin{align}
\frac{\partial_0\left(a^{d-2}J_0\right)}{a^{d-2}}=\partial_iJ_i.
\label{ChargeConservaion_Cosmology}
\end{align}
Moreover, eqs.~\eqref{Spin1_WaveEq_Cosmology} and \eqref{EMPoisson_Cosmology} together show that, except the re-scaling $J_{\mu}\to a^2J_\mu$, the theory is conformally invariant when $d=4$, and, like its Minkowski counterpart, only the spin-1 photon field $\alpha_i$ admits dynamical wave solutions, with the scalar $\Phi$ still obeying a Poisson-type equation.

{\bf Spin-1 Photons} \qquad  To solve for the transverse spin-1 field $\alpha_i$ in the cosmological system, we first re-write its wave equation \eqref{Spin1_WaveEq_Cosmology} as
\begin{align}
\left\{\partial^2+\frac{(d-4)}{4}\left((d-4)\mathcal H^2+2\dot{\mathcal H}\right)\right\}
\left(a^{\frac{d-4}2}\alpha_i\right)=-a^{\frac d2}\Gamma_i,
\label{Spin1_WaveEq2_Cosmology}
\end{align}
where $\mathcal H\equiv\dot a/a$, denoting the conformal Hubble parameter. Then, following the same manipulations in Fourier space performed in Minkowski spacetime, we are able to express the spin-1 field $\alpha_i$ in terms of the following convolution based on eq.~\eqref{Spin1_WaveEq2_Cosmology},
\begin{align}
a[\eta]^{\frac{d-4}2}\alpha_i[\eta,\vec x]&=-\int_{\mathbb{R}^{d-1}} \dd^{d-1}\vec x'\int_{\eta_{\text p}}^{\eta_{\text f}} \dd\eta' \,a[\eta']^{\frac d2}G^{(\gamma,+)}_{ij}[\eta,\eta';R]\,J_j[\eta',\vec x'],
\label{Spin1_Convolution_Cosmology}
\end{align}
where the time interval of integration $(\eta_{\text p},\eta_{\text f})$ covers all the possible values of $\eta$ for an expanding universe, and the spin-1 effective Green's function $G^{(\gamma,+)}_{ij}$ is given by
\begin{align}
G^{(\gamma,+)}_{ij}[\eta,\eta';\vec R]&=-\Theta[T]C^{(\gamma)}_{ij}[\eta,\eta';\vec R], \notag \\
C^{(\gamma)}_{ij}[\eta,\eta';\vec R]&=\delta_{ij}C^{(\gamma)}_{1,d}[\eta,\eta';R]+\partial_i\partial_jC^{(\gamma)}_{2,d}[\eta,\eta';R],
\label{Cij_Cosmology}
\end{align}
with $C^{(\gamma)}_{1,d}$ and $C^{(\gamma)}_{2,d}$, respectively, defined by
{\allowdisplaybreaks
\begin{align}
&C^{(\gamma)}_{1,d}[\eta,\eta';R]\equiv\int_{\mathbb{R}^{d-1}}\frac{\dd^{d-1}\vec{k}}{(2\pi)^{d-1}}\,\widetilde C^{(\gamma)}_{1,d}\left[\eta,\eta';|\vec k|\right]\,e^{i\vec{k}\cdot\vec{R}},
\label{C1_Photon_Cosmology}\\
&C^{(\gamma)}_{2,d}[\eta,\eta';R]\equiv\int_{\mathbb{R}^{d-1}}\frac{\dd^{d-1}\vec{k}}{(2\pi)^{d-1}}\,\widetilde C^{(\gamma)}_{1,d}\left[\eta,\eta';|\vec k|\right]\,\frac{e^{i\vec{k}\cdot\vec{R}}}{\vec k^2};
\label{C2_Photon_Cosmology}
\end{align}}%
the Fourier transform of $C^{(\gamma)}_{1,d}$ can be equivalently expressed in terms of the following decomposition,
\begin{align}
\widetilde C^{(\gamma)}_{1,d}\left[\eta,\eta';|\vec k|\right]=i\left(v^{(\gamma)}_{|\vec k|}[\eta]\,v^{(\gamma)*}_{|\vec k|}[\eta']-v^{(\gamma)*}_{|\vec k|}[\eta]\,v^{(\gamma)}_{|\vec k|}[\eta']\vphantom{\dot A}\right),
\label{C1_Photon_Cosmology_ModeDecomposition}
\end{align}
where $v^{(\gamma)}_{|\vec k|}$ are in fact the mode functions of the massless scalar field that satisfies the homogeneous (i.e., $\Gamma_i=0$) form of the wave equation \eqref{Spin1_WaveEq2_Cosmology}, the Fourier-space version of which is therefore obeyed by $v^{(\gamma)}_{|\vec k|}$ itself and $\widetilde C^{(\gamma)}_{1,d}$; moreover, $v^{(\gamma)}_{|\vec k|}$ have been normalized so that the initial condition imposed on the time derivative of $\widetilde C^{(\gamma)}_{1,d}$, namely $\dot{\widetilde C}^{(\gamma)}_{1,d}\big|_{\eta=\eta'}=1$, coincides with the Wronskian condition for the mode functions, $v^{(\gamma)}_{|\vec k|}\dot{v}^{(\gamma)*}_{|\vec k|}-v^{(\gamma)*}_{|\vec k|}\dot{v}^{(\gamma)}_{|\vec k|}=i$. In this language, the properties of $C^{(\gamma)}_{1,d}$ and $C^{(\gamma)}_{2,d}$ become more transparent. Specifically, both of the scalar functions obey the homogeneous wave equation associated with the wave operator in eq.~\eqref{Spin1_WaveEq2_Cosmology}, and the equal-time initial conditions for $C^{(\gamma)}_{1,d}$, $C^{(\gamma)}_{2,d}$, and their velocities, can be immediately read off,
{\allowdisplaybreaks
\begin{align}
&C^{(\gamma)}_{1,d}\big|_{\eta=\eta'}=C^{(\gamma)}_{2,d}\big|_{\eta=\eta'}=0,  \label{C123_Photon_InitialCondition}\\
&\dot C^{(\gamma)}_{1,d}\big|_{\eta=\eta'}=-\partial_{\eta'} C^{(\gamma)}_{1,d}\big|_{\eta=\eta'}=\delta^{(d-1)}[\vec x-\vec x'],  \label{dotC1_Photon_InitialCondition} \\
&\dot C^{(\gamma)}_{2,d}\big|_{\eta=\eta'}=-\partial_{\eta'} C^{(\gamma)}_{2,d}\big|_{\eta=\eta'}=-G^{\mathrm{(E)}}_d[R].
\label{dotC2_Photon_InitialCondition}
\end{align}}%
It turns out that $C^{(\gamma)}_{1,d}$ and $C^{(\gamma)}_{2,d}$ are the cosmological generalization of their Minkowski counterparts \eqref{C1_Minkowski} and \eqref{C2_Minkowski}, and $C^{(\gamma)}_{ij}$ is connected to the commutator of the massless spin-1 photons in the cosmological background,
\begin{align}
\big[\alpha_i[x], \alpha_j[x']\big] = -i \big(a[\eta]a[\eta']\big)^{-\frac{d-4}2} C^{(\gamma)}_{ij}[\eta,\eta';\vec R].
\label{Spin1_Commutator_Cosmology}
\end{align}

When specializing to the constant-$w$ cosmologies considered in this section, where $\mathcal H=2/(q_w\eta)$ with $(\eta_{\text p},\eta_{\text f})=(-\infty,0)$ for $w<-(d-3)/(d-1)$ and $(\eta_{\text p},\eta_{\text f})=(0,\infty)$ for $w>-(d-3)/(d-1)$, the analytic solution of the massless scalar Green's function $G^{(\gamma,+)}_d=-\Theta[T]C^{(\gamma)}_{1,d}$ for $d\geq3$ has been derived in \cite{Chu:2016ngc} via Nariai's ansatz (see eqs.~(205) and (206) of \cite{Chu:2016ngc}), instead of computing eq.~\eqref{C1_Photon_Cosmology}; the result shows that $G^{(\gamma,+)}_d$ contains pure causal signals propagating either on or inside the light cone. Since $C^{(\gamma)}_{1,d}$ is known, we may employ the time integral method to compute $C^{(\gamma)}_{2,d}$ without resorting to tackling the integral in eq.~\eqref{C2_Photon_Cosmology} directly. The homogeneous counterpart of eqs.~\eqref{Spin1_WaveEq2_Cosmology} together with the relation $-\vec{\nabla}^2 C_{2,d}^{(\gamma)} = C_{1,d}^{(\gamma)}$ leads us to
{\allowdisplaybreaks
	\begin{align}
	\frac{-\partial_\eta\bigg(a^{2\alpha}\,\partial_\eta\left(a^{-\alpha}\, C_{2,d}^{(\gamma)}  \vphantom{\dot A}\right)\bigg)}{a^\alpha}&=C_{1,d}^{(\gamma)},
	\qquad\qquad
	\alpha=\frac{d-4}2 .
	\end{align}}%
With the initial conditions \eqref{C123_Photon_InitialCondition} and \eqref{dotC2_Photon_InitialCondition}, we are able to write down
{\allowdisplaybreaks
\begin{align}
C^{(\gamma)}_{2,d}[\eta,\eta';R]=\,&-a[\eta]^{\frac{d-4}2}\int^\eta_{\eta'}\dd\eta_2\,a[\eta_2]^{-(d-4)}
\int^{\eta_2}_{\eta'}\dd\eta_1\,a[\eta_1]^{\frac{d-4}2}C^{(\gamma)}_{1,d}[\eta_1,\eta';R]  \notag\\
&-G^{\mathrm{(E)}}_{d}[R]\left(\frac{a[\eta]}{a[\eta']}\right)^{\frac{d-4}2}\int^\eta_{\eta'}\dd\eta_1\,\left(\frac{a[\eta']}{a[\eta_1]}
\right)^{d-4},
\label{C2_Photon_TimeIntegral_Cosmology}
\end{align}}%
If we only consider the retarded piece ($\eta>\eta'$) of $C^{(\gamma)}_{2,d}$, then as explained in \S \eqref{MinkowskiSpacetime} since $C^{(\gamma)}_{1,d}[\eta,\eta';R]$ is strictly causal (i.e., its retarded part is proportional to $\Theta[T-R^-]$), the first term of eq.~\eqref{C2_Photon_TimeIntegral_Cosmology} would in turn yield a strictly causal contribution. Whereas the second term is a smooth function of spacetime, consisting of the signals that pervade all physical spacetime points with $\eta>\eta'$, including the region outside the light cone. Hence, the time-integral method has cleanly elucidated the causal structure of $C^{(\gamma)}_{2,d}$, even if the integrals cannot be performed in closed form: acausality is present for all $d\geq3$ and is encoded only in the second term of eq.~\eqref{C2_Photon_TimeIntegral_Cosmology}. Moreover, it also implies the spin-1 quantum operator violates micro-causality in constant-$w$ cosmologies (cf.~eq.~\eqref{Spin1_Commutator_Cosmology}).

{\bf Scalar } \qquad  The Poisson's equation \eqref{EMPoisson_Cosmology} for scalar $\Phi$ can be solved immediately by utilizing the Euclidean Green's function,
\begin{align}
\Phi[\eta,\vec x]=\int_{\mathbb{R}^{d-1}} \dd^{d-1}\vec{x}'\,a[\eta]^2G^{(\mathrm{E})}_{d}[R]\, J_0[\eta,\vec x'],
\label{EMPhi_Cosmology}
\end{align}
which, except for the factor $a[\eta]^2$, is precisely the same as its Minkowski counterpart \eqref{EMPhi_Minkowski}. This in turn implies that, despite the distinct waveforms of the spin-1 field in cosmology, the acausal portion of its velocity must take such a simple form to ensure a causal electric field, as we will demonstrate below.

{\bf  Faraday Tensor} \qquad  Let us now turn to the causal structure of the Faraday tensor $F_{\mu\nu}$ within the cosmological context, where $F_{\mu\nu}$ in the gauge-invariant formalism still takes the form of eq.~\eqref{Maxwell_FaradayTensor}. To make the causality analysis more transparent, we first re-cast the spin-1 photon field \eqref{Spin1_Convolution_Cosmology} in the following convolution that involves both time and spatial components of the electric current $J_\mu$,
\begin{align}
\alpha_i[\eta,\vec x]=-\int_{\mathbb{R}^{d-1}} \dd^{d-1}\vec x'\int_{\eta_{\text p}}^{\eta_{\text f}} \dd  &\eta'  \,a[\eta]^{-\frac{d-4}2}a[\eta']^{\frac d2}\bigg(G^{(\gamma,+)}_{d}J_i[\eta',\vec x']  \notag\\
& \qquad +\Theta[T]\,a[\eta']^{\frac{d-4}2}\partial_{\eta'}\left(a[\eta']^{-\frac {d-4}2}\partial_iC^{(\gamma)}_{2,d}\right)J_0[\eta',\vec x']\bigg),
\label{Spin1_Convolution2_Cosmology}
\end{align}
where we have employed the conservation law \eqref{ChargeConservaion_Cosmology} and removed the boundary terms that arise from integration by parts.\footnote{With the assumption that the electric current is sufficiently localized, the boundary contributions evaluated at spatial infinity are zero, whereas the ones at past infinity, namely at $\eta'=-\infty$ for $w<-(d-3)/(d-1)$ and $\eta' = 0$ for $w>-(d-3)/(d-1)$, still require further justification. However, those boundary terms at past infinity are in fact the surface integrals of $C^{(\gamma)}_{2,d}$ at $\eta'= - \infty$ or $\eta' = 0$, indicating the fact that they satisfy the homogeneous wave equation and do not alter the exact inhomogeneous solution to the spin-1 wave equation. We hope to clarify this issue further in our later work.
} The resulting expression is in fact the cosmological generalization of eq.~\eqref{Spin1_Convolution2_Minkowski}, and notice that, due to the lack of time-translation symmetry, the time derivative of $C^{(\gamma)}_{1,d}$ or $C^{(\gamma)}_{2,d}$ with respect to $\eta'$ cannot simply be replaced with the negative of that with respect to $\eta$.

Start with the $(ij)$ components of $F_{\mu\nu}$. Taking the spatial curl of either eq.~\eqref{Spin1_Convolution_Cosmology} or eq.~\eqref{Spin1_Convolution2_Cosmology} gives rise to the magnetic field in the cosmological background (cf.~eq.~\eqref{Maxwell_FaradayTensor}),
\begin{align}
F_{ij}[\eta,\vec x]&=2\partial_{[i}\alpha_{j]}=-2\int_{\mathbb{R}^{d-1}}\dd^{d-1}\vec x'\int_{\eta_{\text p}}^{\eta_{\text f}} \dd \eta'  \,a[\eta]^{-\frac{d-4}2}a[\eta']^{\frac d2}\partial_{[i}G^{(\gamma,+)}_{d}J_{j]}[\eta',\vec x'],
\end{align}
which is completely causal for $d\geq3$, since the acausal effect of the spin-1 photons, encapsulated in the second term of eq.~\eqref{Spin1_Convolution_Cosmology} or eq.~\eqref{Spin1_Convolution2_Cosmology}, is eliminated by the curl operation here -- as was the case in the Minkowski background. Next, to obtain the electric field $F_{0i}$, we first compute the velocity of the spin-1 field by taking the time derivative of eq.~\eqref{Spin1_Convolution2_Cosmology}, with the expression \eqref{C2_Photon_TimeIntegral_Cosmology} substituted for $C^{(\gamma)}_{2,d}$,
\begin{align}
\dot\alpha_i[\eta,\vec x]=\,&-\int_{\mathbb{R}^{d-1}}\dd^{d-1}\vec x'\int_{\eta_{\text p}}^{\eta_{\text f}} \dd \eta'  \,a[\eta]^{-\frac{d-4}2}a[\eta']^{\frac d2}\Bigg\{\left(\dot G^{(\gamma,+)}_d[\eta,\eta';R]-\frac{(d-4)}{q_w\eta} G^{(\gamma,+)}_d[\eta,\eta';R]\right)J_i[\eta',\vec x'] \notag\\
& +\left(\frac{a[\eta']}{a[\eta]}\right)^{\frac{d-4}2}\partial_{\eta'}\bigg(a[\eta']^{-\frac {d-4}2}\int^{\eta}_{\eta'}\dd\eta_1\,a[\eta_1]^{\frac{d-4}2}
\partial_iG^{(\gamma,+)}_d[\eta_1,\eta';R]\bigg)J_0[\eta',\vec x']\Bigg\} \notag\\
&-\int_{\mathbb{R}^{d-1}} \dd^{d-1}\vec x'\,a[\eta]^{2}\partial_iG^{\mathrm{(E)}}_d[R]J_0[\eta,\vec x'],
\label{DotSpin1_Cosmology}
\end{align}
where the spacetime integral involving $G^{(\gamma,+)}_d$ is purely causal, whereas the last term is an acausal instant-time-surface integral. Then, summing the expression \eqref{DotSpin1_Cosmology} for $\dot\alpha_i$ and the gradient of $\Phi$ in eq.~\eqref{EMPhi_Cosmology} amounts to canceling the acausal term in the last line of eq.~\eqref{DotSpin1_Cosmology}, and yields the purely causal electric field $F_{0i}$,
{\allowdisplaybreaks
\begin{align}
F_{0i}[\eta,\vec x]=\,&\dot\alpha_i[\eta,\vec x]+\partial_i\Phi[\eta,\vec x]  \notag\\
=\,&-\int_{\mathbb{R}^{d-1}}\dd^{d-1}\vec x'\int_{\eta_{\text p}}^{\eta_{\text f}} \dd \eta' \,a[\eta]^{-\frac{d-4}2}a[\eta']^{\frac d2}\Bigg\{\left(\dot G^{(\gamma,+)}_d[\eta,\eta';R]-\frac{(d-4)}{q_w \eta} G^{(\gamma,+)}_d[\eta,\eta';R]\right)J_i[\eta',\vec x'] \notag\\
&+\left(\frac{a[\eta']}{a[\eta]}\right)^{\frac{d-4}2}\partial_{\eta'}\bigg(a[\eta']^{-\frac {d-4}2}\int^{\eta}_{\eta'}\dd\eta_1\,a[\eta_1]^{\frac{d-4}2}
\partial_iG^{(\gamma,+)}_d[\eta_1,\eta';R]\bigg)J_0[\eta',\vec x']\Bigg\}.
\label{F0i_Cosmology}
\end{align}}%
This result extends eq.~\eqref{F0i_CausalSpin1_Minkowski} to the cosmological context; i.e., the electric field is still the causal piece of the velocity of the spin-1 photon $\alpha_i$. Furthermore, as a simple check of consistency, one can show eq.~\eqref{F0i_Cosmology} does reduce to its Minkowski counterpart \eqref{F0i_Minkowski}, by setting $a\to1$ and assuming $G^{(\gamma,+)}_d$ takes its Minkowski form $G^+_d$ with time-translation symmetry.

Although we have shown explicitly that causality is preserved for the electromagnetic observables in constant-$w$ spatially-flat cosmologies, the second line of eq.~\eqref{F0i_Cosmology} still involves a time integral from some initial time $\eta'$ to the present $\eta$. However, by introducing a new Green's function $G^{(\gamma|\text{time},+)}_d$ obeying
\begin{align}
\bigg\{\partial^2  +  \frac{(d-4) \left(q_w+d-4 \right)}{q^2_w \eta^2}   \bigg\}G^{(\gamma|\text{time},+)}_d[\eta,\eta';R]=\delta^{(d)}[x-x'],
\label{Gt+_ElectricField_Cosmology}
\end{align}
the effective Green's function of $F_{0i}$ in eq.~\eqref{F0i_Cosmology} can be further reduced to a localized form,
\begin{align}
F_{0i}[\eta,\vec x]=\,-\int_{\mathbb{R}^{d-1}}\dd^{d-1}\vec x' & \int_{\eta_{\text p}}^{\eta_{\text f}} \dd \eta' \,a[\eta]^{-\frac{d-4}2}a[\eta']^{\frac d2}\Bigg\{\bigg(\dot G^{(\gamma,+)}_d[\eta,\eta';R] \notag\\
&-\frac{(d-4)}{q_w \eta} G^{(\gamma,+)}_d[\eta,\eta';R]\bigg)J_i[\eta',\vec x'] -\partial_iG^{(\gamma|\text{time},+)}_d[\eta,\eta';R] J_0[\eta',\vec x']\Bigg\}.
\label{F0i_Cosmology_Localized}
\end{align}
We will leave the detailed derivation of eq.~\eqref{F0i_Cosmology_Localized} to our future work \cite{ChuLiuInPrep}. It turns out that the result \eqref{F0i_Cosmology_Localized} is exactly consistent with $F_{0i}$ computed directly using the solution of the generalized Lorenz-gauge vector potential $A_\mu$ in \cite{Chu:2016ngc}; the analytic solutions of $G^{(\gamma,+)}_d = -\Theta[T]\mathcal G^{(\gamma|\text{space})}$ and $G^{(\gamma|\text{time},+)}_d = -\Theta[T]\mathcal G^{(\gamma|\text{time})}$ can be found in eqs.~(B38), (B39), (B40), and (B41) of \cite{Chu:2016ngc}.

\subsection{Linearized Gravitation}

Linearized gravity coupled to the isolated astrophysical sources in cosmology is, in fact, a more sophisticated system, due to the additional first-order perturbations of the very fluid driving cosmic expansion. A detailed analysis of this linearized gravitational system, described in terms of the perturbed metric \eqref{PerturbedFLRW}, has been performed in \cite{Chu:2016ngc} within a constant equation-of-state universe, where the background perfect fluid is modeled through an effective field theory description. As explicitly demonstrated in \cite{Chu:2016ngc}, the field equations for the gauge-invariant metric perturbations, deduced from the full linearized Einstein's equations, can be put in a decoupled form with no perturbed fluid quantities involved. The set of resulting equations then reveals that the dynamics of Bardeen scalar potentials varies in different ranges of the equation-of-state $w$. Here, we will focus on the causal structure of the system in the de Sitter ($w=-1$) and relativistic-fluid ($0<w\leq1$) cases for all $d\geq4$.

{\bf Field Equations} \qquad  In terms of the gauge-invariant variables formed in eqs.~\eqref{Bardeen_Psi}, \eqref{Bardeen_Phi}, and \eqref{Bardeen_VandDij}, the relevant equations are derived from Einstein's equations of this cosmological system, linearized about the spatially-flat background with a constant equation-of-state $w$. Based on the results obtained in \cite{Chu:2016ngc}, the character of the decoupled field equation of the Bardeen scalar $\Psi$ for $w=-1$ is distinct from the $0<w\leq1$ case. Hence, in what follows, we will consider these two cases separately.

{\it Field equations for $w=-1$} \qquad  When $w=-1$, there is no fluid and the background geometry is de Sitter spacetime. The corresponding gauge-invariant equations are given by eqs.~\eqref{PhiPsiRelationship} and \eqref{Vi_Poisson_Minkowski}, both of which remain unchanged, and
{\allowdisplaybreaks
\begin{align}
(d-2)\vec\nabla^2\Psi&=8\pi G_\text N\left(\rho+(d-1)\mathcal H\Sigma  \vphantom{\dot h}   \right),  \label{Psi_Poisson_Cosmology} \\
-\ddot D_{ij}-(d-2)\mathcal H\dot D_{ij}+\vec\nabla^2 D_{ij}
&= a^2 \, \overline\Box^{(\text S)} D_{ij} \notag\\
&\equiv \frac{a^2\,\partial_\mu \left( \sqrt{|\overline g|} \overline g^{\mu\nu} \partial_\nu D_{ij} \right)}{\sqrt{|\overline g|}} =-16\pi G_\text N\sigma_{ij}
\label{Spin-2_WaveEquation_Cosmology},
\end{align}}%
where $\overline g$ denotes the determinant of the background metric $\overline g_{\mu\nu}=a^2\eta_{\mu\nu}$, the scale factor is $a[\eta]=-1/(H\eta)$, with $H$ denoting the constant Hubble parameter, and the conformal Hubble parameter reads $\mathcal H=-1/\eta$. Compared with their Minkowski counterparts \eqref{Psi_Poisson_Minkowski} and \eqref{Spin-2_WaveEquation_Minkowski}, both $\Psi$ and $D_{ij}$ retain similar dynamics in the de Sitter case here; the Bardeen scalar $\Psi$ still obeys a Poisson-type equation, sourced not only by the local energy density $\rho$, but also by the non-local longitudinal part $\Sigma$ of $^{(\text{a})}T_{0i}$, while the spin-2 field $D_{ij}$ obeys a tensor wave equation in de Sitter background.

{\it Field equations for $0<w\leq1$} \qquad   For a physical relativistic equation-of-state within the range $0<w\leq1$, the field equations \eqref{PhiPsiRelationship}, \eqref{Vi_Poisson_Minkowski}, and \eqref{Spin-2_WaveEquation_Cosmology} still hold with $\mathcal H=2/(q_w\eta)$ -- recall eq.~\eqref{ScaleFactor} -- but the Bardeen scalar $\Psi$ now obeys a dynamical wave-like equation \cite{Chu:2016ngc}, instead of being governed by a Poisson-type one,
\begin{align}
-\ddot\Psi-\big(q_w+d-2\big)\mathcal H\dot\Psi+w\vec\nabla^2\Psi
=-8\pi G_\text N\bigg(\frac{\partial_0\left(a^{d-2}\Sigma\right)}{(d-2)a^{d-2}}-\frac{w\rho}{(d-2)}+\mathcal H\dot{\Upsilon}\bigg),
\label{Psi_WaveEq_Cosmology}
\end{align}
\footnote{The left-hand side of eq.~\eqref{Psi_WaveEq_Cosmology} can also be re-expressed in terms of the d'Alembertian ${}^{(\Psi)}\Box$ associated with
\begin{align}
{}^{(\Psi)}g_{\mu\nu} \dd x^\mu \dd x^\nu =\left(\frac{\eta}{\eta_0}\right)^{\frac{4\left(q_w+d-2\right)}{(d-2)q_w}}\left(-\dd\eta^2 + w^{-1} \dd\vec{x}\cdot\dd\vec{x} \right) . \notag
\end{align}
See eq.~(128) of \cite{Chu:2016ngc}.}which implies the existence of the acoustic cone, $\sqrt w\,T=R$, on which these scalar gravitational signals propagate at speed $\sqrt w$. It is worth noting that there exists no counterpart of this phenomenon in Minkowski and de Sitter spacetimes; this change in the character of the scalar equation for $\Psi$ is presumably tied to the dynamics of the background fluid.

In addition, as we have already noticed from our previous calculations, the energy-momentum conservation of the astrophysical sources, $\overline\nabla^\mu\,^{(\text{a})}T_{\mu\nu}=0$, where $\overline\nabla_\mu$ is the covariant derivative associated with the background metric, will be crucial in extracting the relevant effective Green's functions and their causal structures. For later convenience, in spatially-flat cosmologies, the conservation law can be re-expressed as
\begin{align}
\partial_i{}^{(\text{a})}T_{ij}&=\frac{\partial_0\left(a^{d-2}\,^{(\text{a})}T_{0j}\right)}{a^{d-2}},  \label{EnergyMomentumConservaion_Time_Cosmology}\\
\partial_j{}^{(\text{a})}T_{0j}&=\frac{\partial_0\left(a^{d-2}\,^{(\text{a})}T_{00}\right)}{a^{d-2}}-\mathcal H\left({}^{(\text{a})}T_{00}-{}^{(\text{a})}T_{ll}\right).
\label{EnergyMomentumConservaion_Spatial_Cosmology}
\end{align}

{\bf Spin-2 Gravitons} \qquad  The spin-2 wave equation for either $w=-1$ or $0<w\leq1$ takes the same form as eq.~\eqref{Spin-2_WaveEquation_Cosmology} with $\mathcal H=2/(q_w\eta)$; therefore, for both cases, the method used in the previous photon computation can be applied directly to solving eq.~\eqref{Spin-2_WaveEquation_Cosmology} via the spin-2 effective Green's function convolved against the local matter sources.

The first step is to re-cast the tensor wave equation \eqref{Spin-2_WaveEquation_Cosmology} into a conformal re-scaled form,
\begin{align}
\left\{\partial^2  +  \frac{(d-2)  \left( d-2-q_w \right)}{q^2_w \eta^2}  \right\}\left(a^{\frac{d-2}2}D_{ij}\right)=-16\pi G_\text Na^{\frac{d-2}2}\sigma_{ij},
\label{Spin-2_WaveEquation2_Cosmology}
\end{align}
and then, a similar procedure of implementing the local Fourier-space projection of eq.~\eqref{Spin-2_WaveEquation2_Cosmology} leads us to the following convolution for the spin-2 gravitons,
\begin{align}
a[\eta]^{\frac{d-2}2}D_{ij}[\eta,\vec x]&=-16\pi G_\text N\int_{\mathbb{R}^{d-1}} \dd^{d-1}\vec x'\int_{\eta_{\text p}}^{\eta_{\text f}} \dd\eta' \,a[\eta']^{\frac{d-2}2}G^{(g,+)}_{ijmn}[\eta,\eta';R]{}^{(\text{a})}T_{mn}[\eta',\vec x'],
\label{Spin2_Convolution_Cosmology}
\end{align}
where the time interval $(\eta_{\text p},\eta_{\text f})$ corresponds to $(-\infty,0)$ for $w=-1$ and $(0,\infty)$ for $0<w\leq1$, and $G^{(g,+)}_{ijmn}$ refers to the spin-2 effective Green's function,
{\allowdisplaybreaks
\begin{align}
G^{(g,+)}_{ijmn}[\eta,\eta';\vec R]&=-\Theta[T]C^{(g)}_{ijmn}[\eta,\eta';\vec R],  \notag\\
C^{(g)}_{ijmn}[\eta,\eta';\vec R]&=\bigg(\delta_{m(i}\delta_{j)n}-\frac{\delta_{ij}\delta_{mn}}{d-2}\bigg)C^{(g)}_{1,d}[\eta,\eta';R]
+\bigg(\delta_{m(i}\partial_{j)}\partial_n+\delta_{n(i}\partial_{j)}\partial_m     \notag\\
&\qquad \qquad -\frac{\delta_{ij}\partial_m\partial_n-\delta_{mn}\partial_i\partial_j}{d-2}
\bigg)C^{(g)}_{2,d}[\eta,\eta';R] +\left(\frac{d-3}{d-2}\right)\partial_i\partial_j\partial_m\partial_nC^{(g)}_{3,d}[\eta,\eta';R],
\label{Cijmn_Cosmology}
\end{align}}%
which has the same tensor structure as eq.~\eqref{Cijmn_Minkowski}, with the scalar functions $C^{(g)}_{1,d}$, $C^{(g)}_{2,d}$, and $C^{(g)}_{3,d}$ generalized to their cosmological versions,
\begin{align}
&C^{(g)}_{1,d}[\eta,\eta';R]\equiv\int_{\mathbb{R}^{d-1}}\frac{\dd^{d-1}\vec{k}}{(2\pi)^{d-1}}\,\widetilde C^{(g)}_{1,d}\left[\eta,\eta';|\vec k|\right]\,e^{i\vec{k}\cdot\vec{R}},
\label{C1_Graviton_Cosmology}\\
&C^{(g)}_{2,d}[\eta,\eta';R]\equiv\int_{\mathbb{R}^{d-1}}\frac{\dd^{d-1}\vec{k}}{(2\pi)^{d-1}}\,\widetilde C^{(g)}_{1,d}\left[\eta,\eta';|\vec k|\right]\,\frac{e^{i\vec{k}\cdot\vec{R}}}{\vec k^2},
\label{C2_Graviton_Cosmology}\\
&C^{(g)}_{3,d}[\eta,\eta';R]\equiv\int_{\mathbb{R}^{d-1}}\frac{\dd^{d-1}\vec{k}}{(2\pi)^{d-1}}\,\widetilde C^{(g)}_{1,d}\left[\eta,\eta';|\vec k|\right]\,\frac{e^{i\vec{k}\cdot\vec{R}}}{\vec k^4};
\label{C3_Graviton_Cosmology}
\end{align}
the Fourier transform of $C^{(g)}_{1,d}$ is denoted by $\widetilde C^{(g)}_{1,d}$. For $G_{ij mn}^{(g,+)}$ to be a Green's function, the $C^{(g)}$s in eqs.~\eqref{Cijmn_Cosmology} must obey the homogeneous version of eq.~\eqref{Spin-2_WaveEquation2_Cosmology}. This implies, for instance,
\begin{align}
\widetilde C^{(g)}_{1,d}\left[\eta,\eta';|\vec k|\right]=i\left(v^{(g)}_{|\vec k|}[\eta]\,v^{(g)*}_{|\vec k|}[\eta']-v^{(g)*}_{|\vec k|}[\eta]\,v^{(g)}_{|\vec k|}[\eta']\vphantom{\dot A}\right),
\label{C1_Graviton_Cosmology_ModeDecomposition}
\end{align}
where $v^{(g)}_{|\vec k|}$ are the mode functions obeying the same homogeneous wave equation as $\widetilde C^{(g)}_{1,d}$, and are normalized to match the Wronskian condition $v^{(g)}_{|\vec k|}\dot{v}^{(g)*}_{|\vec k|}-v^{(g)*}_{|\vec k|}\dot{v}^{(g)}_{|\vec k|}=i$, or the initial condition $\dot{\widetilde C}^{(g)}_{1,d}\big|_{\eta=\eta'}=1$. Furthermore, because their Fourier transforms indicate $-\vec{\nabla}^2 C_{3,d}^{(g)} = C_{2,d}^{(g)}$ and $-\vec{\nabla}^2 C_{2,d}^{(g)} = C_{1,d}^{(g)}$, the homogeneous equations for $C_{1,d}^{(g)}$, $C_{2,d}^{(g)}$, and $C_{3,d}^{(g)}$ translate to the relations:
\begin{align}
\label{Cg12_Cosmology}
\frac{-\partial_\eta\bigg(a^{2\alpha}\,\partial_\eta\left(a^{-\alpha}\, C_{2,d}^{(g)} \vphantom{\dot A}\right)\bigg)}{a^\alpha}&= C_{1,d}^{(g)},
\\
\label{Cg13_Cosmology}
\frac{-\partial_\eta\bigg(a^{2\alpha}\,\partial_\eta\left(a^{-\alpha}\, C_{3,d}^{(g)} \vphantom{\dot A}\right)\bigg)}{a^\alpha}&= C_{2,d}^{(g)},
\qquad \alpha=\frac{d-2}2 .
\end{align}
We may now apply the time-integral method here to relate $C^{(g)}_{2,d}$ and $C^{(g)}_{3,d}$ to $C^{(g)}_{1,d}$, without evaluating their Fourier transform integrals, since $C_{1,d}^{(g)}$ itself has already been derived in \cite{Chu:2016ngc} (cf.~eqs.~(112) and (113) therein). As we shall witness, this will also yield a clean elucidation of their causal structure. To this end, note that the initial conditions for the $C^{(g)}$s at $\eta=\eta'$ may be identified from their Fourier transforms and the anti-symmetric nature of the mode decomposition in eq.~\eqref{C1_Graviton_Cosmology_ModeDecomposition}:
{\allowdisplaybreaks
\begin{align}
&C^{(g)}_{1,d}\big|_{\eta=\eta'}=C^{(g)}_{2,d}\big|_{\eta=\eta'}=C^{(g)}_{3,d}\big|_{\eta=\eta'}=0,  \label{C123_Graviton_InitialCondition}\\
&\dot C^{(g)}_{1,d}\big|_{\eta=\eta'}=-\partial_{\eta'} C^{(g)}_{1,d}\big|_{\eta=\eta'}=\delta^{(d-1)}[\vec x-\vec x'],  \label{dotC1_Graviton_InitialCondition} \\
&\dot C^{(g)}_{2,d}\big|_{\eta=\eta'}=-\partial_{\eta'} C^{(g)}_{2,d}\big|_{\eta=\eta'}=-G^{\mathrm{(E)}}_d[R],
\label{dotC2_Graviton_InitialCondition} \\
&\dot C^{(g)}_{3,d}\big|_{\eta=\eta'}=-\partial_{\eta'} C^{(g)}_{3,d}\big|_{\eta=\eta'}=D_d[R].
\label{dotC3_Graviton_InitialCondition}
\end{align}}%
With eqs.~\eqref{C123_Graviton_InitialCondition}, \eqref{dotC2_Graviton_InitialCondition}, and \eqref{dotC3_Graviton_InitialCondition} imposed, $C^{(g)}_{2,d}$ and $C^{(g)}_{3,d}$ can both be expressed in terms of (the known) $C^{(g)}_{1,d}$ by integrating eqs.~\eqref{Cg12_Cosmology} and \eqref{Cg13_Cosmology},
{\allowdisplaybreaks
\begin{align}
C^{(g)}_{2,d}[\eta,\eta';R]=\,&-a[\eta]^{\frac {d-2}2}\int^\eta_{\eta'}\dd\eta_2\,a[\eta_2]^{-(d-2)}\int^{\eta_2}_{\eta'}\dd\eta_1\,a[\eta_1]^{\frac {d-2}2}C^{(g)}_{1,d}[\eta_1,\eta';R] \notag\\
&-G^{(\mathrm E)}_d[R]\left(\frac{a[\eta]}{a[\eta']}\right)^{\frac {d-2}2}\int^\eta_{\eta'}\dd\eta_1\left(\frac{a[\eta']}{a[\eta_1]}\right)^{d-2},
\label{C2_Graviton_TimeIntegral_Cosmology} \\
C^{(g)}_{3,d}[\eta,\eta';R]=\,&a[\eta]^{\frac {d-2}2}\int^\eta_{\eta'}\dd\eta_4\,a[\eta_4]^{-(d-2)}\int^{\eta_4}_{\eta'}\dd\eta_3\,a[\eta_3]^{d-2}\int^{\eta_3}_{\eta'}\dd\eta_2
\,a[\eta_2]^{-(d-2)}\int^{\eta_2}_{\eta'}\dd\eta_1\,a[\eta_1]^{\frac {d-2}2}C^{(g)}_{1,d}[\eta_1,\eta';R] \notag\\
&+G^{(\mathrm E)}_d[R]\left(\frac{a[\eta]}{a[\eta']}\right)^{\frac{d-2}2}\int^\eta_{\eta'}\dd\eta_3\,a[\eta_3]^{-(d-2)}\int^{\eta_3}_{\eta'}\dd\eta_2
\,a[\eta_2]^{d-2}\int^{\eta_2}_{\eta'}\dd\eta_1\left(\frac{a[\eta']}{a[\eta_1]}\right)^{d-2} \notag\\
&+D_d[R]\left(\frac{a[\eta]}{a[\eta']}\right)^{\frac {d-2}2}\int^\eta_{\eta'}\dd\eta_1\left(\frac{a[\eta']}{a[\eta_1]}\right)^{d-2}.
\label{C3_Graviton_TimeIntegral_Cosmology}
\end{align}}%
Although the integrals in the first lines of eqs.~\eqref{C2_Graviton_TimeIntegral_Cosmology} and \eqref{C3_Graviton_TimeIntegral_Cosmology} can be difficult to carry out, just like eq.~\eqref{C2_Photon_TimeIntegral_Cosmology} for $C^{(\gamma)}_{2,d}$, the causal structures of these expressions can still be readily identified. Firstly, the retarded portion of $C^{(g)}_{1,d}$, or equivalently $G^{(g,+)}_d = - \Theta[T]C^{(g)}_{1,d}$, is composed only of the causal signals \cite{Chu:2016ngc}. Then, as already discussed in \S \eqref{MinkowskiSpacetime}, the first terms of both eqs.~\eqref{C2_Graviton_TimeIntegral_Cosmology} and \eqref{C3_Graviton_TimeIntegral_Cosmology} are causal as well when $\eta>\eta'$; whereas the remaining terms of $C^{(g)}_{2,d}$ and $C^{(g)}_{3,d}$, associated with their initial conditions, being non-zero for all $\eta > \eta'$, admit contributions from outside the light cone. For this reason, we see that, after plugging eqs.~\eqref{C2_Graviton_TimeIntegral_Cosmology} and \eqref{C3_Graviton_TimeIntegral_Cosmology} into eq.~\eqref{Cijmn_Cosmology}, the spin-2 effective Green's function $G^{(g,+)}_{ijmn}$ is acausal for all $d\geq4$.

At the quantum level, therefore, the free massless spin-2 operator $D_{ij}$ necessarily violates micro-causality in spatially flat cosmologies:
\begin{align}
\big[D_{ij}[x], D_{mn}[x']\big] = -i \big(a[\eta]a[\eta']\big)^{-\frac{d-2}2} C_{ijmn}^{(g)}[\eta,\eta';\vec R]  .
\end{align}

For later convenience, the expression \eqref{Spin2_Convolution_Cosmology} for spin-2 gravitons can be re-cast into another form analogous to their Minkowski counterpart \eqref{Spin2_Convolution2_Minkowski},
{\allowdisplaybreaks
\begin{align}
D_{ij}[\eta,\vec x]&=-16\pi G_\text N\int_{\mathbb{R}^{d-1}} \dd^{d-1}\vec x'\int_{\eta_{\text p}}^{\eta_{\text f}} \dd\eta'\,\left(\frac{a[\eta']}{a[\eta]}\right)^{\frac {d-2}2}\Bigg\{G^{(g,+)}_{d}\left({}^{(\text{a})}T_{ij}[\eta',\vec x']-\frac{\delta_{ij}}{d-2}{}^{(\text{a})}T_{ll}[\eta',\vec x']\right) \notag\\
&+2\Theta[T]a[\eta']^{\frac{d-2}2}\partial_{\eta'}\left(a[\eta']^{-\frac{d-2}2}\partial_{(i}C^{(g)}_{2,d}\right){}^{(\text{a})}T_{j)0}[\eta',\vec x']  + \frac{\delta_{ij}}{d-2}\Theta[T]a[\eta']^{\frac{d-2}2} \notag\\
&\times \bigg(\partial^2_{\eta'}\left(a[\eta']^{-\frac{d-2}2}C^{(g)}_{2,d}\right)
{}^{(\text{a})}T_{00}[\eta',\vec x']  +\mathcal H[\eta']\,\partial_{\eta'}\left(a[\eta']^{-\frac{d-2}2}C^{(g)}_{2,d}\right)\left({}^{(\text{a})}T_{00}[\eta',\vec x']-{}^{(\text{a})}T_{ll}[\eta',\vec x']\right)\bigg)
\notag\\
&+\frac1{d-2}\Theta[T]\partial_i\partial_jC^{(g)}_{2,d}{}^{(\text{a})}T_{ll}[\eta',\vec x']-\left(\frac{d-3}{d-2}\right)\Theta[T]a[\eta']^{\frac{d-2}2}
\bigg(\partial^2_{\eta'}\left(a[\eta']^{-\frac{d-2}2}\partial_i\partial_jC^{(g)}_{3,d}\right){}^{(\text{a})}T_{00}[\eta',\vec x']\notag\\
&+\mathcal H[\eta']\,\partial_{\eta'}\left(a[\eta']^{-\frac{d-2}2}\partial_i\partial_jC^{(g)}_{3,d}\right)
\left({}^{(\text{a})}T_{00}[\eta',\vec x']-{}^{(\text{a})}T_{ll}[\eta',\vec x']\right)   \bigg)  \Bigg\} \notag\\
&+\frac{16\pi G_\text N}{d-2}\int_{\mathbb{R}^{d-1}} \dd^{d-1}\vec x'  \bigg(\delta_{ij}G^{\mathrm{(E)}}_d{} ^{(\text{a})}T_{00}[\eta,\vec x']+(d-3)\partial_i\partial_jD_d{}^{(\text{a})}T_{00}[\eta,\vec x']\bigg),
\label{Spin2_Convolution2_Cosmology}
\end{align}}%
where $^{(\text{a})}T_{ll}\equiv\delta^{ij}\,^{(\text{a})}T_{ij}$, and we have used the conservation laws given in eqs.~\eqref{EnergyMomentumConservaion_Time_Cosmology} and \eqref{EnergyMomentumConservaion_Spatial_Cosmology} as well as the initial conditions \eqref{C123_Graviton_InitialCondition}, \eqref{dotC2_Graviton_InitialCondition}, and \eqref{dotC3_Graviton_InitialCondition}, and removed all the boundary contributions that result from integrations by parts.\footnote{Similar to the spin-1 case, the surface integrals upon integration by parts always involve $C^{(g)}_{2,d}$ and $C^{(g)}_{3,d}$; therefore,
when evaluated at past infinity, i.e., $\eta' = - \infty$ for $w=-1$ or $\eta'=0$ for $0 < w \leq1$, those terms obey the spin-2 homogeneous wave equation and will not change the inhomogeneous solution obtained here.
} The convolution in eq.~\eqref{Spin2_Convolution2_Cosmology} now involves different components of ${}^{(\text{a})}T_{\mu\nu}$ from that in eq.~\eqref{Spin2_Convolution_Cosmology}. Below, the former would help us identify how the acausal portions of the spin-2 contribution to the Weyl tensor are canceled by those from other gauge-invariant variables.

{\bf Bardeen Scalars} \qquad  Unlike the spin-2 gravitons, the field equation for the Bardeen scalar $\Psi$ no longer takes an universal form for both $w=-1$ and $0<w\leq1$, which will be solved separately for each case in the following. Once the solution of $\Psi$ is obtained, the other Bardeen scalar potential $\Phi$, related to $\Psi$ via the formula in eq.~\eqref{PhiPsiRelationship}, is given immediately by eq.~\eqref{PhiPsiRelationship_Convolution}.

{\it Solutions for $w=-1$} \qquad     In de Sitter background ($w=-1$), the Poisson-type equation \eqref{Psi_Poisson_Cosmology} obeyed by $\Psi$ involves a non-local function $\Sigma$ of the matter source (cf.~eq.~\eqref{Astro_SVT_IofII}), whose momentum-space counterpart for $\vec k\neq \vec0$ reads
\begin{align}
\widetilde\Sigma[\eta,\vec k]&=\frac{k_j}{i\vec k^2}{}^{(\text{a})}\widetilde T_{0j}[\eta,\vec k].
\label{Sigma_Fourier}
\end{align}
Once again, the solution to eq.~\eqref{Psi_Poisson_Cosmology} can be readily derived by implementing the Fourier transform of eq.~\eqref{Psi_Poisson_Cosmology} with eq.~\eqref{Sigma_Fourier},
\begin{align}
\Psi[\eta,\vec x]=\frac{8\pi G_\text N}{d-2}  \int_{\mathbb{R}^{d-1}} \dd^{d-1}\vec x'  \bigg(G^{\mathrm{(E)}}_d[R] {}^{(\text{a})}T_{00}[\eta,\vec x']+(d-1)\mathcal H[\eta]\, \partial_jD_d[R]  {}^{(\text{a})}T_{0j}[\eta,\vec x']\bigg),
\label{BardeenPsi_deSitter}
\end{align}
which is again the weighted superposition of local source terms over the equal-time hypersurface.

{\it Solutions for $0<w\leq1$} \qquad  When $0<w\leq1$, the Bardeen scalar $\Psi$ becomes dynamical in the sense of being governed by the wave equation \eqref{Psi_WaveEq_Cosmology}, from which we see that the propagation of scalar GWs is in general different than that of the spin-2 ones. However, the strategy of solving the spin-2 wave equation in light of causality still applies here for $\Psi$-waves.

In the same vein, the scalar wave equation \eqref{Psi_WaveEq_Cosmology} can firstly be re-written as a re-scaling form,
\begin{align}
\bigg\{\partial^2_{(w)}  +  &  \frac{ (d-2) \left(q_w + d -2 \right) }{q^2_w\eta^2}  \bigg\}\left(a^{\frac12\left(q_w+d-2\right)}\Psi\right) \notag\\
&\qquad    =-8\pi G_\text N\,a^{\frac12\left(q_w+d-2\right)}\bigg(\frac{\partial_0\left(a^{d-2}\Sigma\right)}{(d-2)a^{d-2}}-\frac{w\rho}{(d-2)}+\mathcal H\dot{\Upsilon}\bigg),
\label{Psi_WaveEq2_Cosmology}
\end{align}
where $\partial^2_{(w)}\equiv-\partial_\eta^2+w\vec\nabla^2$. Then, using eqs.~\eqref{Upsilon_Fourier_Minkowski} and \eqref{Sigma_Fourier} through the same procedure employed for the spin-2 wave equation, we have
{\allowdisplaybreaks
\begin{align}
a[\eta]&^{\frac12\left(q_w+d-2\right)}\Psi[\eta,\vec x]=  \frac{8\pi G_\text N}{d-2} \int_{\mathbb{R}^{d-1}} \dd^{d-1}\vec x'\int_0^\infty \dd\eta'  \,  \Theta[T]a[\eta']^{\frac12\left(q_w+d-2\right)}  \notag\\
&\times w^{-\frac{d-3}2}  \Bigg\{-\partial_jC^{(w)}_{2,d}\left[\eta,\eta';\textstyle{\frac {R}{\sqrt w}}\right] a[\eta']^{-(d-2)} \partial_{\eta'}\left(a[\eta']^{d-2} \,^{(\text{a})}T_{0j}[\eta',\vec x']\right)
-C^{(w)}_{1,d}\left[\eta,\eta';\textstyle{\frac {R}{\sqrt w}}\right] {}^{(\text{a})}T_{00}[\eta',\vec x']       \notag\\
& +\mathcal H[\eta'] \bigg((d-1)w\,\partial_i\partial_jC^{(w)}_{3,d}\left[\eta,\eta';\textstyle{\frac {R}{\sqrt w}}\right]{}^{(\text{a})}\dot{T}_{ij}[\eta',\vec x']
+C^{(w)}_{2,d}\left[\eta,\eta';\textstyle{\frac {R}{\sqrt w}}\right]{}^{(\text{a})}\dot{T}_{ll}[\eta',\vec x']\bigg)\Bigg\},
\label{Psi_Convolution_Cosmology}
\end{align}}%
where the scalar functions $C^{(w)}_{1,d}$, $C^{(w)}_{2,d}$, and $C^{(w)}_{3,d}$, respectively, are defined in a manner similar to eqs.~\eqref{C1_Graviton_Cosmology}, \eqref{C2_Graviton_Cosmology}, and \eqref{C3_Graviton_Cosmology},
{\allowdisplaybreaks
\begin{align}
&C^{(w)}_{1,d}\left[\eta,\eta';\textstyle{\frac {R}{\sqrt w}}\right]\equiv w^{\frac{d-1}2}\int_{\mathbb{R}^{d-1}}\frac{\dd^{d-1}\vec{k}}{(2\pi)^{d-1}}\,\widetilde C^{(w)}_{1,d}\left[\eta,\eta';\sqrt{w}|\vec k|\right]\,e^{i\vec{k}\cdot\vec{R}},
\label{C1_Psi_Cosmology}\\
&C^{(w)}_{2,d}\left[\eta,\eta';\textstyle{\frac {R}{\sqrt w}}\right]\equiv w^{\frac{d-3}2}\int_{\mathbb{R}^{d-1}}\frac{\dd^{d-1}\vec{k}}{(2\pi)^{d-1}}\,\widetilde C^{(w)}_{1,d}\left[\eta,\eta';\sqrt{w}|\vec k|\right]\,\frac{e^{i\vec{k}\cdot\vec{R}}}{\vec k^2},
\label{C2_Psi_Cosmology}\\
&C^{(w)}_{3,d}\left[\eta,\eta';\textstyle{\frac {R}{\sqrt w}}\right]\equiv w^{\frac{d-5}2}\int_{\mathbb{R}^{d-1}}\frac{\dd^{d-1}\vec{k}}{(2\pi)^{d-1}}\,\widetilde C^{(w)}_{1,d}\left[\eta,\eta';\sqrt{w}|\vec k|\right]\,\frac{e^{i\vec{k}\cdot\vec{R}}}{\vec k^4},
\label{C3_Psi_Cosmology}
\end{align}}%
in which $\widetilde C^{(w)}_{1,d}$ represents the Fourier transform of $C^{(w)}_{1,d}$ with respect to $\vec R/\sqrt{w}$, and it obeys the Fourier-transformed homogeneous wave equation of eq.~\eqref{Psi_WaveEq2_Cosmology} with  initial conditions specified by $\widetilde C^{(w)}_{1,d}\big|_{\eta=\eta'}=0$ and $\dot{\widetilde C}^{(w)}_{1,d}\big|_{\eta=\eta'}=-\partial_{\eta'}\widetilde C^{(w)}_{1,d}\big|_{\eta=\eta'}=1$.\footnote{The factor of $w$ appearing in eq.~\eqref{C1_Psi_Cosmology} has been arranged such that the corresponding massless scalar Green's function, $G^{(w,+)}_{d}=-\Theta[T]C^{(w)}_{1,d}$ with re-scaled coordinates $(\eta,\vec y)  \equiv  (\eta,\vec x/\sqrt{w})$, obeys the wave equation
\begin{align}
\bigg\{-\partial^2_\eta+\frac{ (d-2) \left(q_w + d -2 \right) }{q^2_w\eta^2} +\vec\nabla^2_{\vec y}\bigg\}\,G^{(w,+)}_d\left[\eta,\eta';|\vec y-\vec y'|\right]
=   \delta[\eta-\eta']\delta^{(d-1)}[\vec y-\vec y'],                  \notag
\end{align}
where $\vec\nabla^2_{\vec y}$ is the spatial Laplacian with respect to $\vec y$. The analytic solution of $G^{(w,+)}_d$ for $0<w\leq1$ has been obtained in \cite{Chu:2016ngc}.} Analogous to eqs.~\eqref{C1_Photon_Cosmology_ModeDecomposition} and \eqref{C1_Graviton_Cosmology_ModeDecomposition}, $\widetilde C^{(w)}_{1,d}$ admits the following decomposition in terms of the mode functions $v^{(w)}_{\sqrt{w}|\vec k|}$ that obey the same homogeneous wave equation in momentum space,
\begin{align}
\widetilde C^{(w)}_{1,d}\left[\eta,\eta';\sqrt{w}|\vec k|\right]=i\left(v^{(w)}_{\sqrt{w}|\vec k|}[\eta]\,v^{(w)*}_{\sqrt{w}|\vec k|}[\eta']-v^{(w)*}_{\sqrt{w}|\vec k|}[\eta]\,v^{(w)}_{\sqrt{w}|\vec k|}[\eta']\vphantom{\dot A}\right),
\label{C1w_Cosmology_ModeDecomposition}
\end{align}
where the Wronskian condition, $v^{(w)}_{\sqrt{w}|\vec k|}\dot{v}^{(w)*}_{\sqrt{w}|\vec k|}-v^{(w)*}_{\sqrt{w}|\vec k|}\dot{v}^{(w)}_{\sqrt{w}|\vec k|}=i$, is fulfilled to be consistent with the initial condition $\dot{\widetilde C}^{(w)}_{1,d}\big|_{\eta=\eta'}=1$. By construction, these $C^{(w)}$s are solutions to the homogeneous version of eq.~\eqref{Psi_WaveEq2_Cosmology}. Moreover, their equal-time initial conditions may be readily identified,
{\allowdisplaybreaks
\begin{align}
&C^{(w)}_{1,d}\big|_{\eta=\eta'}=C^{(w)}_{2,d}\big|_{\eta=\eta'}=C^{(w)}_{3,d}\big|_{\eta=\eta'}=0,
\label{C123_Psi_InitialCondition}\\
&\dot C^{(w)}_{1,d}\big|_{\eta=\eta'}=-\partial_{\eta'} C^{(w)}_{1,d}\big|_{\eta=\eta'}=w^{\frac{d-1}2}\delta^{(d-1)}[\vec x-\vec x'],  \label{dotC1_Psi_InitialCondition} \\
&\dot C^{(w)}_{2,d}\big|_{\eta=\eta'}=-\partial_{\eta'} C^{(w)}_{2,d}\big|_{\eta=\eta'}=-w^{\frac{d-3}2} G^{\mathrm{(E)}}_d[R],
\label{dotC2_Psi_InitialCondition} \\
&\dot C^{(w)}_{3,d}\big|_{\eta=\eta'}=-\partial_{\eta'} C^{(w)}_{3,d}\big|_{\eta=\eta'}=w^{\frac{d-5}2}D_d[R].
\label{dotC3_Psi_InitialCondition}
\end{align}}%
Exploiting their Fourier representations to observe that $-w\vec{\nabla}^2 C_{2,d}^{(w)} = C_{1,d}^{(w)}$ and $-w\vec{\nabla}^2 C_{3,d}^{(w)} = C_{2,d}^{(w)}$, we see that the homogeneous cousins of eq.~\eqref{Psi_WaveEq2_Cosmology} are
{\allowdisplaybreaks
	\begin{align}
	\frac{-\partial_\eta\bigg(a^{2\alpha}\,\partial_\eta\left(a^{-\alpha}\, C_{2,d}^{(w)}  \vphantom{\dot A}\right)\bigg)}{a^\alpha}&= C_{1,d}^{(w)},\\
	\frac{-\partial_\eta\bigg(a^{2\alpha}\,\partial_\eta\left(a^{-\alpha}\, C_{3,d}^{(w)} \vphantom{\dot A}\right)\bigg)}{a^\alpha}&= C_{2,d}^{(w)},
	\qquad\qquad
	\alpha=-\frac{d-2}2 .
	\end{align}}%
At this point we may integrate these equations to express $C^{(w)}_{2,d}$ and $C^{(w)}_{3,d}$ in terms of $C^{(w)}_{1,d}$, which had been derived analytically in \cite{Chu:2016ngc} (see eqs.~(131) and (132) of \cite{Chu:2016ngc}). The resulting expressions are
{\allowdisplaybreaks
\begin{align}
C_{2,d}^{(w)}\left[\eta,\eta';\textstyle{\frac {R}{\sqrt w}}\right]=\,&-a[\eta]^{-\frac {d-2}2}\int^\eta_{\eta'} \dd \eta_2\,a[\eta_2]^{d-2}\int^{\eta_2}_{\eta'} \dd \eta_1\,a[\eta_1]^{-\frac {d-2}2}C^{(w)}_{1,d}\left[\eta_1,\eta';\textstyle{\frac {R}{\sqrt w}}\right] \notag\\
&-w^{\frac{d-3}2}G^{(\mathrm E)}_d[R]\left(\frac{a[\eta']}{a[\eta]}\right)^{\frac {d-2}2}\int^\eta_{\eta'}\dd\eta_1\left(\frac{a[\eta_1]}{a[\eta']}\right)^{d-2},
\label{C2_Psi_TimeIntegral_Cosmology}   \\
C_{3,d}^{(w)}\left[\eta,\eta';\textstyle{\frac {R}{\sqrt w}}\right]=\,&  a[\eta]^{-\frac {d-2}2}\int^\eta_{\eta'}\dd\eta_4\,a[\eta_4]^{(d-2)}\int^{\eta_4}_{\eta'}\dd\eta_3\,a[\eta_3]^{-(d-2)}  \notag\\
& \qquad \qquad \times\int^{\eta_3}_{\eta'}\dd\eta_2\,a[\eta_2]^{(d-2)}
\int^{\eta_2}_{\eta'}\dd \eta_1\,a[\eta_1]^{-\frac {d-2}2}C^{(w)}_{1,d}\left[\eta_1,\eta';\textstyle{\frac {R}{\sqrt w}}\right]\notag\\
&+w^{\frac{d-3}2}G^{(\mathrm E)}_d[R]\left(\frac{a[\eta']}{a[\eta]}\right)^{\frac{d-2}2}\int^\eta_{\eta'}\dd\eta_3\,a[\eta_3]^{d-2}\int^{\eta_3}_{\eta'}\dd\eta_2
\,a[\eta_2]^{-(d-2)}\int^{\eta_2}_{\eta'}\dd\eta_1\left(\frac{a[\eta_1]}{a[\eta']}\right)^{d-2} \notag\\
&+w^{\frac{d-5}2}D_d[R]\left(\frac{a[\eta']}{a[\eta]}\right)^{\frac {d-2}2}\int^\eta_{\eta'}\dd\eta_1\left(\frac{a[\eta_1]}{a[\eta']}\right)^{d-2},
\label{C3_Psi_TimeIntegral_Cosmology}
\end{align}}%
which bear close resemblance to eqs.~\eqref{C2_Graviton_TimeIntegral_Cosmology} and \eqref{C3_Graviton_TimeIntegral_Cosmology}. The retarded part of $C^{(w)}_{1,d}$, or the massless scalar Green's function $G^{(w,+)}_d=-\Theta[T]C^{(w)}_{1,d}$, has been shown to contain only the causal scalar GW signals propagating either on or within the acoustic cone \cite{Chu:2016ngc}. Applying the arguments in \S \eqref{MinkowskiSpacetime} -- the first terms of eqs.~\eqref{C2_Psi_TimeIntegral_Cosmology} and \eqref{C3_Psi_TimeIntegral_Cosmology} are causal when $\eta>\eta'$ because $C_{1,d}^{(w)}$ is; while the rest of the terms are non-zero both inside and outside the light cone of $(\eta',\vec{x}')$. The Bardeen scalar $\Psi$ is therefore acausal for all relevant spacetime dimensions (cf.~eq.~\eqref{Psi_Convolution_Cosmology}).

Alternatively, we can perform integration-by-parts and employ the energy-momentum conservation laws \eqref{EnergyMomentumConservaion_Time_Cosmology} and \eqref{EnergyMomentumConservaion_Spatial_Cosmology}, as well as the properties of $C^{(w)}_{2,d}$ and $C^{(w)}_{3,d}$, to re-express the effective Green's function of $\Psi$ in eq.~\eqref{Psi_Convolution_Cosmology} as
{\allowdisplaybreaks
\begin{align}
\Psi[\eta,\vec x &] =  \frac{8\pi G_\text N}{d-2} \int_{\mathbb{R}^{d-1}} \dd^{d-1}\vec x'\int_0^\infty \dd\eta' \, \left(\frac{a[\eta']}{a[\eta]}\right)^{\frac12\left(q_w+d-2\right)}   w^{-\frac{d-3}2}\,\Bigg\{G^{(w,+)}_d
{}^{(\text{a})}T_{00}[\eta',\vec x']\notag\\
&-\Theta[T]a[\eta']^{-\frac12\left(q_w+d-2\right)}\partial_{\eta'}\left(a[\eta']^{\frac12\left(q_w+d-2\right)}\mathcal H[\eta']\,C^{(w)}_{2,d}\right){}^{(\text{a})}T_{ll}[\eta',\vec x'] + (d-2)\Theta[T]a[\eta']^{-\frac12\left(q_w-d+2\right)} \notag\\
&\times \partial_{\eta'}\left(a[\eta']^{\frac12\left(q_w-d+2\right)}\mathcal H[\eta']\,C^{(w)}_{2,d}\right){}^{(\text{a})}T_{00}[\eta',\vec x']  +  (d-2)\Theta[T]\mathcal H[\eta']^2\,C^{(w)}_{2,d}\,\left({}^{(\text{a})}T_{00}[\eta',\vec x']-{}^{(\text{a})}T_{ll}[\eta',\vec x']  \right) \notag\\
&-\Theta[T]a[\eta']^{-\frac12\left(q_w-d+2\right)}\partial_{\eta'}\bigg(a[\eta']^{-(d-2)}\partial_{\eta'}
\left(a[\eta']^{\frac12\left(q_w+d-2\right)}\,C^{(w)}_{2,d}\right)\bigg){}^{(\text{a})}T_{00}[\eta',\vec x']\notag\\
&-\Theta[T]a[\eta']^{-\frac12\left(q_w+d-2\right)}\mathcal H[\eta']\partial_{\eta'}\left(a[\eta']^{\frac12\left(q_w+d-2\right)}\,C^{(w)}_{2,d}\right)\left({}^{(\text{a})}T_{00}[\eta',\vec x']
-{}^{(\text{a})}T_{ll}[\eta',\vec x']  \right) \notag\\
&-w(d-1)\Theta[T]a[\eta']^{-\frac12\left(q_w-d+2\right)}\partial_{\eta'}^2\bigg(a[\eta']^{-(d-2)}\partial_{\eta'}
\left(a[\eta']^{\frac12\left(q_w+d-2\right)}\mathcal H[\eta']\,C^{(w)}_{3,d}\right)\bigg){}^{(\text{a})}T_{00}[\eta',\vec x'] \notag\\
&-w(d-1)\Theta[T]a[\eta']^{-\frac12\left(q_w-d+2\right)}\mathcal H[\eta']\partial_{\eta'}\bigg(a[\eta']^{-(d-2)}\partial_{\eta'}
\left(a[\eta']^{\frac12\left(q_w+d-2\right)}\mathcal H[\eta']\,C^{(w)}_{3,d}\right)\bigg)\left({}^{(\text{a})}T_{00}[\eta',\vec x'] \right.\notag\\
& \left.-{}^{(\text{a})}T_{ll}[\eta',\vec x']\right) \Bigg\}  + \frac{8\pi G_\text N}{d-2}  \int_{\mathbb{R}^{d-1}}  \dd^{d-1}\vec x'\,\bigg( G^{(\mathrm E)}_d {}^{(\text{a})}T_{00}[\eta,\vec x']+(d-1)\mathcal H[\eta]\partial_jD_d {}^{(\text{a})}T_{0j}[\eta,\vec x']\bigg),
\label{Psi_Convolution2_Cosmology}
\end{align}}%
where all the boundary terms that arise from the integrations by parts have been discarded.\footnote{As previously reasoned in the spin-1 and spin-2 cases, discarding the boundary contributions at past infinity does not affect the inhomogeneous solutions of the Bardeen scalars, since the corresponding homogeneous wave equation is obeyed by those surface terms, which correspond to evaluating the surface integrals of $C^{(w)}_{2,d}$ and $C^{(w)}_{3,d}$ at $\eta'=-\infty$ (for $w = -1$) or $\eta' = 0$ (for $0 < w \leq 1$).
} This form shows more transparently the convolution with the local matter stress-energy tensor, and will be used for our later analysis of the physical observables.

{\bf Vector Potential} \qquad  According to \cite{Chu:2016ngc}, the vector mode $V_i$, in de Sitter space ($w=-1$), obeys the Poisson-type equation \eqref{Vi_Poisson_Minkowski}, while for $0<w\leq1$, if perturbations are assumed to be negligible in the far past, then  the same vector equation, i.e., eq.~\eqref{Vi_Poisson_Minkowski}, is satisfied as well. Therefore, in both cases, the solution of $V_i$ is that in eq.~\eqref{Vi_Minkowski}.

{\bf Linearized Weyl Tensor} \qquad   As we have discussed in \S\eqref{Chapter_Observables}, the linearized Riemann tensor in the cosmological background is no longer gauge invariant due to its non-zero background value. In cosmological spacetimes, which are conformally flat, the causal and gauge-invariant counterpart to the linearized Riemann in flat spacetimes is the linearized Weyl tensor $\delta_1 C^\mu{}_{\nu \alpha\beta}$. More specifically, since the Weyl tensor $C^\mu_{\phantom{\mu}\nu \alpha\beta}$ is conformally invariant, it is zero when evaluated on the unperturbed cosmological geometry $g_{\mu\nu} = a^2 \eta_{\mu\nu}$ and must therefore be gauge-invariant at first order in $\chi_{\mu\nu}$. Furthermore, its exact wave equation is simply the traceless part of eq.~\eqref{Riemann_WaveEq}; but since it is zero at zeroth order, the first order Weyl tensor $\delta_1 C^\mu_{\phantom{\mu}\nu \alpha\beta}$ must therefore obey an equation involving the wave operator with respect to the background FLRW metric.

Motivated by these considerations, we shall proceed to calculate
\begin{align}
\delta_1 C^i{}_{0j0}=\,&\left(\frac{d-3}{d-2}\right)\Bigg\{\left(\partial_i\partial_j-\frac{\delta_{ij}}{d-1}\vec\nabla^2\right)
\left(\Phi+\Psi\right)+\partial_{(i}\dot{V}_{j)}-\frac12\bigg(\ddot{D}_{ij}+\frac1{d-3}\vec\nabla^2D_{ij}\bigg)\Bigg\}.
\label{LinearizedWeyl_GaugeInvariant}
\end{align}
It is likely that $\delta_1 C^i{}_{0j0}$ encodes the dominant contributions to the first-order tidal forces described in eq.~\eqref{GeodesicDeviation_FreeFall}; but we shall leave this analysis to future work \cite{ChuLiuInPrep}. Here, we will instead focus on the causal structure of this quantity with respect to the background spacetime.

{\it Linearized Weyl tensor for $w=-1$} \qquad  Within the de Sitter case, plugging into eq.~\eqref{LinearizedWeyl_GaugeInvariant} the solution of $D_{ij}$ in eq.~\eqref{Spin2_Convolution2_Cosmology}, those of $\Psi$ and $\Phi$ in eqs.~\eqref{BardeenPsi_deSitter} and \eqref{PhiPsiRelationship_Convolution}, and that of $V_i$ in eq.~\eqref{Vi_Minkowski}, with $\mathcal H[\eta]=-1/\eta$ and $(\eta_{\text p},\eta_{\text f})=(-\infty,0)$, we find that, after employing the conservation conditions \eqref{EnergyMomentumConservaion_Time_Cosmology} and \eqref{EnergyMomentumConservaion_Spatial_Cosmology}, the scalars and vector act to cancel the acausal signals from the tensor contributions to Weyl. In more detail,
\begin{align}
\delta_1 C^i{}_{0j0}=-\frac12\left(\frac{d-3}{d-2}\right)&\left(\ddot{D}_{ij}+\frac1{d-3}\vec\nabla^2D_{ij}\right)_{\mathrm{causal}}\notag\\
&+ \frac{8\pi G_\text N}{d-2} \left({}^{(\text{a})}T_{ij}-\frac{\delta_{ij}}{d-1}\left((d-3){}^{(\text{a})}T_{00}
+2{}^{(\text{a})}T_{ll}\right)\right) ;
\label{LinearizedWeyl_deSitter}
\end{align}
where the first line of eq.~\eqref{LinearizedWeyl_deSitter} denotes the causal part of the spin-2 contributions that depend exclusively on the retarded Green's function $G^{(g,+)}_d[\eta,\eta';R]$,
{\allowdisplaybreaks
\begin{align}
&\bigg(\ddot{D}_{ij}[\eta,\vec x]+\frac1{d-3}\vec\nabla^2D_{ij}[\eta,\vec x]\bigg)_{\mathrm{causal}}\notag\\
=\,&-16\pi G_\text N\left(\frac{d-2}{d-3}\right)\int_{\mathbb{R}^{d-1}} \dd^{d-1}\vec x'\int_{\eta_{\text p}}^{\eta_{\text f}}  \dd\eta' \,\left(\frac{a[\eta']}{a[\eta]}\right)^{\frac {d-2}2}  \Bigg\{\bigg(\ddot G^{(g,+)}_{d}- \frac{2(d-3)}{q_w\eta} \dot G^{(g,+)}_{d}     \notag\\
&+ \frac{(d-2)\left(q_w+d-4\right)}{q_w^2 \eta^2} \, G^{(g,+)}_{d}\bigg)\left({}^{(\text{a})}T_{ij}[\eta',\vec x']-\frac{\delta_{ij}}{d-2}{}^{(\text{a})}T_{ll}[\eta',\vec x']\right)\notag\\
&+2a[\eta']^{\frac{d-2}2}\partial_{\eta'}\Bigg(a[\eta']^{-\frac{d-2}2}\bigg(\partial_{(i}G^{(g,+)}_{d}  -  \frac{2(d-3)}{q_w\eta} a[\eta]^{-\frac{d-2}2}
\int^\eta_{\eta'}\dd\eta_1\,a[\eta_1]^{\frac{d-2}2}\partial_{(i}G^{(g,+)}_{d}\bigg)\Bigg){}^{(\text{a})}T_{j)0}[\eta',\vec x']   \notag\\
&+\frac{\delta_{ij}}{d-2}a[\eta']^{\frac{d-2}2}\Bigg(\partial^2_{\eta'}\bigg(a[\eta']^{-\frac{d-2}2}
\bigg(G^{(g,+)}_{d}- \frac{2(d-3)}{q_w\eta} a[\eta]^{-\frac{d-2}2} \int^\eta_{\eta'}\dd\eta_1\,a[\eta_1]^{\frac{d-2}2}G^{(g,+)}_{d}\bigg)\bigg){}^{(\text{a})}T_{00}[\eta',\vec x'] \notag\\
&+\frac{2}{q_w\eta'}\partial_{\eta'}\bigg(a[\eta']^{-\frac{d-2}2}\bigg(G^{(g,+)}_{d}-\frac{2(d-3)}{q_w\eta}a[\eta]^{-\frac{d-2}2}
\int^\eta_{\eta'}\dd\eta_1\,a[\eta_1]^{\frac{d-2}2}G^{(g,+)}_{d}\bigg)\bigg)\left({}^{(\text{a})}T_{00}[\eta',\vec x'] - {}^{(\text{a})}T_{ll}[\eta',\vec x']\right)\Bigg) \notag\\
&+\frac1{d-2}\bigg(\partial_i\partial_jG^{(g,+)}_{d}-\frac{2(d-3)}{q_w\eta}a[\eta]^{-\frac{d-2}2}
\int^\eta_{\eta'}\dd\eta_1\,a[\eta_1]^{\frac{d-2}2}\partial_i\partial_jG^{(g,+)}_{d}\bigg){}^{(\text{a})}T_{ll}[\eta',\vec x'] \notag\\
&+\left(\frac{d-3}{d-2}\right)a[\eta']^{\frac{d-2}2}a[\eta]^{\frac{d-2}2}\Bigg(\partial_{\eta'}^2\bigg(a[\eta']^{-\frac{d-2}2}
\bigg(\int^\eta_{\eta'}\dd\eta_2\,a[\eta_2]^{-(d-2)}
\int^{\eta_2}_{\eta'}\dd\eta_1\,a[\eta_1]^{\frac{d-2}2}\partial_i\partial_jG^{(g,+)}_{d}\notag\\
&-\frac{2(d-3)}{q_w\eta}a[\eta]^{-(d-2)}\int^\eta_{\eta'}\dd\eta_3\,a[\eta_3]^{d-2}\int_{\eta'}^{\eta_3}\dd\eta_2\,a[\eta_2]^{-(d-2)}
\int^{\eta_2}_{\eta'}\dd\eta_1\,a[\eta_1]^{\frac{d-2}2}\partial_i\partial_jG^{(g,+)}_{d}\bigg)\bigg){}^{(\text{a})}T_{00}[\eta',\vec x'] \notag\\
&+\frac{2}{q_w\eta'}\partial_{\eta'}\bigg(a[\eta']^{-\frac{d-2}2}\bigg(\int^\eta_{\eta'}\dd\eta_2\,a[\eta_2]^{-(d-2)}
\int^{\eta_2}_{\eta'}\dd\eta_1\,a[\eta_1]^{\frac{d-2}2}\partial_i\partial_jG^{(g,+)}_{d}\notag\\
&-\frac{2(d-3)}{q_w\eta}a[\eta]^{-(d-2)}\int^\eta_{\eta'}\dd\eta_3\,a[\eta_3]^{d-2}\int_{\eta'}^{\eta_3}\dd\eta_2\,a[\eta_2]^{-(d-2)}
\int^{\eta_2}_{\eta'}\dd\eta_1\,a[\eta_1]^{\frac{d-2}2}\partial_i\partial_jG^{(g,+)}_{d}\bigg)\bigg) \notag\\
&\times \left({}^{(\text{a})}T_{00}[\eta',\vec x']-{}^{(\text{a})}T_{ll}[\eta',\vec x']\right)\Bigg)\Bigg\},
\label{WeylSpin2Sector_Retarded}
\end{align}}%
while the second line of eq.~\eqref{LinearizedWeyl_deSitter} consists solely of the stress-energy tensor of the GW source evaluated at the observer location. (Recall that $a[\eta]=-1/(H\eta)$ and $q_w=-2$ in de Sitter spacetime.) As long as the observer at $(\eta,\vec{x})$ is not located at the source, these ${}^{(\text{a})}T_{\mu\nu}[\eta,\vec{x}]$ terms in eq.~\eqref{LinearizedWeyl_deSitter} are zero.\footnote{This calculation is greatly simplified by first using the commutator $C^{(g)}_{1,d}$, and only re-expressing the final result in terms of the massless scalar Green's function via $G^{(g,+)}_d=-\Theta[T]C^{(g)}_{1,d}$ at the very end. In particular, we notice that a local term will show up in the conversion involving a second time derivative, namely $\ddot G^{(g,+)}_{d}=-\delta^{(d)}[x-x']-\Theta[T]\ddot C^{(g)}_{1,d}$ or  $\partial_{\eta'}^2 G^{(g,+)}_{d}=-\delta^{(d)}[x-x']-\Theta[T]\partial^2_{\eta'} C^{(g)}_{1,d}$. \newline A simple check of eq.~\eqref{LinearizedWeyl_deSitter} can be made by taking the limit of eq.~\eqref{WeylSpin2Sector_Retarded} as $a\to1$ and assuming $G^{(g,+)}_d$ takes the form of $G^+_d$ in Minkowski spacetime. We may then show explicitly that the first line of eq.~\eqref{LinearizedWeyl_deSitter} reduces to $\delta_1 R_{0i0j}$ given in eq.~\eqref{LinearizedRiemann_Sol_Minkowski}, and the resulting $\delta_1 C^i{}_{0j0}$ is consistent with its Minkowski counterpart obtained from the solutions derived in the last section. Moreover, the Minkowski form of $\delta_1 C^i{}_{0j0}$ also agrees with the relationship between the Riemann and the Weyl tensor,
\begin{align}
C^{\rho}{}_{\sigma\mu\nu}=\,& R^{\rho}{}_{\sigma\mu\nu} - \frac{16\pi G_\text N}{d-2}\left( \delta^\rho_{[\mu}T_{\nu]\sigma}-g_{\sigma[\mu}T_{\nu]}{}^\rho  -  \delta^\rho_{[\mu}g_{\nu]\sigma}
\frac{2g^{\alpha\beta}T_{\alpha\beta}}{d-1}\right),
\label{RiemannWeyl}
\end{align}
linearized about the Minkowski background; where $T_{\mu\nu}$ is the total energy-momentum tensor of matter and Einstein's equation has been imposed on the trace parts of the Riemann tensor.}

To sum: the result in eq.~\eqref{LinearizedWeyl_deSitter} reveals that $\delta_1 C^i{}_{0j0}$ on a de Sitter background receives only signals from the spin-2 sector, as long as the observer is away from the isolated matter source(s) of GWs.

Analogous to the localization of the effective Green's function of $F_{0i}$ shown in eq.~\eqref{F0i_Cosmology_Localized}, the expression of eq.~\eqref{WeylSpin2Sector_Retarded} in de Sitter spacetime can be further simplified in a localized manner by introducing two additional massless scalar Green's functions $G^{(\text V,+)}_d$ and $G^{(\text{Tr},+)}_d$ that, respectively, obey the following wave equations,
\begin{align}
&\left\{\partial^2+\frac{(d-4)(d-2)}{4\eta^2}\right\}G^{(\text V,+)}_d[\eta,\eta';R]=\delta^{(d)}[x-x'],  \label{GV+_deSitter} \\
&\left\{\partial^2+\frac{(d-6)(d-4)}{4\eta^2}\right\}G^{(\text{Tr},+)}_d[\eta,\eta';R]=\delta^{(d)}[x-x'].
\label{GS+_deSitter}
\end{align}
The effective Green's function of $\delta_1 C^i{}_{0j0}$ in eq.~\eqref{LinearizedWeyl_deSitter} can then be localized accordingly in terms of $G^{(g,+)}_d$, $G^{(\text V,+)}_d$, and $G^{(\text{Tr},+)}_d$,
{\allowdisplaybreaks
\begin{align}
\delta_1 C^i{}_{0j0}&[\eta,\vec x]=8\pi G_\text N\int_{\mathbb{R}^{d-1}} \dd^{d-1}\vec x'\int_{-\infty}^{0}  \dd\eta' \,
\left(\frac\eta{\eta'}\right)^{\frac {d-2}2}\Bigg\{ \bigg(\ddot G^{(g,+)}_{d}+\frac{(d-3)}{\eta}\dot G^{(g,+)}_{d}+\frac{(d-2)(d-6)}{4\eta^2}G^{(g,+)}_{d}\bigg) \notag\\
&\times \left({}^{(\text{a})}T_{ij}[\eta',\vec x']-\frac{\delta_{ij}}{d-2}{}^{(\text{a})}T_{ll}[\eta',\vec x']\right)-2\eta^{-\frac{d-4}2}\partial_\eta\left(\eta^{\frac{d-4}2}\partial_{(i}G^{(\text V,+)}_d\right){}^{(\text{a})}T_{j)0}[\eta',\vec x'] \notag\\
&-\frac{\delta_{ij}}{d-2}\Bigg(\left(\eta\eta'\right)^{-\frac{d-4}2}\partial_\eta\partial_{\eta'}
\left(\left(\eta\eta'\right)^{\frac{d-4}2}G^{(\text V,+)}_d\right)
{}^{(\text{a})}T_{00}[\eta',\vec x']+\eta'^{-1}\eta^{-\frac{d-4}2}\partial_\eta\left(\eta^{\frac{d-4}2}G^{(\text V,+)}_d\right){}^{(\text{a})}T_{ll}[\eta',\vec x']\Bigg) \notag\\ &+\frac1{d-2}\partial_i\partial_jG^{(\text{Tr},+)}_d\left((d-3){}^{(\text{a})}T_{00}[\eta',\vec x']+{}^{(\text{a})}T_{ll}[\eta',\vec x']\right)\Bigg\}    \notag\\
&+ \frac{8\pi G_\text N}{d-2} \left({}^{(\text{a})}T_{ij}[\eta,\vec x]-\frac{\delta_{ij}}{d-1}\left((d-3){}^{(\text{a})}T_{00}[\eta,\vec x]
+2{}^{(\text{a})}T_{ll}[\eta,\vec x]\right)\right),
\label{LinearizedWeyl_deSitter_Localized}
\end{align}}%
which will be explained in more detail in \cite{ChuLiuInPrep}. Eq.~\eqref{LinearizedWeyl_deSitter_Localized} turns out to be consistent with $\delta_1 C^i{}_{0j0}$ computed from the generalized de Donder gauge $\overline \chi_{\mu\nu}$ solution obtained in \cite{Chu:2016qxp}; the analytic solutions of $G^{(g,+)}_d = -\Theta[T] \mathcal G^{(\text T)}$, $G^{(\text V,+)}_d = -\Theta[T] \mathcal G^{(\text V)}$, and $G^{(\text{Tr},+)}_d = -\Theta[T] \mathcal G^{(\text{Tr})}$ can be found in eqs.~(28), (29), (33), (34), (38), and (39) of \cite{Chu:2016qxp}.

{\it Linearized Weyl tensor for $0<w\leq1$} \qquad   To obtain $\delta_1 C^i{}_{0j0}$ for a relativistic equation-of-state $w$ within $0<w\leq1$, we insert into eq.~\eqref{LinearizedWeyl_GaugeInvariant} eq.~\eqref{Spin2_Convolution2_Cosmology} for $D_{ij}$, eqs.~\eqref{Psi_Convolution2_Cosmology} and \eqref{PhiPsiRelationship_Convolution} for $\Psi$ and $\Phi$, and eq.~\eqref{Vi_Minkowski} for $V_i$; recalling that $\mathcal H[\eta]=2/(q_w\eta)$ and $(\eta_{\text p},\eta_{\text f})=(0,\infty)$. A direct computation then reveals that an exact cancellation of the acausal signals takes place again in eq.~\eqref{LinearizedWeyl_GaugeInvariant}, so that
\begin{align}
\delta_1 C^i{}_{0j0}=\,&\left(\frac{d-3}{d-2}\right)\Bigg\{\Bigg(\left(\partial_i\partial_j-\frac{\delta_{ij}}{d-1}\vec\nabla^2\right)
\left(\Phi+\Psi\right)\Bigg)_{\mathrm{causal}}-\frac12\bigg(\ddot{D}_{ij}+\frac1{d-3}\vec\nabla^2D_{ij}\bigg)_{\mathrm{causal}}\Bigg\}    \notag\\
&+ \frac{8\pi G_\text N}{d-2} \left({}^{(\text{a})}T_{ij}-\frac{\delta_{ij}}{d-1}\left((d-3){}^{(\text{a})}T_{00}
+2{}^{(\text{a})}T_{ll}\right)\right) .
\label{LinearizedWeyl_RelativisticEoS}
\end{align}
The causal portion of the spin-2 sector takes precisely the same form as eq.~\eqref{WeylSpin2Sector_Retarded} but with scale factor given in eq.~\eqref{ScaleFactor} and $(\eta_{\text p},\eta_{\text f})=(0,\infty)$. On the other hand, the causal contributions of the Bardeen scalar potentials in the first line are given by
{\allowdisplaybreaks
\begin{align}
&\Bigg(\left(\partial_i\partial_j-\frac{\delta_{ij}}{d-1}\vec\nabla^2\right)\big(\Phi[\eta,\vec x]+\Psi[\eta,\vec x]\big)\Bigg)_{\mathrm{causal}} \notag\\
&=-8\pi G_\text N \int_{\mathbb{R}^{d-1}} \dd^{d-1}\vec x'\int_0^\infty \dd\eta' \, \left(\frac{a[\eta']}{a[\eta]}\right)^{\frac12\left(q_w+d-2\right)}\!\!w^{-\frac{d-1}2}\Bigg\{\frac{\delta_{ij}}{d-1}
\Bigg(   \bigg(   \ddot G^{(w,+)}_d   -\frac{ (d-2) \left(q_w + d -2 \right) }{q^2_w\eta^2} \, G^{(w,+)}_d  \notag\\
&-a[\eta']^{-\frac12\left(q_w-3d+8\right)}\partial_{\eta'}\bigg(a[\eta']^{-(2d-5)}\partial_{\eta'}
\left(a[\eta']^{\frac12\left(q_w+d-2\right)}\,G^{(w,+)}_d\right)\bigg) -(d-2) \frac{2q_w + 4(d-3)}{q^2_w\eta'^2} \, G^{(w,+)}_d \bigg){}^{(\text{a})}T_{00}[\eta',\vec x']  \notag\\
& + \frac{2q_w-4(d-2)}{q^2_w \eta'^2} \, G^{(w,+)}_d{}^{(\text{a})}T_{ll}[\eta',\vec x'] \Bigg)
+\delta_{ij}w\,a[\eta']^{-\frac12\left(q_w-d+2\right)}\Bigg(\partial_{\eta'}^2\bigg(a[\eta']^{-(d-2)}\partial_{\eta'}
\bigg(\frac{2}{q_w \eta'} a[\eta']^{\frac12\left(q_w+d-2\right)} a[\eta]^{-\frac{d-2}2} \notag\\
& \times \int^\eta_{\eta'}\dd\eta_2\,a[\eta_2]^{(d-2)}   \int^{\eta_2}_{\eta'}\dd\eta_1\,a[\eta_1]^{-\frac {d-2}2}G^{(w,+)}_d\bigg)\bigg){}^{(\text{a})}T_{00}[\eta',\vec x'] + \frac{2}{q_w\eta'}\partial_{\eta'}\bigg(a[\eta']^{-(d-2)}\partial_{\eta'}
\bigg( \frac{2}{q_w\eta'} a[\eta']^{\frac12\left(q_w+d-2\right)}  \notag\\
& \times a[\eta]^{-\frac{d-2}2} \int^\eta_{\eta'}\dd\eta_2\,a[\eta_2]^{(d-2)} \int^{\eta_2}_{\eta'}\dd\eta_1\,a[\eta_1]^{-\frac {d-2}2}G^{(w,+)}_d\bigg)\bigg)\left({}^{(\text{a})}T_{00}[\eta',\vec x']-{}^{(\text{a})}T_{ll}[\eta',\vec x']\right)\Bigg) \notag\\
&+w \Bigg( \! \bigg(a[\eta']^{-\frac12\left(q_w-3d+8\right)}\partial_{\eta'}\bigg(a[\eta']^{-(2d-5)}\partial_{\eta'}
\bigg(a[\eta']^{\frac12\left(q_w+d-2\right)}
a[\eta]^{-\frac{d-2}2}  \!\! \int^\eta_{\eta'}\dd\eta_2\,a[\eta_2]^{(d-2)} \!\! \int^{\eta_2}_{\eta'}\dd\eta_1\,a[\eta_1]^{-\frac {d-2}2}\partial_i\partial_jG^{(w,+)}_d \! \bigg) \! \bigg) \notag\\
& +(d-2) \frac{2q_w + 4(d-3)}{q^2_w\eta'^2} a[\eta]^{-\frac {d-2}2}\int^\eta_{\eta'}\dd\eta_2\,a[\eta_2]^{(d-2)}\int^{\eta_2}_{\eta'}\dd\eta_1\,a[\eta_1]^{-\frac {d-2}2}\partial_i\partial_jG^{(w,+)}_d\bigg){}^{(\text{a})}T_{00}[\eta',\vec x'] \notag\\
&- \frac{2q_w-4(d-2)}{q^2_w \eta'^2} a[\eta]^{-\frac {d-2}2}\int^\eta_{\eta'}\dd\eta_2\,a[\eta_2]^{(d-2)}\int^{\eta_2}_{\eta'}\dd\eta_1\,a[\eta_1]^{-\frac {d-2}2}\partial_i\partial_jG^{(w,+)}_d {}^{(\text{a})}T_{ll}[\eta',\vec x']\Bigg) \notag\\
&-w\,\partial_i\partial_jG^{(w,+)}_d {}^{(\text{a})}T_{00}[\eta',\vec x']
-(d-1)w^2a[\eta']^{-\frac12\left(q_w-d+2\right)}\Bigg(\partial_{\eta'}^2\bigg(a[\eta']^{-(d-2)}\partial_{\eta'}
\bigg(\frac{2}{q_w\eta'}\,a[\eta']^{\frac12\left(q_w+d-2\right)} a[\eta]^{-\frac{d-2}2}\notag\\
&\times \int^\eta_{\eta'}\dd\eta_4\,a[\eta_4]^{(d-2)}\int^{\eta_4}_{\eta'}\dd\eta_3\,a[\eta_3]^{-(d-2)}\int^{\eta_3}_{\eta'}\dd\eta_2\,a[\eta_2]^{(d-2)}
\int^{\eta_2}_{\eta'}\dd\eta_1\,a[\eta_1]^{-\frac {d-2}2}\partial_i\partial_jG^{(w,+)}_d\bigg)\bigg) {}^{(\text{a})}T_{00}[\eta',\vec x'] \notag\\
&+\frac{2}{q_w\eta'}\partial_{\eta'}\bigg(a[\eta']^{-(d-2)}\partial_{\eta'}
\bigg( \frac{2}{q_w\eta'}\,a[\eta']^{\frac12\left(q_w+d-2\right)} a[\eta]^{-\frac{d-2}2}\int^\eta_{\eta'}\dd\eta_4\,a[\eta_4]^{(d-2)}\int^{\eta_4}_{\eta'}\dd\eta_3\,a[\eta_3]^{-(d-2)}
\int^{\eta_3}_{\eta'}\dd\eta_2\,a[\eta_2]^{(d-2)}\notag\\
&\times
\int^{\eta_2}_{\eta'}\dd\eta_1\,a[\eta_1]^{-\frac {d-2}2}\partial_i\partial_jG^{(w,+)}_d\bigg)\bigg)\left( {}^{(\text{a})}T_{00}[\eta',\vec x']- {}^{(\text{a})}T_{ll}[\eta',\vec x']\right)  \Bigg)  \Bigg\} .
\label{WeylScalarSector_Retarded}
\end{align}}%
Observe that eq.~\eqref{WeylScalarSector_Retarded} is fully determined by the massless scalar Green's function $G^{(w,+)}_d\big[\eta,\eta';\textstyle{\frac {R}{\sqrt w}}\big]$, and whose contributions to $\delta_1 C^i{}_{0j0}$ are therefore restricted either on or inside the acoustic cone.\footnote{Once again, notice that in eq.~\eqref{WeylScalarSector_Retarded}, we have switched from $C^{(w)}_{1,d}$ to $G^{(w,+)}_d$, where the local terms are incurred in the conversion between their second time derivatives, $\ddot G^{(w,+)}_{d}=-w^{\frac{d-1}2}\delta^{(d)}[x-x']-\Theta[T]\ddot C^{(w)}_{1,d}$ and $\partial_{\eta'}^2G^{(w,+)}_{d}=-w^{\frac{d-1}2}\delta^{(d)}[x-x']-\Theta[T]\partial_{\eta'}^2C^{(w)}_{1,d}$. Similarly, the Minkowski counterpart of $\delta_1 C^i{}_{0j0}$ is again recovered by letting $a\to1$, replacing $G^{(g,+)}_d$ with $G^+_d$ in eq.~\eqref{WeylSpin2Sector_Retarded}, and assuming no scalar contributions of eq.~\eqref{WeylScalarSector_Retarded} to eq.~\eqref{LinearizedWeyl_RelativisticEoS}.}

The physically intriguing feature of the relativistic $w$ result in eq.~\eqref{LinearizedWeyl_RelativisticEoS} is that, not only do spin-2 gravitons contribute to $\delta_1 C^i{}_{0j0}$, it appears the Bardeen scalars do so as well. To be sure, however, it would be prudent to obtain a more explicit expression for eq.~\eqref{WeylScalarSector_Retarded}. In an upcoming work, we hope to tackle this important step towards a more comprehensive understanding of gravitational tidal forces within a cosmological setting.

\section{Summary, Discussions, and Future Directions}
\label{Section_Summary}

In this paper, we have sought to clarify the physical roles played by the TT and tt gravitational perturbations; as well as the analogous issues for the spin-1 photon. Even though the TT GW is gauge-invariant -- it remains un-altered under an infinitesimal change in coordinates -- it is acausal. Since the bulk of the paper involves heavy mathematical analysis for arbitrary dimensions and cosmological equations-of-state, we summarize here the 4D Minkowski case for the reader's convenience.

Let us begin with the electromagnetic sector. The gauge-invariant 4D transverse photon, which obeys $\partial_i \alpha_i = 0$, cannot be a standalone observable because its solution
\begin{align}
\alpha_i[\eta,\vec{x}]
= - \int_{\mathbb{R}^{3,1}} \dd^4 x' G_{ij}^+[T,\vec{R}] J_j[\eta',\vec{x}']
\qquad\qquad (T \equiv \eta - \eta' \text{ and } \vec{R} \equiv \vec{x}-\vec{x}')
\end{align}
receives contributions from portions of the electric current $J_j[\eta',\vec{x}']$ lying outside the past lightcone of the observer at $(\eta,\vec{x})$. This is because, the photon retarded Green's function
\begin{align}
G_{ij}^+[T,\vec{R}]
&= - \delta_{ij} \frac{\delta[T-R]}{4\pi R} - \frac{1}{4\pi} \partial_{i} \partial_{j} \left( \Theta[T-R] + \frac{T}{R} \Theta[T] \Theta[R-T] \right)
\end{align}
contains an acausal portion: $G_{ij}^{(+,\text{acausal})} = - (4\pi)^{-1} \Theta[T] \Theta[R-T] T \partial_{i} \partial_{j} R^{-1}$. However, since this acausal term of the photon Green's function is part of a pure gradient, namely $-(4\pi)^{-1} \partial_i \partial_j \big(\Theta[T] \Theta[R-T] T/R\big)$, the magnetic field $F_{ij} = \partial_i \alpha_j - \partial_j \alpha_i$ -- which involves its curl -- is therefore entirely causal. The electric field, on the other hand, is the sum of the photon velocity $\dot{\alpha}_i$ and the gradient of the gauge-invariant scalar potential,
\begin{align}
\label{4D_F0i}
F_{0i} = \dot{\alpha}_i + \partial_i \Phi .
\end{align}
In detail, integration-by-parts (IBPs) and the conservation of the electric current yield
\begin{align}
\dot{\alpha}_i[\eta,\vec x] &= \big(\dot{\alpha}_i\big)_\text{causal} + \int_{\mathbb{R}^{3}} \dd^3 \vec{x}' \partial_i \frac{J_0[\eta,\vec{x}']}{4\pi R} ;
\label{4D_PhotonVelocity}
\end{align}
where we have denoted the causal part of the photon velocity as
\begin{align}
\label{4D_PhotonVelocity_Causal}
\big(\dot{\alpha}_i\big)_\text{causal}
&\equiv \int_{\mathbb{R}^{3,1}} \dd^4 x' \left( \partial_\eta \frac{\delta[T-R]}{4\pi R} J_i[\eta',\vec x'] - \partial_i \frac{\delta[T-R]}{4\pi R} J_0[\eta',\vec x'] \right) .
\end{align}
Whereas, the gradient of the scalar potential is
\begin{align}
\label{4D_GradScalar}
\partial_i \Phi &= -\int_{\mathbb{R}^{3}} \dd^3 \vec{x}' \partial_i \frac{J_0[\eta,\vec{x}']}{4\pi R} .
\end{align}
Adding equations \eqref{4D_PhotonVelocity} and \eqref{4D_GradScalar} to obtain eq.~\eqref{4D_F0i}, we see that the sole purpose of $\Phi$ -- as far as electromagnetic fields are concerned -- is to cancel the acausal part of the photon velocity. This in turn ensures the electric field of eq.~\eqref{4D_F0i}, in a given inertial frame, is the causal part of the latter; namely, $F_{0i} = (\dot{\alpha}_i)_\text{causal}$.

We have also pointed out: upon quantization, these transverse massless spin-1 photon operators violate microcausality, because their Green's functions do not vanish at spacelike intervals.

The transverse-traceless graviton, which obeys $\partial_i D_{ij} = 0 =  \delta^{ij} D_{ij}$, also cannot be a standalone observable -- for very similar reasons as its acausal transverse photon counterpart. Its solution
\begin{align}
D_{ij}[\eta,\vec{x}]
= -16\pi \GN \int_{\mathbb{R}^{3,1}} \dd^4 x' G^+_{ij mn}[T,\vec{R}] T_{mn}[\eta',\vec{x}']
\end{align}
receives signals from regions of the stress tensor $T_{mn}[\eta',\vec{x}']$ outside the past null cone of the observer at $(\eta,\vec{x})$. For, this TT graviton Green's function reads
{\allowdisplaybreaks\begin{align}
G^+_{ijmn}[T,\vec R]=\,& -\bigg(\delta_{m(i}\delta_{j)n}-\frac{\delta_{ij}\delta_{mn}}{2}\bigg)\frac{\delta[T-R]}{4\pi R} \nonumber\\
&\qquad
- \frac1{4\pi}\bigg(\delta_{m(i}\partial_{j)}\partial_n+
\delta_{n(i}\partial_{j)}\partial_m -\frac{\delta_{ij}\partial_m\partial_n-\delta_{mn}\partial_i\partial_j}{2}
\bigg) \left(\Theta[T-R] + \frac{T}{R} \Theta[T]\Theta[R-T] \right) \nonumber\\ &\qquad
+\frac1{48\pi} \partial_i\partial_j\partial_m\partial_n \left( \Theta[T-R]\left(R^2 + 3T^2\right)+ \Theta[T] \Theta[R-T]\,\frac{3R^2T + T^3}{ R} \right) ;
\end{align}}%
which contains the acausal terms $ - \Theta[T]\Theta[R-T]  T \big( \delta_{m(i}\partial_{j)}\partial_n + \delta_{n(i}\partial_{j)}\partial_m -(1/2) ( \delta_{ij}\partial_m\partial_n-\delta_{mn}\partial_i\partial_j ) \big) (4\pi R)^{-1}$ and $\Theta[T]\Theta[R-T] \partial_i\partial_j\partial_m\partial_n \big\{ (3R^2T + T^3) (48 \pi R)^{-1} \big\}$.

On the other hand, the dominant physical tidal forces ought to be encoded within the linearized Riemann tensor, which in turn involves all the gauge-invariant variables, not just the spin-2 graviton. In particular, its $0i0j$ components -- which are usually associated with the spatial tidal forces in a given nearly-Lorentzian inertial frame -- are
\begin{align}
\delta_1 R_{0i 0j}
= \delta_{ij} \ddot{\Psi} + \partial_i\partial_j\Phi + \partial_{(i}\dot{V}_{j)} - \frac12\ddot{D}_{ij} .
\end{align}
As one may expect from the preceding discussion for the spin-1 photon, the linearized Riemann $\delta_1 R_{0i0j}$ really only depends on the causal part of the spin-2 graviton acceleration:
\begin{align}
\delta_1 R_{0i 0j} = - \frac12 \big(\ddot{D}_{ij}\big)_\text{causal} ;
\end{align}
where, upon IBPs and invoking the conservation of the energy-momentum-shear-stress tensor, we have
{\allowdisplaybreaks
	\begin{align}
	\big(\ddot D_{ij}\big)_\text{causal} = \,& 4 G_\text N\int_{\mathbb{R}^{3,1}} \dd^4 x'\,\Bigg\{ \partial_\eta^2 \frac{\delta[T-R]}{ R} \left(^{(\text{a})}T_{ij}[\eta',\vec x']+\frac{\delta_{ij}}{2}\left({}^{(\text{a})}T_{00}[\eta',\vec x']-{}^{(\text{a})}T_{ll}[\eta',\vec x']\right)\right) \nonumber\\
	&
	-2\partial_\eta \partial_{(i} \frac{\delta[T-R]}{ R} {}^{(\text{a})}T_{j)0}[\eta',\vec x']
	+\frac{1}{2}\partial_i\partial_j \frac{\delta[T-R]}{ R} \left(\,{}^{(\text{a})}T_{00}[\eta',\vec x']+{}^{(\text{a})}T_{ll}[\eta',\vec x']\right)\Bigg\} .
	\end{align}}%
The sole purpose of the rest of the gauge-invariant variables $(\Psi,\Phi,V_i)$, as far as the $\delta_1 R_{0i0j}$ components are concerned, is to cancel the acausal part of the graviton acceleration. Moreover, all of them are needed to ensure causality. We may verify these claims by simply comparing the following expressions.
{\allowdisplaybreaks
	\begin{align}
	\ddot D_{ij}[\eta,\vec x]&= \big(\ddot D_{ij}\big)_\text{causal} \nonumber\\
	&- 4 G_\text N\int_{\mathbb{R}^{3,1}} \dd^{3}\vec{x}'\,\Bigg\{\! -2\partial_m\partial_{(i} \frac{1}{ R}
	{}^{(\text{a})}T_{j)m}[\eta,\vec{x}']+\frac1{2}\bigg(\delta_{ij}\partial_m\partial_n  \frac{1}{ R}{}^{(\text{a})}T_{mn}[\eta,\vec{x}'] \notag\\
	&+\partial_i\partial_j  \frac{1}{ R} \left({}^{(\text{a})}T_{00}[\eta,\vec{x}']+{}^{(\text{a})}T_{ll}[\eta,\vec{x}']\right)\bigg) + \frac{1}{4}\partial_i\partial_j\partial_m\partial_n R {}^{(\text{a})}T_{mn}[\eta,\vec{x}']\Bigg\}
	\end{align}}%
{\allowdisplaybreaks
	\begin{align}
	\ddot \Psi[\eta,\vec x]=\,& - 4 G_\text N \int_{\mathbb{R}^{3}} \dd^{3}\vec{x}'\,  \frac{1}{4} \partial_m \partial_n \frac{1}{ R} {}^{(\text{a})}T_{mn}[\eta,\vec x'],  \nonumber \\
	\partial_i\partial_j\Phi[\eta,\vec x]=\,& -4 G_\text N \int_{\mathbb{R}^{3}} \dd^{3}\vec{x}'\, \frac{1}{4} \bigg(\partial_i\partial_j \frac{1}{ R} {}^{(\text{a})}T_{00}[\eta,\vec{x}'] +
	\partial_i\partial_j \frac{1}{ R} {}^{(\text{a})}T_{ll}[\eta,\vec{x}']
	- \frac32 \partial_i\partial_j\partial_m\partial_n R {}^{(\text{a})}T_{mn}[\eta,\vec{x}'] \bigg), \nonumber\\
	\partial_{(i}\dot{V}_{j)}[\eta,\vec x]  =\,&  4 G_\text N \int_{\mathbb{R}^{3}} \dd^{3}\vec{x}'\left( \partial_m\partial_{(i} \frac{1}{ R}  {}^{(\text{a})}T_{j)m}[\eta,\vec{x}']
	- \frac12\partial_i\partial_j\partial_m\partial_n
	R {}^{(\text{a})}T_{mn}[\eta,\vec{x}']\right) .
	\end{align}}%

These massless spin-2 graviton fields, upon quantization, would violate microcausality, because their Green's functions do not vanish at spacelike intervals.

It is worth highlighting, we are not asserting that relativists are computing gravitational wave-forms wrongly. In the far zone, $|\omega| r \gg 1$, we have shown that the distortion of space due to GWs (at finite frequencies) do reduce to the tt ones gotten by performing a local-in-space projection of the (de Donder gauge) spatial perturbations $\chi_{ij}$. These tt GWs, as opposed to their TT counterparts, are in fact the ones computed in the gravitational literature. On the other hand, within this far zone, these tt GWs in fact coincide with the TT ones, because the acausal parts of the latter begin at higher orders in $1/(\omega r)$.

In a cosmology driven by a relativistic fluid, we have uncovered tentative evidence that the Bardeen scalar potentials contribute to gravitational tidal forces, and their wave-like solutions could therefore be legitimately dubbed `scalar gravitational waves' in this sense. More work would be required to confirm or deny this \cite{ChuLiuInPrep}. Nonetheless, if the Bardeen $\Psi$ and $\Phi$ are indeed an integral part of cosmological GWs, we hope this work constitutes the first step towards illuminating not only their associated GW patterns but also potential scalar GW memory effects.

Let us end on a more speculative note. Even though the TT graviton is acausal and cannot be a standalone observable within classical physics, it may be produced quantum mechanically -- and independently of other gauge-invariant perturbations -- during a (still hypothetical) exponentially expanding phase of the early universe.\footnote{We note in passing: when constructing the quantum field theory of photons and gravitons in 4D flat spacetime, Weinberg \cite{Weinberg:1965rz} had to add non-local terms to the Hamiltonian of these massless spin-1 and spin-2 gauge theories in order to preserve the Lorentz invariance of the $S-$matrix.} On the other hand, we have also pointed out that the quantum operators associated with both the free massless spin-1 and spin-2 particles violate micro-causality. Is it possible to exploit this violation to ascertain whether $B-$modes in the Cosmic Microwave Background, if we ever detect them, were truly engendered by {\it quantum} fluctuations of spacetime itself? Or, for the photon case, are there laboratory experiments involving quantum generation of photons that could not only serve as analogs to the inflationary scenario, but also allow the quantum nature of their production mechanism to be probed directly?

\section{Acknowledgments}

YZC is supported by the Ministry of Science and Technology of the R.O.C. under the grant 106-2112-M-008-024-MY3. He also thanks the organizers of the 2018 TGWG Conference (at Tamkang University, Taipei), where he gave a talk about this work; as well as Lior Burko and Feng-Li Lin for questions and Jan Steinhoff for constructive comments. YWL wishes to gratefully acknowledge the support of Shih-Chang Lee and the Department of Physics at National Central University during the completion of the most part of this work. He is supported by the Ministry of Science and Technology of the R.O.C. under Project No.~MOST 108-2811-M-008-503. They would also like to acknowledge Bonga, R\'{a}cz, Flanagan and Hughes for e-mail exchanges.



\begin{thebibliography}{99}

\bibitem{Racz:2009nq}
I.~R\'{a}cz,
``Gravitational radiation and isotropic change of the spatial geometry,''
arXiv:0912.0128 [gr-qc].
	
\bibitem{Ashtekar:2017wgq}
A.~Ashtekar and B.~Bonga,
``On the ambiguity in the notion of transverse traceless modes of gravitational waves,''
Gen.\ Rel.\ Grav.\  {\bf 49}, no. 9, 122 (2017)
doi:10.1007/s10714-017-2290-z
[arXiv:1707.09914 [gr-qc]].

\bibitem{Ashtekar:2017ydh}
A.~Ashtekar and B.~Bonga,
``On a basic conceptual confusion in gravitational radiation theory,''
Class.\ Quant.\ Grav.\  {\bf 34}, no. 20, 20LT01 (2017)
doi:10.1088/1361-6382/aa88e2
[arXiv:1707.07729 [gr-qc]].

\bibitem{Frenkel:2014cra}
A.~Frenkel and I.~R\'{a}cz,
``On the use of projection operators in electrodynamics,''
Eur.\ J.\ Phys.\  {\bf 36}, no. 1, 015022 (2015)
doi:10.1088/0143-0807/36/1/015022
[arXiv:1407.7396 [math-ph]].

\bibitem{BrillGoodman}
O.~L.~Brill and B.~Goodman,
``Causality in the Coulomb Gauge,"
Am. J. Phys. {\bf 35}, 832 (1967)


\bibitem{ThorneBlandfordBook}
Thorne, Kip S.; Blandford, Roger D.,
``Modern Classical Physics Optics, Fluids, Plasmas, Elasticity, Relativity, and Statistical Physics,"
Princeton University Press, 2017

\bibitem{Flanagan:2005yc}
E.~E.~Flanagan and S.~A.~Hughes,
``The Basics of gravitational wave theory,''
New J.\ Phys.\  {\bf 7}, 204 (2005)
doi:10.1088/1367-2630/7/1/204
[gr-qc/0501041].

\bibitem{Misner:1974qy}
C.~W.~Misner, K.~S.~Thorne and J.~A.~Wheeler,
``Gravitation,''
San Francisco 1973, 1279p

\bibitem{Weinberg:1972kfs}
S.~Weinberg,
``Gravitation and Cosmology : Principles and Applications of the General Theory of Relativity,''
Wiley, 1972

\bibitem{Schutz:1985jx}
B.~F.~Schutz,
``A First Course In General Relativity,''
Cambridge University Press, 1985

\bibitem{Maggiore:1900zz}
M.~Maggiore,
``Gravitational Waves. Vol. 1: Theory and Experiments,''
Oxford University Press, 2007


\bibitem{Chu:2016ngc}
Y.~Z.~Chu,
``More On Cosmological Gravitational Waves And Their Memories,''
Class.\ Quant.\ Grav.\  {\bf 34}, no. 19, 194001 (2017)
doi:10.1088/1361-6382/aa8392
[arXiv:1611.00018 [gr-qc]].

\bibitem{ChuLiuInPrep}
Y.-Z.~Chu and Y.-W.~Liu, {\it in progress}


\bibitem{Chu:2016qxp}
Y.~Z.~Chu,
``Gravitational Wave Memory In dS$_{4+2n}$ and 4D Cosmology,''
Class.\ Quant.\ Grav.\  {\bf 34}, no. 3, 035009 (2017)
doi:10.1088/1361-6382/34/3/035009
[arXiv:1603.00151 [gr-qc]].

\bibitem{Weinberg:1965rz}
S.~Weinberg,
``Photons and gravitons in perturbation theory: Derivation of Maxwell's and Einstein's equations,''
Phys.\ Rev.\  {\bf 138}, B988 (1965).
doi:10.1103/PhysRev.138.B988

\end{thebibliography}
\end{document}